%% file: thesis.tex
\documentclass[12pt,a4paper]{iiscthes}  

\usepackage{psfig}
\usepackage{eceps}
\usepackage{epsfig}
\usepackage{rotating}
\usepackage{graphicx}  
\usepackage{dsfont}
\usepackage{amsmath}
\usepackage{amsbsy}
\usepackage{amssymb}
\usepackage{amsfonts}
\usepackage{dsfont}

\newif\ifpdf\ifx\pdfoutput\undefined
\pdffalse\else\pdfoutput=1\pdftrue\fi

\newtheorem{theorem}{Theorem}[section]
\newtheorem{lemma}{Lemma}[section]
\newtheorem{corollary}{Corollary}[section]
\newtheorem{definition}{Definition}[section]

\newtheorem{property}{Property}[section]

\def\QED{\mbox{\rule[0pt]{1.5ex}{1.5ex}}}

\def\tw{\textwidth}

\def\Q{\hspace*{\fill}~\QED\par\endtrivlist\unskip}

\begin{document}

\begin{frontmatter}
\title{Scheduling for Stable and Reliable Communication \\
over Multiaccess Channels \\ and Degraded Broadcast Channels}
\author{KCV Kalyanarama Sesha Sayee}
\submitdate{July 2006}
\degree{phd}
\dept{Department of Electrical Communication Engineering}
\enggfaculty
\phd
\iisclogotrue 
\figurespagetrue
\maketitle
\input{declaration}
\prefacesection{Publications}
\input{publication}

\prefacesection{Abstract}
\input{abstract}
\prefacesection{Acknowledgements}
\input{ack}
\tableofcontents
\listoffigures
\end{frontmatter}

\pagestyle{bfheadings}
\setcounter{page}{1}
\input{introduction}
\input{chapter1}

\input{independentDECODING}
\input{jointDECODING}

\input{broadcastChannel}
\input{conclusion}

\appendix
\input{Appendix-A}

\input{Appendix-B}
\bibliographystyle{plain}
\bibliography{info}

\end{document}

%% file: declaration.tex

\hspace{-1.5cm}\letterhead{\logo}

\begin{center}
\textbf{DECLARATION}
\end{center}

I hereby declare that the work reported in this thesis is entirely
original. It was carried out by me in the Department of Electrical
Communication Engineering, Indian Institute of Science, Bangalore, under
the supervision of Professor Utpal Mukherji.  I further declare that it
has not formed the basis of any degree, diploma, membership,
associateship or similar title of any University or Institution.

\vspace{2cm}

$\hfill \mbox{KCV Kalyanarama Sesha Sayee}$

\vspace{-0.3cm}
$\hfill \mbox{Dated: \ 24 \ July, 2006}$

\vspace{2cm}
\begin{flushleft}
$\mbox{Dr. Utpal Mukherji}$\\
\noindent
$\mbox{ Associate Professor} $

\end{flushleft}
\newpage

%% file: publication.tex
\begin{enumerate}
\item	``Multi-access Poisson Traffic Communication with Random Coding,
Independent Decoding and Unequal Powers", \emph{Proceedings of
Information Theory Workshop}, page 220, October 2002.

\item	``Stability of Scheduled Multi-access Communication over
Quasi-static Flat Fading Channels with Random Coding and Independent
Decoding," \emph{2005 IEEE International Symposium on Information
Theory}, pages 2261-2265, September 2005.

\item	``Stability of Scheduled Message Communication over Degraded
Broadcast Channels", \emph{2006 IEEE International Symposium on Information
Theory}, pages 2764-2768, July 2006.

\item	``A Multiclass Discrete-Time Processor-Sharing Queueing Model
for Scheduled Message Communication over Multiaccess Channels with Joint
Maximum-Likelihood Decoding'', {\bf Submitted to 2006 Allerton
Conference.}

\item	``Scheduling for Stable and Reliable Communication over Multiaccess
Channels and Degraded Broadcast Channels,'' {\bf To be communicated to IEEE
Transactions on Information Theory.} 
\end{enumerate}

%% file: abstract.tex
Information-theoretic arguments  focus on modeling the reliability of
information transmission, assuming availability of infinite data at
sources, thus ignoring randomness in message generation  times at the
respective sources.  However, in information transport networks, not
only is reliable transmission important, but also stability, i.e.,
finiteness of mean delay incurred by messages from the time of
generation to the time of successful reception. Usually, delay analysis
is done separately using queueing-theoretic arguments, whereas reliable
information transmission is studied using information theory.  In this
thesis, we investigate these two important aspects of data communication
jointly by suitably combining models from these two fields. In
particular, we model scheduled communication of messages , that arrive
in a random process, (i) over multiaccess channels, with  either
independent decoding or joint decoding, and (ii) over degraded broadcast
channels. The scheduling policies proposed permit up to a certain
maximum number of messages for simultaneous transmission.

In the first part of the thesis, we develop a multi-class discrete-time
processor-sharing queueing model, and then investigate the stability of
this queue. In particular, we model the queue by a discrete-time Markov
chain defined on a countable state space, and then establish (i) a
sufficient condition for $c$-regularity of the chain, and hence positive
recurrence and finiteness of stationary mean of the function $c$ of the
state, and (ii) a sufficient condition for transience of the chain.
These stability results form the basis for the conclusions drawn in the
thesis.

The second part of the thesis is on multiaccess communication with
random message arrivals. In the context of independent decoding, we
assume that messages can be classified into a fixed number of classes,
each of which specifies a combination of received signal power, message
length, and target probability of decoding error.  Each message is
encoded independently and decoded independently. In the context of joint
decoding, we assume that messages can be classified into a fixed number
of classes, each of which specifies a message length, and for each of
which there is a message queue. From each queue, some number of messages
are encoded jointly, and received at a signal power corresponding to the
queue. The messages are decoded jointly across all queues with a target
probability of joint decoding error.

For both independent decoding and joint decoding, we derive respective
discrete-time multiclass processor-sharing queueing models assuming the
corresponding information-theoretic models for the underlying
communication process.  Then, for both the decoding schemes, we (i)
derive respective outer bounds to the stability region of message
arrival rate vectors achievable by the class of stationary scheduling
policies, (ii) show for any message arrival rate vector that satisfies
the outer bound, that there exists a stationary ``state-independent''
policy that results in a stable system for the corresponding message
arrival process, and  (iii) show that the stability region of
information arrival rate vectors, in the limit of large message lengths,
equals an appropriate information-theoretic capacity region for
independent decoding, and equals the information-theoretic capacity
region for joint decoding.  For independent decoding, we identify a
class of stationary scheduling policies, for which we show that the
stability region in the limit of large maximum number of simultaneous
transmissions is independent of the received signal powers, and each of
which achieves a spectral efficiency of 1 nat/s/Hz in the limit of large
message lengths.

In the third and last part of the thesis, we show that the queueing
model developed for multiaccess channels with joint decoding can be used
to model communication over degraded broadcast channels, with
superposition encoding and successive decoding across all queues. We
then show respective results (i), (ii), and (iii), stated above.

%% file: ack.tex
I wish to thank Prof. Utpal Mukherji for having kindly agreed to
supervise my Ph.D. thesis, for the complete freedom given to me after
the initial part of my research work, and for the countless hours of
time that he gave me during the initial phase of my Ph.D. work. The
research discussions with him were so insightful that, after each
discussion I was left with wondering how come I didn't think the way he
thought. He has asked all the right questions and often saved me from
slipping into mathematical obscurities.

Another individual who deeply influenced me is Prof. Anurag Kumar, who
supervised my M.Sc (Engg.) thesis. I wish to thank him for the courses
he taught and for having provided  excellent lab facilities. His remarks
during one of my departmental talks has in fact lead to the title of one
of the chapters of this thesis. His advise that I should fill my head
with research even while I play tennis has done both good and bad!. I am
especially grateful to him for the employment he provided me for over six
months in 2001.

Over the last 7 years, I had the fortune of attending many good courses
offered by the Mathematics department and TIFR. I have especially
enjoyed various courses offered by Prof. Vittal Rao and the Real
Analysis course offered by Prof. Mythily Ramaswamy.

My friendship with G. Manjunath, Munish Goyal, C. Venkat (paavi), and
Pattu has been the longest and they all have made  my stay in the campus
more enjoyable. 

I wish to thank my Andhra friends Syam Prasad, Praveen, Ravi, Srinu,
Krishna, Hema, and recently, Suresh, Moorthy, and N. Gangadhar for
having provided wonderful company in the campus. I thank Syam Prasad
especially for the moral support he gave me during my not so good times.

I take this opportunity to thank my past lab-mates Dr. Arzad Alam
Kherani, Dr.  Aditya Karnik, Dr. Munish Goyal, and the present
lab-mates Avijit Chakraborthy, R. Venkat, K. Prem Kumar, Mallesh,
and Manoj.

Right from the beginning of my stay here in the campus, playing tennis
became an integral part of my daily routine. I thank Prof. Narasimhan,
Probal, Venkat and Kulkarni for making my evenings more enjoyable.

I also thank SVR Anand, Chandrika and Manjunath. My frequent visits to
coffee board with Anand have been more memorable. I take this opportunity
to thank R. Srinivasa Murthy, ECE office, for being available to me
whenever I needed his assistance.

I am grateful to my wife Sobha for putting up with a moody and nocturnal
graduate student for the last one year. Besides supporting me
financially, she has given me all the understanding that I could desire.
It is my father who inspired me and provided the right environment for
my academic growth. His unfailing confidence in my abilities has in fact
made this thesis a reality. My mother, brother, sister, and
brother-in-law all have extended their valuable support to me during the
last 7 years of my stay here.

%% file: introduction.tex
\chapter{Introduction}
\label{ch:introduction}

Information-theoretic arguments  focus on modeling the reliability of
information transmission, assuming availability of infinite data at
sources, thus ignoring randomness in message generation  times at the
respective sources.  However, in information transport networks, not
only is reliable transmission important, but also stability, i.e.,
finiteness of mean delay incurred by messages from the time of
generation to the time of successful reception. Usually, delay analysis
is done separately using queueing-theoretic arguments, whereas reliable
information transmission is studied using information theory.  In his
seminal paper~\cite{GAL-JRN-ITTRAN} published in 1985, Gallager
explains:

\begin{quote}
For the last ten years there have been at least three bodies of research
on multiaccess channels, each proceeding in virtual isolation from the
others and each using totally different models. The objective here is to
contrast these bodies of work and to give some perspective on what is
needed to provide some unification between the areas. We shall refer to
the three areas as collision resolution, multiaccess information theory,
and spread spectrum.
\end{quote}
Then he goes on to say that $\cdots$
\begin{quote}
Collision resolution research has always focused on the bursty arrivals
of messages and the interference between transmitters, but has generally
ignored the noise. More generally, this approach ignores the underlying
communication process, assuming only that a message transmission is
correctly received in the absence of collision and incorrectly received
otherwise.
\end{quote}
\begin{quote}
$\cdots$ In this approach (multiaccess information theory), the noise
and interference aspects of the multiaccess channel are appropriately
modeled, but the random arrivals of the messages are ignored. 
\end{quote}
In this thesis, we investigate these two important aspects of data
communication jointly by suitably combining models from these two
fields. In particular, we model scheduled communication of messages ,
that arrive in a random process, (i) over multiaccess channels, with
either independent decoding or joint decoding, and (ii) over degraded
broadcast channels. The scheduling policies proposed permit up to a
certain maximum number of messages for simultaneous transmission.

\section{Problem Formulation}
The following three multiuser communication scenarios {\bf S1}, {\bf
S2}, and {\bf S3}, are investigated in the dissertation.
\begin{enumerate}
\item[({\bf S1})]       There are $J \geq 1$ transmitting stations
communicating to a central receiver. We assume that the transmitting
stations and central receiver are time synchronized, and that there
exists an error-free feedback channel over which the central receiver
broadcasts pertinent control information to the transmitting stations.
For $1 \leq j \leq J$ and integers $M_{j} \geq 2$, let messages of
length $\ln M_{j}$ nats arrive at the $j$th station in a batch arrival
process with i.i.d. batch sizes.  The transmitter at the $j$th
transmitting station is assigned an average transmit power $P_{j}$.  At
transmitting station $j$, there is a block encoder that \emph{jointly}
encodes at most $s_{j} \geq 1$ packets into a code word. The central
receiver decodes the received word using joint maximum-likelihood
decoding.  It is required that the received word be decoded with an
expected error probability of at most $p_{e}$. Then, we ask the
question: for what message arrival rates at the respective transmitting
stations is the message communication system stable, i.e., messages are
decoded in finite mean time?.

\item[({\bf S2})]       There is a base station and potentially an
unlimited number of terminals communicating to the base station. We say
that a terminal is active if it has a packet to transmit, otherwise the
terminal is said to be inactive. Terminals become active at random
times.  We assume independent and identically distributed quasi-static
flat fades from the active terminals to the base station in the
respective channels. With this assumption, there is an i.i.d.
multiplicative gain in the channel from each terminal to the base
station.  Thus, for a given multiplicative gain $\gamma$, a message
signal of average transmit power $P$ will be received at the signal
power $|\gamma|^{2}P$.  We assume that the multiplicative gain $\gamma$
is known to the base station, and is a random variable that has $J$
possible values for magnitude. Each message has to be decoded with
expected error probability at most $p_{e}$. Then, again we ask the
question: at what rates can the terminals become active so that, when
joint maximum-likelihood decoding is performed at the base station,
messages are decoded in finite mean time.

\item[({\bf S3})]       There are $J$ message sources co-located with a
transmitter, and an equal number of receivers. Each source wishes to
communicate information to its receiver such that the expected decoding
error probability at the $j$th receiver is at most $p_{ej}$. The
transmitter encodes messages from these sources using superposition
encoding, and broadcasts the encoded signal over a degraded broadcast
channel (DBC). At each receiver, the decoder maps its received signal
into an estimate of the message intended for it.  Messages are generated
at random times at each source.  Again we ask the question: at what
rates can these sources communicate reliably and stably to their
respective receivers.  
\end{enumerate}

\section{Summary of Related Work}
The first effort, in the direction pointed out by Gallager in his
seminal work~\cite{GAL-JRN-ITTRAN}, that the random generation of
messages and the subsequent reliable information transmission must be
understood in a unified framework, was reported in
~\cite{TelGal-JRN-JSAC} and~\cite{Tel-THESIS}. The framework considered
therein is as follows.  Consider a multiaccess message communication
system.  Requests for message transmissions over a flat bandpass
additive white Gaussian noise (AWGN) channel arrive according to  a
Poisson process.  Messages, upon arrival, are given immediate access,
i.e., each transmitter transmits its signal, starting at its message
arrival time.  Existence of an errorless, delayless, control channel in
each direction is assumed.  Upon noticing the presence of a message
request, the receiver and the transmitter agree upon a Gaussian codebook
with Gaussian codewords of zero mean, equal power $P$, and uniform power
spectral density over a narrow frequency band of width $W$, following
the random coding principle. Messages are selected from a finite message
alphabet of size $M$. Each message has to be transmitted reliably with
reliability quantified by the tolerable message decoding error
probability, $p_{e}$.

Signal propagation delays in the system are assumed to be negligible. It
is assumed that the receiver operates with full knowledge of the message
alphabet sizes and received signal powers of all transmitters in the
system.  The receiver decodes the message of a transmitter by treating
the signals from other transmitters as independent additive noise. This
is the independent decoding assumption for decoding of a message at the
receiver. The receiver uses the codebook of a transmitter in maximum
likelihood decoding of the message of the transmitter.  Each message
transmits its signal for a random duration determined by the receiver. A
stopping rule is used by the receiver to stop transmission of the signal
for a message.  The stopping rule  ensures that the expected probability
of error in decoding a message in the system is less than the tolerable
value $p_e$.

In ~\cite{TelGal-JRN-JSAC},~\cite{Tel-THESIS} this random-coded
multi-access system is then modelled as a continuous-time
processor-sharing queue in which the transmitters are ``customers'' that
are ``served'' by the receiver.  The processor-sharing model is then
analyzed to determine the stability condition and the mean delays
experienced by the incoming messages, by determining steady-state
probabilities.

\section{Modelling}
In this thesis, we first generalize the
framework~\cite{TelGal-JRN-JSAC},~\cite{Tel-THESIS} that models both the
random message arrivals and the subsequent reliable communication by
suitably combining techniques from queueing theory and information
theory. We then investigate message communication over (i) multiaccess
channels with independent decoding and joint maximum-likelihood
decoding, and (ii) degraded broadcast channels, in that general
framework. In the following, we point out the ways in which our model
differs from the model in~\cite{TelGal-JRN-JSAC},~\cite{Tel-THESIS}, and
then summarize the contributions made in the thesis.

\noindent
\begin{enumerate}
\item	Signal transmissions from different transmitters may be
received at different signal powers at the receiver

\begin{itemize}
\item[]	Unlike in the model~\cite{TelGal-JRN-JSAC},~\cite{Tel-THESIS}, we
allow independent and identically distributed quasi-static flat fades
from the transmitters to the receiver in the respective channels. With
this assumption, there is an i.i.d.  multiplicative gain in the channel
from each transmitter to the receiver.  Thus, for a given multiplicative
gain $\gamma$, a message signal of average power $P$ will be received at
the signal power $|\gamma|^{2}P$.  We assume that the multiplicative
gain $\gamma$ is known to the receiver, and is modelled as a random
variable that has a finite number of finite possible magnitudes.
\end{itemize}

\noindent
\item	The receiver schedules message transmissions

\begin{itemize}
\item[]	We assume that messages can be classified into a fixed number of
classes each of which specifies a combination of received signal power,
message length, and target probability of decoding error. The notion of
message classes naturally leads to scheduling, i.e., the question of how
many messages of each class are to be scheduled at a given time.  Due to
the complexity involved in joint maximum-likelihood decoding of an
arbitrary number of messages, we restrict the receiver to schedule upto
at most some finite number of messages at a time. Also, in the case of
DBC, the complexity involved in joint superposition encoding of an
arbitrary number of messages again leads us to the same restriction.
Specifically, the scheduling policies proposed in this thesis permit up
to a certain maximum number $\mathsf{K} \geq 1$ of messages for
simultaneous transmission.  
\end{itemize}

\noindent
\item	Decoding techniques

\begin{itemize}
\item[]	In~\cite{TelGal-JRN-JSAC},~\cite{Tel-THESIS}, independent
maximum-likelihood decoding of signal transmissions is proposed. In
independent decoding, a message signal is decoded treating all other
signal transmissions, if any, as interference. Thus the effective noise
is the sum of additive Gaussian noise plus other active signal
transmissions present in the system. We should observe here that
scheduling at most  a finite number $\mathsf{K}$ of messages for
simultaneous transmission has the effect of limiting the interference as
seen by any message transmission, i.e., $\mathsf{K}-1$ transmissions can
interfere.  Since independent decoding is suboptimal, we also consider
\emph{joint maximum-likelihood decoding} of signal transmissions across
all message classes with a common target probability of joint decoding
error. Some previous work with joint decoding is reported
in~\cite{Tel-CONF-ITW1995}. But, to our knowledge, the details of this
work have not been published elsewhere. We believe that the decoding
technique proposed in~\cite{Tel-CONF-ITW1995} is complicated for the
following reason: to decode $n$ active transmitters, one has to create
$(2^{n}-1)$ joint decoders, one for each non-empty subset of the set of
active transmitters, and this number increases exponentially with $n$.
With scheduling being made part of our model and with the restriction on
the maximum number of simultaneous message transmissions, a message is
decoded by only one joint decoder.
\end{itemize}
\end{enumerate}

In our model, the communication channel is a quasi-static flat bandpass
AWGN channel of bandwidth $W$.  Formally, $y(t) = \gamma x(t) + N(t)$,
where the input $x(t)$ is a band-limited zero-mean Gaussian process of
bandwidth $W$ and average power $P$, $\gamma$ is a finite valued real
random variable, and $N(t)$ is a white Gaussian noise process
\emph{independent} of the input $x(t)$ with noise power spectral density
$\frac{N_{0}}{2}$. The analysis of the model starts with first replacing
this continuous-time model by an equivalent discrete-time model. This is
done by first replacing the continuous-time model by an \emph{equivalent
continuous-time complex low-pass} model. In this model, the inputs and
outputs are continuous-time complex low-pass signals of bandwidth
$\frac{W}{2}$, and the channel is a low-pass filter of bandwidth
$\frac{W}{2}$. Then using the sampling theorem for low-pass signals, we
sample the input and output at the rate of $W$ complex samples per
second, or $2W$ real samples per second.  Thus we reduce the
continuous-time AWGN channel to a sequence of independent \emph{complex
baseband} channels $i$ such that the model for the $i$th channel is
$y_{i} = \gamma x_{i} + n_{i}$. The input $x_{i} = \left( x_{i}^{(I)},
x_{i}^{(Q)} \right)$ is circular symmetric complex Gaussian random
variable with the distribution $\mathcal{C}\mathcal{N}\left(0,
\frac{P}{2W} \right)$, and noise $n_{i} = \left(n_{i}^{(I)}, n_{i}^{(Q)}
\right)$ is circular symmetric complex Gaussian random variable with the
distribution $\mathcal{C} \mathcal{N}\left(0, \frac{N_{0}}{2} \right)$.
In this thesis, we analyze communication over stationary discrete
memoryless channel (DMC) with complex inputs and outputs.

\section{Contributions}
\noindent
For multiaccess communication with independent decoding, we show the
following.
\begin{enumerate}
\item   For finite message lengths, inner bounds and outer bounds to the
message arrival rate stability region are derived. For arrival rates
within the inner bounds, we show finiteness of the stationary mean for
the number of messages in the system and hence for message delay. For
the case of equal received signal powers, with sufficiently large SNR,
the stability threshold increases with decreasing maximum number of
simultaneous transmissions.

\item   When message lengths are large , the information arrival rate
stability region has an interpretation in terms of interference-limited
information-theoretic capacities. For the case of equal received powers,
this stability threshold is the interference-limited
information-theoretic capacity.

\item   We propose a class of stationary policies called
state-independent scheduling policies, and then show that they achieve
this asymptotic information arrival rate stability region.

\item   In the asymptotic limit corresponding to immediate access, the
stability region for non-idling scheduling policies is shown to be
identical irrespective of received signal powers. This observation
essentially shows that transmit power control is not needed.  We show
that, in the asymptotic limit corresponding to immediate access and
large message lengths, a spectral efficiency of 1 nat/s/Hz is achievable
with non-idling scheduling policies.
\end{enumerate}
For multiaccess communication with joint maximum-likelihood decoding and
degraded broadcast channels with joint superposition encoding and
successive decoding, we show the following.
\begin{enumerate}
\item	For scheduled message communication over  (i) multiaccess
channels with joint maximum-likelihood decoding, and (ii) degraded
broadcast channel, we derive outerbounds to the respective stability
region of message arrival rate vectors achievable by the class of
stationary scheduling policies.  Then we show for any message arrival
rate vector that satisfies the outer bound, that there exists a
stationary ``state-independent'' scheduling policy that results in a
stable system for the corresponding message arrival processes.

\item	We show that the stability region of information arrival rate
vectors for (i) multiaccess communication with joint maximum-likelihood
decoding, and (ii) message communication over degraded broadcast
channels, with superposition encoding and successive decoding, are the
information-theoretic capacity regions, respectively.  For example,
consider a rate vector $r = (r_{1}, r_{2})$ in the two-user multiaccess
achievable rate region corresponding to an arbitrary product probability
distribution $Q_{1}(x_{1})Q_{2}(x_{2})$.  Then we show that there exists
a scheduling strategy that tells us how many messages of what length
from each information source must be scheduled together so that, when
the $j$th source, $j=1,2$, generates information at the rate $r_{j}$
information units/time unit, the corresponding message communication
system is stable, i.e., messages are decoded in finite mean time.
\end{enumerate}

\section{A Note to the Reader}
Chapter~\ref{ch:chapter1} can be read independent of everything else in
this thesis. But the purposes of the model introduced and the results
obtained in that chapter become apparent in subsequent chapters.
Chapter~\ref{ch:independent decoding} and Chapter~\ref{ch:joint
decoding} can be read to a large extent independently. Except for
Section~\ref{section:information theoretic model degraded broadcast
channel}, Chapter~\ref{ch:degraded broadcast channel} should be read
only after Chapter~\ref{ch:joint decoding} is read.

%% file: chapter1.tex
\chapter{A MultiClass Discrete-Time Processor-Sharing Queue}
\label{ch:chapter1}

In this chapter, we develop a multi-class discrete-time
processor-sharing queueing model, and then investigate the stability of
this queue. In particular, we model the queue by a discrete-time Markov
chain defined on a countable state space, and then establish (i) a
sufficient condition for $c$-regularity~\cite{Mey-paper} of the chain,
and hence positive recurrence and finiteness of stationary mean of the
function $c$ of the state, and (ii) a sufficient condition for
transience of the chain.  These stability results form the basis for the
conclusions drawn in the following chapters.

\section{The Queueing Model}
\label{section:The Queueing Model chapter 1}
Consider a queueing system consisting of $J$ queues operating in
discrete-time.  Time is divided into equal length time intervals called
time-slots. Each queue is fed by an independent, stationary, batch
arrival process with i.i.d. batch sizes for different time-slots. Let
the random variable $A_{j}$ represent the number of customers that
arrive in any time-slot to the $j$th queue.  Assume that the pmf
$\Pr(A_{j} = k) = p_{j}(k), k \geq 0$, has finite moments
$\mathds{E}A_{j}$ and $\mathds{E}A_{j}^{2}$. $\{A_{j}; 1 \leq j \leq J
\}$ are independent random variables. Let $\mathds{E}A =
(\mathds{E}A_{1}, \mathds{E}A_{2}, \ldots, \mathds{E}A_{J}) \in
\mathbb{R}_{+}^{J}$ be the vector of arrival rates of the arrival
processes.

We assume that a customer that arrives at the system has associated with
it a class that gives sufficient information about the customer.  A
customer requires an amount of service and the service requirement is
modeled as a \emph{constant} quantity.  Let $S_{j}$ denote the service
requirement of a class-$j$ customer.  When the cumulative service
quantum that a customer has received equals or exceeds its service
requirement, the customer leaves the system.  To define the state of the
system we keep track of the residual service requirement of each
customer present in the system. We shall define by $\alpha_{j} = \left(
x_{j}(1), x_{j}(2), \ldots, x_{j}\left( n_{j}(\alpha) \right) \right)$
the state of queue $j$, where $n_{j}(\alpha)$ denotes the number of
class-$j$ customers in state $\alpha$ and $x_{j}(k)$ gives the residual
service requirement of $k$th customer of class-$j$ in state $\alpha$,
and by 
\begin{eqnarray}
\label{eq:state vector definition}
\alpha &=& (\alpha_{1}, \alpha_{2}, \ldots, \alpha_{J}) 
\end{eqnarray}
the state of the system.  Obviously, $n(\alpha) = \sum_{j=1}^{J}
n_{j}(\alpha)$ is the total number of customers in the system state
$\alpha$. 

Further, we assume that the server schedules certain numbers of
customers of the various classes for providing simultaneous service in
each time-slot using a preemptive resume scheduling policy.  We define a
schedule by a non-negative integer vector $s = (s_{1}, s_{2},\ldots,
s_{J}).$  For an integer $\mathsf{K} \geq 1$, we define the set
$\mathcal{S}_{\mathsf{K}} = \left\{ s: 0 \leq \sum_{j=1}^{J} s_{j} \leq
\mathsf{K} \right\}$ to be the set of all schedules that schedule at
most $\mathsf{K}$ customers in each time-slot.  We say that schedule $s$
is feasible in state $\alpha$ if $s_{j} \leq n_{j}(\alpha)$, for $j = 1,
2, \ldots, J$. We implement a feasible schedule $s$ by serving the first
$s_{j}$ customers at the head of queue-$j$, for $1 \leq j \leq J$. A
schedule $s$ such that $s_{j} = 0$ for $1 \leq j \leq J$ is called the
\emph{empty} schedule.

In this thesis we consider only stationary scheduling policies. We
define a stationary deterministic scheduling policy $\omega$ as a
mapping $\left\{ \omega: \mathcal{X} \rightarrow
\mathcal{S}_{\mathsf{K}} \right\}$ for which the schedule
$\omega(\alpha)$  is feasible in state $\alpha$ for all $\alpha$. For a
stationary randomized policy $\omega$, $\omega(\alpha)$ is then a random
variable taking values in $\mathcal{S}_{\mathsf{K}}$ with some
probability distribution $\left\{ p_{\alpha}^{\omega}(s); s \in
\mathcal{S}_{\mathsf{K}} \right\}$. We note here that deterministic
policies are special cases of randomized scheduling policies.

Define $\phi_{j}(s) \geq 0$ to be the service quantum~\footnote{We use
the convention that $\phi_{j}(s) = 0$ if $s_{j} = 0$.} that a class-$j$
customer is \emph{eligible} to receive under the schedule $s$.  We allow
for the possibility that the service quantum made available to a
customer in a time-slot may be more than the residual service
requirement of the customer, and in that case, the amount by which the
offered service quantum is in excess of the customer residual
requirement goes unused. Since $s_{j}$ customers of class-$j$ are
provided service under the schedule $s$, a total service quantum upto
$s_{j}\phi_{j}(s)$ can be provided to class-$j$ customers. But, this
could be interpreted as being equivalent to completing service of up to
$\frac{s_{j}\phi_{j}(s)}{S_{j}}$ customers in a time-slot under the
schedule $s$. Thus, for each $s \in \mathcal{S}_{\mathsf{K}}$, we define
a rate vector $r(s) = \left( r_{1}(s), r_{2}(s), \ldots, r_{J}(s)
\right)$, in units of customers/time-slot, where $r_{j}(s) =
\frac{s_{j}\phi_{j}(s)}{S_{j}}$ for $1 \leq j \leq J$.

Here we make the observation that the service quantum $\phi_{j}(s)$ made
available to a class-$j$ customer can vary with the schedule $s$, and
also, the fraction of the total service quantum made available to
class-$j$ under the schedule $s$ can vary over the set $\{1, 2, \ldots,
J\}$ of customer classes. In other words, the server is modeled as a
possibly non-uniform processor-sharing server.

Let $\mathcal{X}$ be the countable set of all state vectors $\alpha$.
Countability of the state space $\mathcal{X}$ follows from the fact that
the residual service requirement variable $x$ for any customer class can
take only finitely many values.  Let $\{X_{n}; n \geq 0 \}$ be a
discrete-time Markov chain defined over the state space $\mathcal{X}$
with $\left\{ p_{\alpha \alpha^{\prime}}^{\omega}; \alpha,
\alpha^{\prime} \in \mathcal{X} \right\}$ as the state transition
probability matrix under the scheduling policy $\omega$.  In each
time-slot three events take place.  Just after the beginning of a
time-slot, first, the system state $\alpha$ is read, next the schedule
$\omega(\alpha)$ is implemented and finally, new arrivals, if any,  are
admitted into the system.

\section{Stability for the Underlying Markov Chain}
Let $\{ X_{n}; n \geq 0 \}$ be a positive recurrent discrete-time Markov
chain defined on a countable state space $\mathcal{X}$ with stationary
probability measure $\{ \mu(\alpha); \alpha \in \mathcal{X} \}$. Let $c$
be a \emph{bounded} function on $\mathcal{X}$. Then the ensemble average
of $c$, $\mathds{E}^{\mu}(c) = \sum_{\alpha} c(\alpha) \mu(\alpha)$,
exists and for every initial condition $\alpha \in \mathcal{X}$,
\[
\lim_{n \rightarrow \infty} \mathds{E}_{\alpha} \left[ c \left( X_{n}
\right) \right] = \mathds{E}^{\mu}(c)
\]
We can relax the boundedness assumption made on $c$ and still have the
ensemble average $\mathds{E}^{\mu}(c)$ exist if the Markov chain under
consideration is ``$c$-regular''~\cite{MeyTwe-BOOK}~\cite{Mey-paper}.
\begin{definition}[$c$-\emph{Regularity}]
Let $c: \mathcal{X} \rightarrow [1, \infty ]$ be a function defined on
the state space $\mathcal{X}$. A set $Y \in \mathcal{X}$ is called
$c$-\emph{regular} if, for each non-empty subset $Y^{\prime} \in
\mathcal{X}$,
\[
\sup_{\alpha \in Y} \mathds{E}_{\alpha} \left[
\sum_{n=0}^{\tau_{Y^{\prime}}-1} c \left( X_{n} \right) \right] <
\infty,
\]
where $\tau_{Y^{\prime}}$ is the first passage time to the set
$Y^{\prime}$.  The Markov-chain chain $\{ X_{n}; n \geq 0 \}$ itself is
called $c$-\emph{regular} if there is a countable cover of $\mathcal{X}$
with $c$-regular sets.  
\end{definition}

A $c$-regular chain is positive recurrent and possesses an invariant
probability measure $\mu$ satisfying $\mathds{E}^{\mu}(c) < \infty$.  An
approach to establish $c$-regularity for a Markov chain with transition
probability matrix $\{p^{\omega}_{\alpha, \alpha^{\prime}}; \alpha,
\alpha^{\prime} \in \mathcal{X} \}$ is to (i) construct a Lyapunov
function $V: \mathcal{X} \rightarrow \mathbb{R}_{+}$, (ii) find a $c$
function that is \emph{near-monotone}, i.e., $\{\alpha \in \mathcal{X}:
c(\alpha) \leq \eta \}$ is finite for any $\eta < \sup_{\alpha}
c(\alpha)$, and (iii) find a constant $\mathsf{J} < \sup_{\alpha}
c(\alpha)$ such that 
\begin{eqnarray*}
\Delta V(\alpha) &\equiv& \sum_{\alpha^{\prime} \in \mathcal{X}} 
V \left( \alpha^{\prime} \right) p^{\omega}_{\alpha, \alpha^{\prime}}
- V(\alpha) \leq -c(\alpha) + \mathsf{J}
\end{eqnarray*} 
Then, under the above assumptions, Theorem~10.3 in~\cite{Mey-paper}
guarantees that the Markov chain $\{X_{n}; n \geq 0 \}$ is $c$-regular.
The notion of stability that we consider in this thesis, for a
discrete-time Markov chain defined on a countable state space, and
underlying the queueing model, is given in the following definition.
\begin{definition}
\label{def:stability}
We say that a discrete-time countable-state Markov chain $\{X_{n}; n
\geq 0 \}$ under a stationary scheduling policy $\omega$ is (i) stable
if it is positive recurrent and has finite stationary mean for the
number of customers in the system, and (ii) unstable if it is transient.
\Q
\end{definition}

\section{Sufficient Conditions for $c$-Regularity and Transience for the
Queueing Model}
\subsection{A Sufficient Condition for $c$-Regularity}
In what follows in the present chapter and in subsequent chapters, we
will need to consider non-negative real valued functions defined on the
state space $\mathcal{X}$ that possess the property~\ref{"c" Property}
stated below.  To state that property, we first fix a scheduling policy
$\omega$. Let $a_{j}$ (a sample value for the random variable $A_{j}$)
new customers arrive to the $j$th queue in any time-slot, and let $a =
(a_{1}, a_{2}, \ldots, a_{J}) \in \mathbb{Z}_{+}^{J}$.  In this thesis,
we assume customer arrival processes in the future to be independent of
the current state of the system.  For each customer-class $j$, we assume
the existence of a real-valued deterministic function $h_{j}^{\omega}:
\mathcal{X} \rightarrow \mathbb{R}_{+}$, defined on the state space
$\mathcal{X}$ and with the following property: assume that $a_{j}$
class-$j$ customers arrive in state $\alpha$ and that the feasible
schedule $s$ is implemented in the state $\alpha$. As a result, assume
that the chain moves to the state $\alpha^{\prime}$. Then
$h_{j}^{\omega} \left( \alpha^{\prime}
\right)$ can be written as
\begin{eqnarray}
\label{eq:h-function}
h_{j}^{\omega}\left( \alpha^{\prime} \right) &=& h_{j}^{\omega}(\alpha) + 
f_{j}(a) - g_{j}(\alpha, s)
\end{eqnarray}
where $f_{j}(a) $ and $g_{j}(\alpha, s)$ are non-negative numbers.  When
this property holds we say that, as the chain makes the transition
$\alpha \rightarrow \alpha^{\prime}$, $h_{j}^{\omega}(\alpha)$ first
decreases by $g_{j}(\alpha, s)$, due to the delivery of service quantum,
and then increases by $f_{j}(a)$ , due to new customer arrivals, thus
increasing by the net amount $f_{j}(a) - g_{j}( \alpha, s)$.  
\begin{property}
\label{"c" Property}
Let $h_{j}^{\omega}: \mathcal{X} \rightarrow \mathbb{R}_{+}$. For a
given stationary scheduling policy $\omega$, customer arrival processes
$ \left\{ A_{j}; 1 \leq j \leq J \right\}$ , the function
$h_{j}^{\omega}$ satisfies
\[
h_{j}^{\omega} \left( \alpha^{\prime} \right) = h_{j}^{\omega}(\alpha) + 
f_{j}(a) - g_{j}(\alpha, s),
\]
where $f_{j}(a) \geq 0$, and $g_{j}(\alpha, s)  \geq 0$ depends on the
precise specification of the scheduling policy $\omega$. \Q
\end{property}
Define $p^{\omega}_{\alpha} = \left\{ p_{\alpha}^{\omega} (s); \alpha
\in \mathcal{X}\; \mbox{and}\; s \in \mathcal{S}_{\mathsf{K}} \right\}$
to be a probability distribution on the set of schedules
$\mathcal{S}_{\mathsf{K}}$, and indexed by the state $\alpha$. The
interpretation for $p_{\alpha}^{\omega} (s)$ is that the schedule $s$
gets implemented in the state $\alpha$ with probability
$p_{\alpha}^{\omega}(s)$.  

Further, we assume that, for each class-$j$, a partition $\left\{
\mathcal{H}_{j}, \mathcal{H}_{j}^{c} \right\}$ of the state space
$\mathcal{X}$ exists such that $\sup_{\alpha \in \mathcal{H}_{j}}
h_{j}^{\omega}(\alpha)$ is finite.  Define the following set of
partitions: for $1 \leq j \leq J$,
\begin{eqnarray*}
\Xi_{j}^{\omega} &=& \left\{ 
\left\{ \mathcal{H}_{j}, \mathcal{H}_{j}^{c} \right\} : 
\sup_{\alpha \in \mathcal{H}_{j}} h_{j}^{\omega}(\alpha) \;\mbox{is finite}
\right\}
\end{eqnarray*}
Define $g_{j} (\alpha) = \sum_{s \in \mathcal{S}_{\mathsf{K}}} g_{j}
(\alpha, s)  p_{\alpha}^{\omega} (s)$ , and the following
two quantities
\begin{eqnarray}
\label{eq:inner bound}
g_{j}^{\omega} &=& \sup_{\Xi_{j}^{\omega}} \inf_{\alpha \in \mathcal{H}_{j}^{c}}
g_{j} (\alpha) \\
\label{eq:outer bound}
G_{j}^{\omega} &=& \inf_{\Xi_{j}^{\omega}} \sup_{\alpha \in \mathcal{H}_{j}^{c}}
g_{j} (\alpha)
\end{eqnarray}
Equivalently, for any arbitrarily small $\epsilon_{j} > 0$, there exists
a partition $\left\{ \mathcal{H}_{j}, \mathcal{H}_{j}^{c} \right\} \in
\Xi_{j}^{\omega}$ such that $\inf_{\alpha \in \mathcal{H}_{j}^{c}}
g_{j}(\alpha) > g_{j}^{\omega} - \epsilon_{j}$.  That is,
$g_{j}^{\omega} - \epsilon_{j} < g_{j}(\alpha)$ for $\alpha \in
\mathcal{H}_{j}^{c}$.  Similarly, for an arbitrarily small $\delta_{j} >
0$, there exists a partition $\left\{ \mathcal{H}_{j},
\mathcal{H}_{j}^{c} \right\} \in \Xi_{j}^{\omega}$ such that
$g_{j}(\alpha) < G_{j}^{\omega} + \delta_{j}$ for $\alpha \in
\mathcal{H}_{j}^{c}$. Define the expected increase in the function
$h_{j}^{\omega}$, in any state $\alpha$, due to customer arrivals as
\begin{eqnarray} 
\label{eq:expected increase}
\mathds{E}f_{j} &= \sum_{a}f_{j}(a) p(a),
\end{eqnarray}
where $p(a) = \prod_{j=1}^{J}p_{j}(a_{j})$, 
and assume that $\mathds{E}f_{j}$ and the second moment
$\mathds{E}f_{j}^{2}$ are finite. We assume that \newline $\sup_{\alpha
\in \mathcal{X}} \sum_{s \in \mathcal{S}_{\mathsf{K}}} g_{j}^{2}(\alpha,
s) p^{\omega}_{\alpha}(s) < \infty$. This assumption is valid in most
practical situations, because the total service quantum available to any
queue in any time-slot is bounded.  Assume that, for each real number
$\eta$ and for each $j$, $1 \leq j \leq J$, the set
$Z_{j}^{\omega}(\eta) = \{ \alpha: h_{j}^{\omega}(\alpha) \leq \eta \}$
is such that $n_{j}(\alpha)$ is bounded on $Z_{j}(\eta)$.  Then we prove
the following simple observation.
\begin{lemma}
\label{lemma:near-monotone property}
Define $h^{\omega}(\alpha) = \sum_{j=1}^{J} h_{j}^{\omega}(\alpha)$.
Then for each real number $\eta$, the set $Z^{\omega}(\eta) = \{ \alpha:
h^{\omega}(\alpha) \leq \eta \}$ is a finite set. Hence the
function $h^{\omega}$ is near-monotone.
\end{lemma}
\begin{proof}
Since $h^{\omega}(\alpha) \leq \eta \Rightarrow h_{j}^{\omega}(\alpha)
\leq \eta$, for each $j$, we have that $n_{j}(\alpha)$, for each $j$, is
bounded on the set $Z^{\omega}(\eta)$. From the definition of state
$\alpha$, since each residual service requirement variable can assume
only finitely many values, it follows that $Z^{\omega}(\eta)$ is a
finite set.
\end{proof}
Let $V: \mathcal{X} \rightarrow \mathbb{R}_{+}$ be a Lyapunov function
defined on $\mathcal{X}$. Let $\mathcal{R}^{\omega} \subset
\mathbb{R}_{+}^{J}$ be the set of customer arrival rate vectors
$\mathds{E}A$ such that, for $\mathds{E}A \in \mathcal{R}^{\omega}$, the
Markov chain under the scheduling policy $\omega$ is stable.  Define the
set $\mathcal{R}_{in}^{\omega} \in \mathbb{R}_{+}^{J}$ such that
$\mathcal{R}_{in}^{\omega} \subseteq \mathcal{R}^{\omega}$.
\begin{lemma}
\label{lemma:c-regularity}
For $1 \leq j \leq J$, assume that (i) $h_{j}^{\omega}: \mathcal{X}
\rightarrow \mathbb{R}_{+}$ is a real-valued function defined on the
state space $\mathcal{X}$, and (ii) $h_{j}^{\omega}$ possesses
property~\ref{"c" Property}.  Assume that the function \footnote{It is
possible that a near-monotone function $c$ can arise as a sum of
\emph{non} near-monotone functions $h_{j}^{\omega}$.} $c(\alpha) = 1 +
\sum_{j=1}^{J} h_{j}^{\omega}(\alpha)$ is near-monotone and
\[
V(\alpha) = \sum_{j=1}^{J} \frac{ \left[ h_{j}^{\omega}(\alpha)
\right]^{2}}{2 \left( g_{j}^{\omega} - \mathds{E}f_{j} \right)},
\]
where $g_{j}^{\omega}$ and $\mathds{E}f_{j}$ are as defined
in~(\ref{eq:inner bound}) and~(\ref{eq:expected increase}) respectively.
Then the Markov chain $\{ X_{n}; n \geq 0 \}$ for the queueing model is
$c$-regular if, for each $j$, $\mathds{E}f_{j} < g_{j}^{\omega}$.
\Q
\end{lemma}
\begin{proof}
For each $j$, define a function $V_{j}^{\prime}$, as
$V_{j}^{\prime}(\alpha) = \left[ h_{j}^{\omega} (\alpha)\right]^{2}$.
The expected drift in $V_{j}^{\prime}$ in an arbitrary state $\alpha$,
conditioned on the schedule $s$ to be implemented in the state $\alpha$,
is 
\begin{eqnarray*}
\Delta V_{j}^{\prime} (\alpha|s) &=& 
\sum_{a} \left( 
\left[ h_{j}^{\omega}\left( \alpha^{\prime} \right) \right]^{2} - 
\left[ h_{j}^{\omega}(\alpha) \right]^{2} \right) 
p(a) \\
&=& h_{j}^{\omega}(\alpha) \sum_{\alpha^{\prime}}  
-2\left( g_{j}(\alpha,s) - f_{j}(a) \right) p(a) + \sum_{a} 
\left( f_{j}(a) - g_{j}(\alpha,s) \right)^{2} p(a) \\
&=& -2 \left( g_{j}(\alpha, s) - \mathds{E}f_{j} \right)
h_{j}^{\omega}(\alpha) + \left( \mathds{E}f_{j}^{2} - 
2 g_{j}(\alpha, s) \mathds{E}f_{j} +  g_{j}^{2}(\alpha, s) \right)
\end{eqnarray*}
The unconditional expected drift $\Delta V_{j}^{\prime}(\alpha)$ is then
written as
\begin{eqnarray*}
\Delta V_{j}^{\prime}(\alpha) &=& \sum_{s \in \mathcal{S}_{\mathsf{K}}} 
\Delta V_{j}^{\prime} (\alpha |s) p_{\alpha}^{\omega} (s)
= -2 \left( g_{j}(\alpha) - \mathds{E}f_{j} \right) 
h_{j}^{\omega}(\alpha) +  \left( \mathds{E}f_{j}^{2} - 
2 g_{j} (\alpha) \mathds{E}f_{j} +  g_{j}^{\prime}(\alpha) \right),
\end{eqnarray*}
where $g_{j}^{\prime}(\alpha) = \sum_{s \in \mathcal{S}_{\mathsf{K}}}
g_{j}^{2}(\alpha, s) p^{\omega}_{\alpha}(s) < \infty$.

Let $\epsilon_{j}$ be an arbitrary small positive real number. Then
there exists a partition $\{\mathcal{H}_{j}, \mathcal{H}_{j}^{c}\} \in
\Xi_{j}^{\omega}$ such that, for $\alpha \in \mathcal{H}^{c}_{j}$, the
unconditional expected drift is bounded above as 
\begin{eqnarray*}
\Delta V_{j}^{\prime}(\alpha) & \leq & -2 \left( 
g_{j}^{\omega} - \epsilon_{j} - \mathds{E}f_{j}
\right) h_{j}^{\omega}(\alpha) +  
\left( \mathds{E}f_{j}^{2} - 2 g_{j} (\alpha) \mathds{E}f_{j}
+  g_{j}^{\prime}(\alpha) \right)
\end{eqnarray*}
Assume that $\mathds{E}f_{j} < g_{j}^{\omega} - \epsilon_{j}$ , and then
scale the function $V_{j}^{\prime}(\alpha)$ as
\[
V_{j}(\alpha) = \frac{V_{j}^{\prime}(\alpha)}
{2 \left( g_{j}^{\omega} - \epsilon_{j} -\mathds{E}f_{j} \right)}
\]
Then, for $\alpha \in \mathcal{H}^{c}_{j}$ the expected drift in
$V_{j}(\alpha)$ can be bounded above as
\[
\Delta V_{j}(\alpha) \leq  -h_{j}^{\omega}(\alpha) +
\mathsf{J}_{j}(1),\;\mbox{for}\;
\mathsf{J}_{j}(1) = 
\frac{ 
\left[ \mathds{E}f_{j}^{2} - 2 g_{j} (\alpha) \mathds{E}f_{j}
+  g_{j}^{\prime}(\alpha) \right]
} 
{ 2 \left( g_{j}^{\omega} - \epsilon_{j} - \mathds{E}f_{j} \right)}
\]
Since $h_{j}^{\omega}(\alpha)$ is bounded for $\alpha \in
\mathcal{H}_{j}$, and $\mathds{E}f_{j}^{2}$ and $g_{j}^{\prime}(\alpha)$
are finite for $1 \leq j \leq J$, therefore, for $\alpha \in
\mathcal{H}_{j}$, $\Delta V_{j}(\alpha) \leq -h_{j}^{\omega}(\alpha) +
\mathsf{J}_{j}(2)$, where $\mathsf{J}_{j}(2)$ is a finite constant.
Hence, for all $\alpha \in \mathcal{X}$, $\Delta V_{j}(\alpha)  \leq
-h_{j}^{\omega}(\alpha) + \mathsf{J}_{j}$, where $\mathsf{J}_{j} = \max
\{ \mathsf{J}_{j}(1), \mathsf{J}_{j}(2) \}$. Define $V(\alpha) =
\sum_{j=1}^{J} V_{j}(\alpha)$. Then, for $\alpha \in \mathcal{X}$,
\[
\Delta V(\alpha) = \sum_{j=1}^{J} \Delta V_{j} (\alpha)
\leq  \sum_{j=1}^{J} \left( -h_{j}^{\omega}(\alpha) + 
\mathsf{J}_{j} \right) =  -c(\alpha) + \mathsf{J},
\]
where $\mathsf{J} = 1 + \sum_{j=1}^{J} \mathsf{J}_{j}$. Since the arguments
presented above are valid for any arbitrarily small $\epsilon_{j} > 0$,
we conclude that the Markov chain is $c$-regular when $\mathds{E}f_{j} <
g_{j}^{\omega}$ for $1 \leq j \leq J$. As a consequence, the Markov chain
is positive recurrent and the function $c(\alpha)$ of the state $\alpha$
has finite stationary mean.  
\end{proof}

From Lemma~\ref{lemma:c-regularity}, we see that
$\mathcal{R}^{\omega}_{in} = \left\{ \mathds{E}A: \mathds{E}f_{j} <
g^{\omega}_{j}\; \mbox{for}\; 1 \leq j \leq J \right\}$ is an innerbound
to the stability region $\mathcal{R}^{\omega}$ of message arrival rate
vectors $\mathds{E}A$.

\noindent
{\bf Remark:} Under the conditions in the statement of
Lemma~\ref{lemma:c-regularity}, Foster's criterion~\cite{MeyTwe-BOOK}
also holds. To see this, we first observe that the drift $\Delta
V(\alpha)$ is negative when $c(\alpha) > \mathsf{J}$. Due to
near-monotone property of the $c$-function
(Lemma~\ref{lemma:near-monotone property}), the set of states for which
$c(\alpha) \leq \mathsf{J}$ is a finite set. Hence the drift is strictly
negative except possibly on a finite subset of the state space.

\subsection{A Sufficient Condition for Transience}
In the following theorem, we prove sufficiency of a condition for
transience of the Markov chain $\{X_{n}; n \geq 0\}$ for the queueing
model by showing the existence of a Lyapunov function that satisfies the
theorem for transience stated in Appendix~\ref{appendix:drift theorems}.
\begin{lemma}
\label{lemma:transience}
Let $\omega$ be a stationary scheduling policy. For $1 \leq j \leq J$,
let $h_{j}^{\omega}: \mathcal{X} \rightarrow \mathbb{R}_{+}$ be a
non-negative unbounded function defined on $\mathcal{X}$ such that
$h_{j}^{\omega}$ satisfies property~\ref{"c" Property}.  Then the
Markov chain $\{ X_{n}; n \geq 0 \}$ is transient if $\mathds{E}f_{j} >
G_{j}^{\omega}$ for at least one $j$, where $G_{j}^{\omega}$ is as defined
in~(\ref{eq:outer bound}). \Q
\end{lemma}
\begin{proof}
Define a Lyapunov function $V_{j}$, of the form $V_{j}(\alpha) = 1 -
\theta^{h_{j}^{\omega}(\alpha)}$, where $0 < \theta < 1$. It can be
easily seen that with this choice of $V_{j}$, $V_{j}$ is bounded for all
$\alpha \in \mathcal{X}$. We now show the existence of $\theta =
\theta_{0}$ for which the Lyapunov function satisfies the conditions for
the theorem for transience. For $\alpha \in \mathcal{X}$, the
conditional expected drift $\Delta V_{j} (\alpha|s)$ can be written as 
\begin{eqnarray*}
\Delta V_{j} (\alpha|s) &=& 
\sum_{a} \left[
\left( 1 - \theta^{h_{j}^{\omega}(\alpha^{\prime})} \right) -  
\left( 1 - \theta^{h_{j}^{\omega}(\alpha)} \right) \right] p(a) 
= \theta^{h_{j}^{\omega}(\alpha)} \left[
1 - \sum_{a} \theta^{f_{j}(a) - g_{j}(\alpha, s) }
p(a)  \right]
\end{eqnarray*}
The unconditional expected drift $\Delta V_{j} (\alpha)$, in state
$\alpha$, then becomes
\begin{eqnarray*}
\Delta V_{j}(\alpha) &=& \sum_{s \in \mathcal{S}_{\mathsf{K}}} 
\Delta V_{j} (\alpha| s) p_{\alpha}^{\omega} (s)
= \theta^{h_{j}^{\omega}(\alpha)} \left[
1 - \sum_{s \in \mathcal{S}_{\mathsf{K}}} 
\sum_{\alpha^{\prime}} \theta^{f_{j}(a) - g_{j}(\alpha, s) }
p(a) p_{\alpha}^{\omega} (s) \right]
\end{eqnarray*}
Define $A_{j}(\theta) = \frac{\Delta
V_{j}(\alpha)}{\theta^{h_{j}^{\omega}(\alpha)}} $. We can observe that
$A_{j}(1) = 0$ and 
\begin{eqnarray*}
\left. \frac{dA_{j}(\theta)}{d\theta}\right|_{\theta = 1} &=& 
- \sum_{s \in \mathcal{S}_{\mathsf{K}}} \sum_{a}
\left( f_{j}(a) - g_{j}(\alpha, s) \right) p(a) p_{\alpha}^{\omega}(s)
= g_{j}(\alpha) - \mathds{E}f_{j} \\
\end{eqnarray*}
Given small $\delta_{j} > 0$, there exists a partition $\left\{
\mathcal{H}_{j}, \mathcal{H}_{j}^{c} \right\} \in \Xi_{j}$ such that for
$\alpha \in \mathcal{H}_{j}^{c}$, $g_{j}(\alpha) < G_{j}^{\omega} +
\delta_{j}$ and $\frac{dA_{j}(\theta)}{d\theta}|_{\theta = 1} \leq
G_{j}^{\omega} + \delta_{j} - \mathds{E} f_{j}$.  Let $\mathds{E}f_{j} >
G_{j}^{\omega}+\delta_{j}$.  We then have
$\frac{dA_{j}(\theta)}{d\theta}|_{\theta = 1} < 0$, and hence
$A_{j}(\theta)$ is a decreasing function in $\theta$ at $\theta = 1$.
Therefore, there exists a $0 < \theta_{0} < 1$ such that $\Delta
V_{j}(\alpha) \geq 0$ for $\alpha \in \mathcal{H}_{j}^{c}$. Since
$h_{j}^{\omega}(\alpha)$ is unbounded over the set $\mathcal{H}_{j}^{c}$
and by the choice of the Lyapunov function $V_{j}(\alpha) = 1 -
\theta_{0}^{h_{j}^{\omega}(\alpha)}$, there exists $\alpha^{\prime} \in
\mathcal{H}_{j}^{c}$  such that $ V_{j}\left( \alpha^{\prime} \right) >
\sup_{\alpha \in \mathcal{H}_{j}} V_{j}(\alpha)$.  Thus we have found a
bounded non-negative function $V_{j}(\alpha) = 1 -
\theta_{0}^{h_{j}^{\omega}(\alpha)}$ such that (i) $\Delta V_{j}(\alpha)
\geq 0$ for  $\alpha \in \mathcal{H}_{j}^{c}$, and (ii) there exists an
$\alpha^{\prime} \in \mathcal{H}_{j}^{c}$ such that $V_{j}\left(
\alpha^{\prime} \right) > \sup_{\alpha \in \mathcal{H}_{j}}
V_{j}(\alpha)$.

Since $\delta_{j} > 0$ is an arbitrary small positive number, we
conclude from the theorem for transience stated in the
Appendix~\ref{appendix:drift theorems} that, $\{ X_{n}; n \geq 0 \}$ is
transient for $\mathds{E}f_{j}
> G_{j}^{\omega}$.
\end{proof}
Now, by further assuming that finiteness of stationary mean for
$c(\alpha)$ implies finiteness of stationary mean for the number of
customers $n(\alpha)$ in the system, we state the following theorem on
stability of the queueing model.
\begin{theorem}
\label{th:stability theorem}
For the stationary scheduling policy $\omega$, the Markov chain
$\{X_{n}; n \geq 0\}$ for the queueing model is (i) stable if $
\mathds{E}f_{j} < g_{j}^{\omega}$ for each queue-$j$, and (ii) unstable
if $ \mathds{E}f_{j} > G_{j}^{\omega}$ for at least one queue-$j$. \Q
\end{theorem}
We observe here that the sufficiency result for $c$-regularity stated in
Lemma~\ref{lemma:c-regularity} is defined by $J$ conditions, one for
each customer class. Now we prove a sufficiency result that is defined
by only \emph{one} condition. Assume the existence of a near-monotone
function $c:\mathcal{X} \rightarrow [1,\infty)$ that satisfies
Property~\ref{"c" Property}, i.e., $c\left(\alpha^{\prime}\right) =
c(\alpha) + f_{c}(a)-g_{c}(\alpha, s)$.  Define the expected increase in
$c(\alpha)$ as $\mathds{E}f_{c} = \sum_{a} f_{c}(a)p(a)$, and
$g_{c}(\alpha) = \sum_{s \in \mathcal{S}_{\mathsf{K}}} g_{c}(\alpha, s)
p^{\omega}_{\alpha}(s)$.  Define the set of partitions $\Xi_{c}^{\omega}
= \left\{ \left\{ \mathcal{H}, \mathcal{H}^{c} \right\} : \sup_{\alpha
\in \mathcal{H}} c(\alpha) \;\mbox{is finite} \right\}$, and the two
quantities $g_{c}^{\omega} = \sup_{\Xi^{\omega}_{c}} \inf_{\alpha \in
\mathcal{H}^{c}} g_{c} (\alpha)$ and $G_{j}^{\omega} =
\inf_{\Xi^{\omega}_{c}} \sup_{\alpha \in \mathcal{H}^{c}} g_{c}
(\alpha)$.  Now we state the following Lemma~\ref{lemma:c-regularity
single condition}.

\begin{lemma}
\label{lemma:c-regularity single condition}
Let $\omega$ be a stationary scheduling policy.
\begin{enumerate}
\item[(A)]	Assume the existence of a near-monotone function $c:
\mathcal{X} \rightarrow [1,\infty]$ that satisfies Property~\ref{"c"
Property}. Define the Lyapunov function $V(\alpha) =
\frac{c^{2}(\alpha)}{2\left(g_{c}^{\omega} - \mathds{E}f_{c} \right)}$.
Then the Markov chain $\{X_{n}; n \geq 0\}$ for the queueing model is
$c$-regular if, $\mathds{E}f_{c} < g_{c}^{\omega}$.

\item[(B)]	Let $c$ be a non-negative unbounded function defined on
$\mathcal{X}$ such that $c$ satisfies property~\ref{"c" Property}.  Then
the Markov chain $\{ X_{n}; n \geq 0 \}$ for the queueing model is
transient if $\mathds{E}f_{c} > G_{c}^{\omega}$. \Q
\end{enumerate}
\end{lemma} 
\begin{proof}
Proof of Part$(A)$ is similar to the proof of
Lemma~\ref{lemma:c-regularity} and proof of Part$(B)$ is similar to the
proof of Lemma~\ref{lemma:transience} except that we now have
$V^{\prime}$, $c$, and $\Xi_{c}^{\omega}$ in places of
$V_{j}^{\prime}$, $h^{\omega}_{j}$, and $\Xi_{j}^{\omega}$,
respectively, of Lemma~\ref{lemma:c-regularity}.

\end{proof}

\section{A General Outer Bound to The Stability Region of Customer
Arrival Rate Vectors, $\mathds{E}A$} 
\label{section:outer bound chapter 1}
In this section, we derive an outerbound $\mathcal{R}_{out} \in
\mathbb{R}_{+}^{J}$ to the region $\bigcup_{\omega}
\mathcal{R}^{\omega}$ of customer arrival rate vectors $\mathds{E}A$ for
each of which there exists a stationary scheduling policy such that the
corresponding Markov chain $\{X_{n}; n \geq 0\}$ is stable.  Consider
customer arrival processes $\{A_{j}; 1 \leq j \leq J \}$ and a
stationary scheduling policy $\omega$ that schedules at most
$\mathsf{K}$ messages for simultaneous transmission.  Let
$\mathcal{R}_{out}$ denote the convex hull of the set of rate vectors
$\left\{ r(s); s \in \mathcal{S}_{\mathsf{K}} \right\}$.
\begin{theorem}
\label{th:outer bound chapter 1}
Let the Markov chain $\{ X_{n}; n \geq 0 \}$, for the customer arrival
processes $\{A_{j}; 1 \leq j \leq J\}$ and the stationary scheduling
policy $\omega$, be stable.  Then $\mathds{E}A \in \mathcal{R}_{out}$.
\Q
\end{theorem}
\begin{proof}
We first observe that, for finite $S_{j}$, finiteness of stationary mean
for the total number of customers in the system implies finiteness of
stationary mean for the total residual service requirement in the
system. Hence, for any customer class-$j$ and under stationary
conditions, the average service requirement $S_{j}\mathds{E}A_{j}$ that
arrives in a time-slot equals the average amount by which residual
service requirement decreases in that time-slot due to service received.
Let $\left\{ \pi_{\mathsf{K}}(s); s \in \mathcal{S}_{\mathsf{K}}
\right\}$ be the probability measure induced on
$\mathcal{S}_{\mathsf{K}}$ under stationary conditions, for arrival
processes $\left\{ A_{j}; 1 \leq j \leq J \right\}$ and stationary
scheduling policy $\omega$.  Since each of $s_{j}$ class-$j$ customers
can receive a service quantum up to $\phi_{j}(s)$ under the schedule
$s$, we have $S_{j} \mathds{E}A_{j} \leq \sum_{ s \in
\mathcal{S}_{\mathsf{K}} } \pi_{\mathsf{K}}(s) s_{j}\phi_{j}(s)$ and
hence $\mathds{E}A \in \mathcal{R}_{out}$.
\end{proof}

%% file: independentDECODING.tex
\chapter{Multiaccess Communication with Independent Decoding}
\label{ch:independent decoding}


We derive a multiclass discrete-time processor-sharing queueing model,
of the type developed in Chapter~\ref{ch:chapter1}, for scheduled
message communication over a discrete memoryless multiaccess channel
with independent message decoding at the receiver, when messages are
generated at random times.

\section{The Information-Theoretic Model}
A discrete stationary memoryless channel (DMC) is specified by a finite
input alphabet $\mathcal{X}$, a finite output alphabet $\mathcal{Y}$,
and a probability assignment $\{ p(y|x); x \in \mathcal{X}, y \in
\mathcal{Y}\}$. The property that the channel is memoryless and is used
without feedback implies that, for each positive integer $N$, 
\[
p \left( y^{(N)} | x^{(N)}\right) = \prod_{n=1}^{N} p(y_{n}|x_{n}),
\]
where the $N$-length sequences $x^{(N)} \in \mathcal{X}^{N}$ and
$y^{(N)} \in \mathcal{Y}^{N}$. For a given probability assignment $Q = \{
Q(x); x \in \mathcal{X}\}$ on the input alphabet, we define the average
mutual information between the channel input $\mathcal{X}$ and channel
output $\mathcal{Y}$ of a DMC as
\[
I(X;Y) = \sum_{x \in \mathcal{X}} \sum_{y \in \mathcal{Y}} 
Q(x)p(y|x) \ln \frac{p(x|y)}{p(x)}
\]
Since $I(X;Y)$ is a function of $\{Q(x); x \in \mathcal{X}\}$ for a
given transition probability assignment $\{ p(y|x); x \in \mathcal{X}, y
\in \mathcal{Y}\}$, we define the capacity $C$ of a DMC as the largest
average mutual information $I(X;Y)$ , maximized over all input
probability assignments. Thus
\[
C = \sup_{ \{Q(x); x \in \mathcal{X}\} } I(X;Y)
\]
Consider the situation when $N$-length channel input sequences $x^{(N)}$
are to be transmitted over the channel in $N$ successive channel uses.
For each such transmitted input sequence $x^{(N)}$ the corresponding
received sequence $y^{(N)}$ is determined, letter by letter, according
to the channel transition probability assignment $\{p(y|x); x \in
\mathcal{X}, y \in \mathcal{Y} \}$. A decoder examines the received
word, and maps it to an estimate of the transmitted input sequence.

For $N \geq 1$ and $M \geq 2$, we define a block code $(N, M)$ to be a
set of $M$ channel input sequences $x^{(N)}$. The rate $R$ of the code
in natural units is defined as $R = (\ln M)/N$. A message communication
system can be designed by forming a message source that has $M$ possible
messages to be communicated over the channel. Each $N$ units of time the
source generates a message and the encoder then maps that message to a
code word in the code $(N, M)$.  The Noisy-channel Coding Theorem
(Theorem 5.6.2 in~\cite{Gal-BOOK}) states that, for $R < C$, arbitrarily
reliable communication is possible in the sense that the probability of
block decoding error can be made as small as required, and that, for $R
> C$, arbitrarily reliable communication is not possible. For $R < C$, a
significant issue to consider is the rate of decay of the probability of
message decoding error with the length $N$ of the code word.

An upper bound on block error probability  exists, that decays
exponentially with block length $N$ for all rates $R < C$. This bound is
derived by analyzing an ensemble of codes rather than just one code.
The ensemble of codes is generated by choosing each letter of each code
word independently with the probability distribution $Q$. We state here
Theorem 5.6.2 in~\cite{Gal-BOOK} that gives an upper bound to the
expectation, over the ensemble, of this block error probability.

\begin{theorem}[\cite{Gal-BOOK}]
\label{th:DMC coding theorem}
Let a discrete memoryless channel have transition probabilities $p(y|x)$
and, for any positive integer $N$ and another positive integer $M$,
consider the ensemble of $(N, M)$ block codes  in which each letter of
each code word is independently selected with the probability assignment
$Q$. Then, for each message $m$, $1 \leq m \leq M$, and all $\rho$,
$0 \leq \rho \leq 1$, the ensemble average probability of decoding error
using maximum-likelihood decoding satisfies
\begin{eqnarray}
\label{eq:random coding bound}
\overline{p}_{e, m} & \leq & \exp \left\{ 
\rho \ln M -N E_{o}(\rho, Q) \right\}, \quad \mbox{where} \\
\label{eq:random coding exponent}
E_{o}(\rho, Q) &=& -\ln \sum_{y \in \mathcal{Y}} \left[ \sum_{x \in
\mathcal{X}} Q(x)p(y|x)^{\frac{1}{1+\rho}} \right]^{1+\rho} 
\end{eqnarray} \Q
\end{theorem}
\section{The Queueing-Theoretic Model}
In this section we derive a multiclass discrete-time processor-sharing
queueing model, of the type developed in Chapter~\ref{ch:chapter1}, for
scheduled message communication over a multiaccess channel with independent
decoding being performed at the receiver, when requests for message
transmission arrive at random times. This queueing model is defined as
in~\cite{Tel-THESIS},~\cite{TelGal-JRN-JSAC} by considering messages as
customers in queue, and the combination of communication channel and
decoder as server.

Suppose that a message chosen from a message alphabet of size $M$ is
transmitted using block encoding and maximum-likelihood decoding, and
that the decoding error probability is required to be at most $p_{e}$.
Following the random coding principle, we pick a code book at random
from the ensemble of block codes $(N, M)$. The message is then
communicated by transmitting its assigned code word. We choose the code
word length $N$ to be the \emph{smallest positive integer} satisfying
\begin{eqnarray*}
\exp \left\{ \rho \ln M - NE_{o}(\rho, Q) \right\} \leq p_{e}
\end{eqnarray*}
Then, on an average, the decoded message is in error with a probability
not more than $p_{e}$. Let us rewrite the above inequality as
\begin{eqnarray}
\label{eq: queueing interpretation}
NE_{o}(\rho, Q) \geq -\ln p_{e} + \rho\ln M
\end{eqnarray}
Inequality~(\ref{eq: queueing interpretation}) can be used to interpret
the above message communication scheme in the following way: for any
message to be decoded with an expected error probability not more than
$p_{e}$, the message may be viewed as a customer in a queue with a
``service requirement'' of $-\ln p_{e} + \rho\ln M$ and that is served
by a decoder that provides a ``service quantum'' of $E_{o}(\rho, Q)$ in
a channel use.

The ``service requirement'' and ``service quantum'' interpretation given
above for communication of a single message can be extended to the
context when simultaneous message transmissions are allowed and each
message is decoded independently, i.e.,  signals resulting from other
message transmissions are treated as noise-like interference. In this
extension, we see that the definition of service requirement remains the
same while the definition of available service quantum is suitably
changed to account for the interference seen by a message transmission.

We assume that a request for message transmission can (i) choose its
message value from one of a finite number of message alphabets, and (ii)
specify the expected message decoding error probability.  We assume
independent and identically distributed quasi-static flat fades from the
transmitters to the receiver in the respective channels. With this
assumption, there is an i.i.d.  multiplicative gain in the channel from
each transmitter to the receiver.  Thus, for a given multiplicative gain
$\gamma$, a message signal of average power $P$ will be received at the
signal power $|\gamma|^{2}P$.  We assume that the multiplicative gain
$\gamma$ is known to the receiver, and is a random variable that has a
finite number of finite possible magnitudes.  We define ``class'' for a
message request by specifying the message alphabet $\mathcal{M} =
\left\{ 1, 2, \ldots, M \right\}$, probability of message decoding error
$p_{e}$, and the multiplicative gain of the channel $\gamma$. Thus, a
message request is characterized by a triple of numbers.  For our
purposes, we assume that a message request can assume one of $J \geq 1$
different message classes $( \mathcal{M}_{j}, p_{e, j}, \gamma_{j} )$,
where for $1 \leq j \leq J$, $\mathcal{M}_{j} = \left\{ 1, 2, \ldots,
M_{j} \right\}$ and $\gamma_{j}$ is the $j$th multiplicative gain value. 

\begin{figure}[h!]
\centering
\includegraphics[width=10cm,height=3cm]{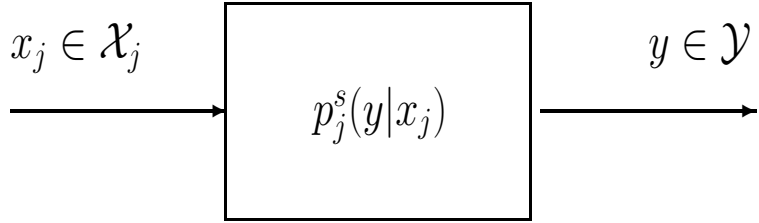}
\caption{Equivalent DMC seen by a class-$j$ message under the schedule
$s$}
\label{fig:effective MAC}
\end{figure}

Next, we allow scheduling of multiple messages for simultaneous
transmission, i.e., signal transmissions from the same message class can
overlap in time.  Let $s \in \mathcal{S}_{\mathsf{K}}$ be as defined in
Chapter~\ref{ch:chapter1}.  Let $\mathcal{X}_{j}$ denote the set of
channel input letters for class-$j$ , and $Q_{j} = \left\{ Q_{j}(x_{j});
x_{j} \in \mathcal{X}_{j} \right\}$ be an arbitrary probability
assignment on $\mathcal{X}_{j}$. Consider a schedule $s \in
\mathcal{S}_{\mathsf{K}}$. Define the channel vector input $x^{s} =
\left( x_{j}^{k};\; \mbox{all $j$},\; 1 \leq j \leq J,\; \mbox{such that
$s_{j} > 0$, and}\; 1 \leq k \leq s_{j} \right)$, where $x_{j}^{k} \in
\mathcal{X}_{j}$. Then, for the schedule $s$, the communication channel
under consideration is the multiaccess channel with the transition
probability law $p^{s}\left(y|x^{s} \right)$.  Assuming random coding,
for each $j$, $1 \leq j \leq J$, such that $s_{j} > 0$, define the
effective channel transition probability law $
\left\{p_{j}^{s}(y|x_{j}); x_{j} \in \mathcal{X}_{j}; y \in \mathcal{Y}
\right\}$ for a class-$j$ message under the schedule $s$ as (see
Fig.~\ref{fig:effective MAC})
\begin{eqnarray*}
p_{j}^{s}(y|x_{j}) &=& \left\{
\begin{array}{ll}
\sum_{\left\{x^{s}:x_{j}^{1}=x_{j} \right\}} 
p^{s}\left(y|x^{s} \right) \left( \prod_{k=2}^{s_{j}} Q_{j}
\left(x_{j}^{k} \right) \right) \left( 
\prod_{\left\{l \neq j: s_{l} > 0 \right\}} \prod_{k=1}^{s_{l}}
Q_{l} \left(x_{l}^{k} \right) \right) & \mbox{if}\; s_{j} > 1\\
\sum_{\left\{x^{s}:x_{j}^{1}=x_{j} \right\}} 
p^{s}\left(y|x^{s} \right) \left( 
\prod_{\left\{l \neq j: s_{l} > 0 \right\}} \prod_{k=1}^{s_{l}}
Q_{l} \left(x_{l}^{k} \right) \right) & \mbox{if}\; s_{j} = 1\\
\end{array}
\right.
\end{eqnarray*}

For an arbitrary $N$-length code word $x_{j}^{(N)} = \left(x_{j}(l); 1
\leq l \leq N \right)$ from the set $\mathcal{X}_{j}$ and an $N$-length
schedule sequence $s^{(N)} = \left(s^{1}, s^{2}, \ldots, s^{N} \right)$,
where $s^{n} \in \left\{s \in \mathcal{S}_{\mathsf{K}}: s_{j} > 0
\right\}$ , we define that $p \left( y^{(N)} \left| x_{j}^{(N)}, s^{(N)}
\right.  \right) = \prod_{n=1}^{N} p_{j}^{s^{n}} \left. \left( y_{n}
\right| x_{j}(n) \right)$ , and the $N$ channel uses may be
non-contiguous.  We can show the following Theorem by extending the
proof of Theorem~\ref{th:DMC coding theorem} (Theorem 5.6.2
in~\cite{Gal-BOOK}).

\begin{theorem}
Let the effective discrete memoryless channel as seen by a class-$j$
message transmission under the schedule $s$ such that $s_{j} > 0$ have
the transition probabilities $p_{j}^{s}(y|x_{j})$.  For any positive
integer $N$ and the message alphabet size $M_{j}$, consider the ensemble
of $\left(N, M_{j}\right)$ block codes  in which each letter of each
code word is independently selected with the probability assignment
$Q_{j}$.  Then, for each message $m_{j}$, $1 \leq m_{j} \leq M_{j}$, and
all $\rho$, $0 \leq \rho \leq 1$, the ensemble average probability of
decoding error using maximum-likelihood decoding satisfies

\begin{eqnarray}
\overline{p}_{e, m_{j}} & \leq & \exp \left\{
\rho \ln M_{j} - \sum_{n=1}^{N} E_{o, j}^{s^{n}}(\rho, Q_{j}) \right\},
\quad \mbox{where} \nonumber \\
\label{eq:interference service quantum}
E_{o, j}^{s}\left(\rho, Q_{j}\right) &=& -\ln \sum_{y \in \mathcal{Y}} \left[ 
\sum_{x_{j} \in \mathcal{X}_{j}} Q_{j} \left(x_{j} \right)
p_{j}^{s}(y|x_{j})^{\frac{1}{1+\rho}} \right]^{1+\rho} 
\end{eqnarray} \Q
\end{theorem}

For independent Gaussian encoding of messages with power $P_{j} = \left|
\gamma_{j} \right|^{2}P$, we can evaluate $E_{o,j}^{s}\left(\rho,
Q_{j}\right)$ in~(\ref{eq:interference service quantum}), and the value
is given below. For $s \in \left\{s \in \mathcal{S}_{\mathsf{K}} : s_{j}
> 0 \right\}$,

\begin{eqnarray}
\label{eq:Gaussian service quantum}
E_{o,j}^{s}\left(\rho, Q_{j}\right) &=& \rho \ln \left(
1 + \frac{P_{j}}{(1+\rho)\left[\left( \sum_{k=1}^{J}s_{k}P_{k}\right) -
P_{j} +
N_{0}W\right]}
\right)
\end{eqnarray}
Suppose that a class-$j$ message signal is scheduled as part of the
schedule $s \in \left\{s \in \mathcal{S}_{\mathsf{K}} : s_{j} > 0
\right\}$ for $d_{j}(s)$ possibly non-contiguous channel uses.  For a
given tolerable decoding error probability $p_{e, j}$, assume that the
upper bound on the expected message decoding error probability satisfies
the inequality $\overline{p}_{e, m_{j}} \leq p_{e, j}$ so that
\begin{eqnarray}
\label{eq: queueing interpretation 2}
\sum_{ s \in \left\{s \in \mathcal{S}_{\mathsf{K}} :
s_{j} > 0 \right\} } d_{j}(s) E_{o, j}^{s}\left(\rho, Q_{j}\right) \geq 
-\ln p_{e, j} + \rho\ln M_{j}
\end{eqnarray}
By extending the interpretation given to inequality~\ref{eq: queueing
interpretation} to the inequality~\ref{eq: queueing interpretation 2},
we can observe that the definition of service requirement remains the
same, whereas a message service quantum now is $E_{o, j}^{s}\left(\rho,
Q_{j}\right)$ thus reflecting interference. We say that the message
code word length is $N_{j} = \sum_{s \in \left\{s \in
\mathcal{S}_{\mathsf{K}} : s_{j} > 0 \right\} } d_{j}(s)$ and that the
message received a cumulative service of $\sum_{ s \in \left\{s \in
\mathcal{S}_{\mathsf{K}} : s_{j} > 0 \right\} } d_{j}(s) E_{o,
j}^{s}\left(\rho, Q_{j}\right)$ over $N_{j}$ channel uses.  We should
observe here that, for a given $p_{e, j}$, there may exist many
different solutions $\left\{d_{j}(s); s \in
\mathcal{S}_{\mathsf{K}}\;\mbox{and}\; s_{j} > 0 \right\}$ such that the
cumulative service equals or exceeds $-\ln p_{e, j} + \rho \ln M_{j}$.
\begin{definition}[Service Requirement]
For $1 \leq j \leq J$, a class-$j$ message service requirement is
denoted by $S_{j} < \infty$ and is defined by $S_{j} = -\ln p_{e, j} +
\rho\ln M_{j}$.
\end{definition}
\begin{definition}[Service Quantum]
For $\mathsf{K} \geq 1$ and $1 \leq j \leq J$, a class-$j$ message under
the schedule $s \in \left\{ s \in \mathcal{S}_{\mathsf{K}}; s_{j} > 0
\right\}$ can receive a service quantum of magnitude $E_{o,
j}^{s}\left(\rho, Q_{j}\right) > 0$ in a channel use.
\end{definition}
In the notation of Chapter~\ref{ch:chapter1}, define the available
service quantum to a class-$j$ message as a function of the schedule $s$
as
\begin{eqnarray}
\label{eq:Quantum}
\phi_{j}(s) &=& \left\{
\begin{array}{ll}
E_{o, j}^{s}\left(\rho, Q_{j}\right) &\mbox{if}\; s_{j} > 0 \\
0 & \mbox{if}\; s_{j} = 0
\end{array}
\right.
\end{eqnarray}
A few remarks on the definitions of service requirement and service
quantum are in order.
\begin{itemize}
\item	A significant difference between a message's service requirement
and its available service quantum is that the former quantity depends
\emph{only} on the message class whereas the available service quantum
depends on the particular schedule $s$ and its message class. This
observation implies that a message can be offered different service
quanta under different schedules.

\item	For a schedule $s \in \mathcal{S}_{\mathsf{K}}$, it is possible
that the total available service quantum $s_{j}\phi_{j}(s)$ to queue-$j$
is different for different queues. Then in that case, we have a multiclass
\emph{non-uniform} processor-sharing queueing model.
\end{itemize}
Having defined a service requirement for a message transmission, and
modeled the decoder by a server, we are now in a position to analyze
this communication scheme when requests for message transmission arrive
at random times. The model for random generation of message requests for
transmission is as given in the Chapter~\ref{ch:chapter1}.  In this
setting, messages transmit their signals over a random duration
(equivalently, code words of random length), determined by the message
arrival processes and the service statistics of the server.

In the rest of this chapter, we consider two classes of stationary
scheduling policies: for an integer $\mathsf{K} \geq 1$, we define (i)
non-idling policies, denoted by $\Omega_{\mathsf{K}}$, and (ii)
``state-independent'' scheduling policies, denoted by
$\Omega^{\mathsf{K}}$. For each scheduling policy $\omega$, we define a
discrete-time Markov chain for the queueing model, evolving on the
countable space $\mathcal{X}$ of states $\alpha$, as defined
in~(\ref{eq:state vector definition}) of Chapter~\ref{ch:chapter1}.  We
then analyze for the stability (Definition~\ref{def:stability}) of the
chain.  These stability results are derived by obtaining appropriate
drift conditions for suitably defined Lyapunov functions $V(\alpha) $ of
the state of the Markov chain. In particular, we prove that the Markov
chain is $c$-regular and stable by applying Theorem 10.3
from~\cite{Mey-paper}. 

\section{Stability Analysis for the Class of Non-Idling \\
Scheduling Policies }
Define $\overline{\mathcal{S}}_{\mathsf{K}}  = \left\{ s \in
\mathcal{S}_{\mathsf{K}}: \sum_{j=1}^{J} s_{j} = \mathsf{K} \right\}$,
and $\left\{ \mathcal{H}_{\mathsf{K}}, \mathcal{H}_{\mathsf{K}}^{c}
\right\}$ to be a partition of $\mathcal{X}$ such that
$\mathcal{H}_{\mathsf{K}}^{c} = \left\{ \alpha \in \mathcal{X}:
n(\alpha) \geq \mathsf{K} \right\}$.  Define
$\overline{\mathcal{S}}_{\mathsf{K}}^{\alpha} = \left\{ s \in
\overline{\mathcal{S}}_{\mathsf{K}}: s_{j} \leq n_{j}(\alpha)\;
\mbox{for}\; 1 \leq j \leq J \right\}$ to be the set of all feasible
schedules in state $\alpha$ that schedule exactly $\mathsf{K}$ messages
for simultaneous transmission.  A scheduling policy $\omega \in
\Omega_{\mathsf{K}}$ is defined by (i) the mapping $\omega: \mathcal{X}
\rightarrow \mathcal{S}_{\mathsf{K}}$, and (ii) a probability
distribution $\left\{ p^{\omega}_{\alpha}(s); \alpha \in
\mathcal{X}\;\mbox{and}\; s \in \mathcal{S}_{\mathsf{K}} \right\}$ with
the following two properties: (i) $p^{\omega}_{\alpha}(s) = 0$ if $s$ is
an infeasible schedule in state $\alpha$, and (ii) $\sum_{s \in
\overline{\mathcal{S}}_{\mathsf{K}}^{\alpha}} p^{\omega}_{\alpha}(s) =
1$ for $\alpha \in \mathcal{H}_{\mathsf{K}}^{c}$.  Thus the policy
$\omega$ ensures that some group of $\mathsf{K}$ messages are scheduled
for transmission whenever there are at least $\mathsf{K}$ messages
present in the system.  Define $\underline{\phi}_{j} = \min_{ \left\{ s
\in \overline{\mathcal{S}}_{\mathsf{K}}: s_{j} > 0 \right\}}
\phi_{j}(s)$, $\overline{\phi}_{j} = \max_{\left\{ s \in
\mathcal{S}_{\mathsf{K}} \right\} } \phi_{j}(s)$.  We introduce the notation that,
for any $x>0$ and $q>0$, $\lceil x \rceil_{q} = \min (n \geq 1: x \leq
nq)q$.

\begin{lemma} 
\label{lemma:unequal1}
Let $\mathsf{K} \geq 1$, $J \geq 1$ and $\omega \in
\Omega_{\mathsf{K}}$. For $\alpha \in \mathcal{X}$, let $c(\alpha) = 1 +
\sum_{j=1}^{J} \sum_{k=1}^{n_{j}(\alpha)} \left\lceil
\frac{x_{j}(k)}{\underline{\phi}_{j}} \right\rceil$ and 
\[
V(\alpha) = \frac{c^{2}(\alpha)}{2\left( \mathsf{K}
- \sum_{j=1}^{J} \mathds{E}A_{j}  \left\lceil
\frac{S_{j}}{\underline{\phi}_{j}} \right\rceil \right)}. 
\]
Then the Markov chain is $c$-regular and stable if $\sum_{j=1}^{J}
\mathds{E}A_{j} \left\lceil \frac{S_{j}}{\underline{\phi}_{j}}
\right\rceil < \mathsf{K}$. \Q
\end{lemma}
\begin{proof}
Let $a_{j}$ class-$j$ messages arrive in state $\alpha$ and that the
feasible schedule $s$ is implemented in the state $\alpha$. Assuming
that the chain moves to state $\alpha^{\prime}$, we have
\begin{eqnarray*}
c\left(\alpha^{\prime}\right) &=& c(\alpha) + f_{c}(a) -
g_{c}(\alpha, s),\; \mbox{where} \\
f_{c}(a) &=&  \sum_{j=1}^{J} a_{j}
\left\lceil \frac{S_{j}}{\underline{\phi}_{j}} \right\rceil \\
g_{c}(\alpha, s) &=& \sum_{j=1}^{J} \sum_{k=1}^{s_{j}}  \left\{
\left\lceil \frac{x_{j}(k)}{\underline{\phi}_{j}} \right\rceil
I_{\left\{ x_{j}(k) \leq \phi_{j}(s) \right\}} + 
\left( \left\lceil \frac{x_{j}(k)}
{\underline{\phi}_{j}} \right\rceil - 
\left\lceil  \frac{x_{j}(k)-\phi_{j}(s)} 
{\underline{\phi}_{j}} \right\rceil \right) 
I_{\left\{ x_{j}(k) > \phi_{j}(s) \right\}} \right\},
\end{eqnarray*}

We now consider $\alpha \in \mathcal{H}_{\mathsf{K}}^{c}$.  We show that
$\sum_{s \in \mathcal{S}_{\mathsf{K}}} g_{c}(\alpha, s)
p^{\omega}_{\alpha}(s) \geq \mathsf{K}$.  We first observe that $
\left\lceil \frac{x_{j}(k)} {\underline{\phi}_{j}} \right\rceil \geq 1$
since $x_{j}(k) > 0$. Consider
$\overline{\mathcal{S}}_{\mathsf{K}}^{\alpha}$.  Hence $\phi_{j}(s) \geq
\underline{\phi}_{j}$. For $x_{j}(k) > \phi_{j}(s)$, since $\phi_{j}(s)
\geq \underline{\phi}_{j}$, $\left\lceil \frac{x_{j}(k)}
{\underline{\phi}_{j}} \right\rceil - \left\lceil \frac{x_{j}(k)-
\phi_{j}(s)} {\underline{\phi}_{j}} \right\rceil \geq 1$. Hence
$g_{c}(\alpha, s) \geq \mathsf{K}$ and $\sum_{s \in
\mathcal{S}_{\mathsf{K}}} g_{c}(\alpha, s) p^{\omega}_{\alpha}(s) \geq
\mathsf{K}$ for $\alpha \in \mathcal{H}_{\mathsf{K}}^{c}$. But the
expected increase $\mathds{E}f_{c}$ in $c(\alpha)$ is $\sum_{j=1}^{J}
\mathds{E}A_{j} \left\lceil \frac{S_{j}}{\underline{\phi}_{j}}
\right\rceil$.



Assuming $\sum_{j=1}^{J} \mathds{E}A_{j} \left\lceil
\frac{S_{j}}{\underline{\phi}_{j}} \right\rceil < \mathsf{K}$, and then
applying Part$(A)$ of Lemma~\ref{lemma:c-regularity single condition} to
$c(\alpha)$ and $V(\alpha)$ as defined in the statement of
Lemma~\ref{lemma:unequal1}, we find that the Markov chain is
$c$-regular.  Since $c(\alpha) > n(\alpha)$ for every $\alpha$,
existence of finite stationary mean for $c(\alpha)$ implies existence of
finite stationary mean for $n(\alpha)$.  Hence the Markov-chain
$\{X_{n}; n\geq 0\}$ is stable.

\end{proof}
\noindent
{\bf Remark:} For Gaussian encoding of messages, we can see that
$\left\lceil \frac{x_{j}(k)}{\underline{\phi}_{j}} \right\rceil$ is the
maximum number of code symbols that a message with residual service
requirement $x_{j}(k)$ would possibly transmit. Thus $c(\alpha)$ gives
the maximum total outstanding number of code symbols in the system
still to be transmitted in state $\alpha$.

\begin{lemma}
\label{lemma:unequal2}
Let $\mathsf{K} \geq 1$, $J \geq 1$ and $\omega \in
\Omega_{\mathsf{K}}$. For $\alpha \in \mathcal{X}$ , define 
\begin{eqnarray*}
c(\alpha) &=& 1 + \sum_{j=1}^{J} \sum_{k=1}^{n_{j}(\alpha)} 
\left( x_{j}(k) + \overline{\phi}_{j} \right), \quad \mbox{and} \\
V(\alpha) &=& \frac{c^{2}(\alpha)}{2 \left( 
\min_{s \in \overline{\mathcal{S}}_{\mathsf{K}} } \sum_{j=1}^{J} 
s_{j} \phi_{j}(s) - \sum_{j=1}^{J} \mathds{E}A_{j}
\left(S_{j}+\overline{\phi}_{j}\right) \right)}.
\end{eqnarray*}
Then the Markov chain is $c$-regular and stable if $\sum_{j=1}^{J}
\mathds{E}A_{j} \left( S_{j}+\overline{\phi}_{j} \right) < \min_{s \in
\overline{\mathcal{S}}_{\mathsf{K}} } \sum_{j=1}^{J} s_{j} \phi_{j}(s)$.
\Q
\end{lemma}

\begin{proof}
Let $a_{j}$ class-$j$ messages arrive in state $\alpha$ and that the
feasible schedule $s$ is implemented in the state $\alpha$. Assuming
that the chain moves to state $\alpha^{\prime}$, we have
\begin{eqnarray}
c\left(\alpha^{\prime} \right) &=& c(\alpha) + f_{c}(a) -
g_{c}(\alpha, s),\; \mbox{where} \nonumber \\
f_{c}(a) &=& \sum_{j=1}^{J} a_{j} \left( S_{j} + \overline{\phi}_{j}
\right), \quad \mbox{and} \nonumber \\
\label{eq:decrease 2}
g_{c}(\alpha, s) &=& \sum_{j=1}^{J} \sum_{k=1}^{s_{j}} \left[ \left(
x_{j}(k) + \overline{\phi}_{j} \right) I_{\left\{ x_{j}(k) \leq
\phi_{j}(s) \right\}} + \phi_{j}(s)
I_{\left\{ x_{j}(k) > \phi_{j}(s) \right\}} \right]
\end{eqnarray}
We now consider $\alpha \in \mathcal{H}_{\mathsf{K}}^{c}$.  We show that
$\sum_{s \in \mathcal{S}_{\mathsf{K}}} g_{c}(\alpha, s)
p^{\omega}_{\alpha}(s) \geq \min _{s \in
\overline{\mathcal{S}}_{\mathsf{K}}} \sum_{j=1}^{J} s_{j} \phi_{j}(s)$
for $\alpha \in \mathcal{H}_{\mathsf{K}}^{c}$. Since $x_{j}(k) +
\overline{\phi}_{j}  > \phi_{j}(s)$, we can see from~(\ref{eq:decrease
2}) that $g_{c}(\alpha, s) > \sum_{j=1}^{J} s_{j} \phi_{j}(s)$. Since
$p^{\omega}_{\alpha}(s) = 0$ for $s \notin
\overline{\mathcal{S}}_{\mathsf{K}}$ and $\alpha \in
\mathcal{H}_{\mathsf{K}}^{c}$, we have that for $\alpha \in
\mathcal{H}_{\mathsf{K}}^{c}$,
\[ 
\sum_{s \in \mathcal{S}_{\mathsf{K}}} g_{c}(\alpha, s)
p^{\omega}_{\alpha}(s) = \sum_{s \in
\overline{\mathcal{S}}_{\mathsf{K}}} g_{c}(\alpha, s) p^{\omega}_{\alpha}(s)
> \sum_{s \in \overline{\mathcal{S}}_{\mathsf{K}}} 
\left( \sum_{j=1}^{J} s_{j} \phi_{j}(s) \right) p^{\omega}_{\alpha}(s) 
\geq \min_{s \in \overline{\mathcal{S}}_{\mathsf{K}}}
\sum_{j=1}^{J} s_{j} \phi_{j}(s).
\]
But the expected increase $\mathds{E}f_{c}$ in $c(\alpha)$ is
$\sum_{j=1}^{J} \mathds{E}A_{j} \left(S_{j} + \overline{\phi}_{j}
\right)$.  Assuming that $\mathds{E}A_{j} \left( S_{j} +
\underline{\phi}_{j} \right) < \min_{s \in
\overline{\mathcal{S}}_{\mathsf{K}}} \sum_{j=1}^{J} s_{j} \phi_{j}(s)$,
and then applying Part$(A)$ of Lemma~\ref{lemma:c-regularity single
condition} to $c(\alpha)$ and
$V(\alpha)$ as defined in the statement of Lemma~\ref{lemma:unequal2},
we find that the Markov chain is $c$-regular.  Since $c(\alpha) > 1 +
\left(\min_{j} \overline{\phi}_{j} \right) n(\alpha)$ for every
$\alpha$, existence of finite stationary mean for $c(\alpha)$ implies
finite stationary mean for $n(\alpha)$.  Hence the queueing model
$\{X_{n}; n\geq 0\}$ is stable.  
\end{proof}
\begin{lemma}
\label{lemma:K>2J>2 transience}
For $\mathsf{K} \geq 1$, $J \geq 1$, and for any non-empty subset $B$ of
the set $\{ 1, 2, \ldots, J \}$, the Markov chain is unstable if,
$\sum_{j \in B} S_{j} \mathds{E}A_{j} \geq
\max_{s \in \overline{\mathcal{S}}_{\mathsf{K}} }
\sum_{j \in B} s_{j} \phi_{j}(s)$. \Q
\end{lemma}
\begin{proof}
For each non-empty subset $B$ of the set $\{ 1, 2, \ldots, J \}$, define
the function $c^{B}(\alpha) = \sum_{j \in B} \sum_{k=1}^{n_{j}(\alpha)}
x_{j}(k)$, and then the  Lyapunov function $V_{B} = 1 -
\theta^{c^{B}(\alpha)}$ for $0 < \theta < 1$ on the state space
$\mathcal{X}$.  Then we have the following:
\begin{eqnarray*}
c^{B} \left( \alpha^{\prime} \right) &=& c^{B}(\alpha) + 
f_{c^{B}}(a) - g_{c^{B}}(\alpha, s), \quad \mbox{where}  \\
f_{c^{B}}(a) &=& \sum_{j \in B} a_{j}, \quad \mbox{and} \\
g_{c^{B}}(\alpha, s) &=& \sum_{j \in B} \sum_{k=1}^{s_{j}} 
\min \left\{ x_{j}(k), \phi_{j}(s) \right\}
\end{eqnarray*}
Since $p^{\omega}_{\alpha}(s) = 0$ for $\alpha \in
\mathcal{H}^{c}_{\mathsf{K}}$ and $s \notin
\overline{\mathcal{S}}_{\mathsf{K}}$, we have the following inequalities
for $\alpha \in \mathcal{H}^{c}_{\mathsf{K}}$:
\[
\sum_{s \in \mathcal{S}_{\mathsf{K}}} g_{c^{B}}(\alpha, s)
p^{\omega}_{\alpha}(s) = \sum_{s \in
\overline{\mathcal{S}}_{\mathsf{K}}} g_{c^{B}}(\alpha, s) p^{\omega}_{\alpha}(s)
< \sum_{s \in \overline{\mathcal{S}}_{\mathsf{K}}}
\left( \sum_{j\in B} s_{j} \phi_{j}(s) \right) p^{\omega}_{\alpha}(s)
\leq \max_{s \in \overline{\mathcal{S}}_{\mathsf{K}}}
\sum_{j \in B} s_{j} \phi_{j}(s).
\]
By applying Part $(B)$ of Lemma~\ref{lemma:c-regularity single
condition} to $V_{B}(\alpha)$, we find that the Markov chain is unstable
if $\sum_{j \in B} S_{j} \mathds{E}A_{j} \geq \max_{s \in
\overline{\mathcal{S}}_{\mathsf{K}} } \sum_{j \in B} s_{j} \phi_{j}(s)$.
\end{proof}
For certain specific values of $\mathsf{K}$ and $J$, \emph{exact}
characterization of message arrival rate stability region can be found.
\begin{theorem}
\label{th:unequal1}
Let $\omega \in \Omega_{\mathsf{K}}$.
\begin{itemize}
\item[(A)] Let either $\mathsf{K}=1$ and $J \geq 1$, or $\mathsf{K} \geq
1$ and $J=1$. Then the Markov chain is (i) stable if $\sum_{j=1}^{J}
\mathds{E}A_{j} \left\lceil \frac{S_{j}} {\underline{\phi}_{j}}
\right\rceil < \mathsf{K}$, and (ii) unstable if $ \sum_{j=1}^{J}
\mathds{E}A_{j} \left\lceil \frac{S_{j}}{\underline{\phi}_{j}}
\right\rceil > \mathsf{K} $.

\item[(B)]	Let $J \geq 1$. Then, for Gaussian encoding of messages
and in the limit $\mathsf{K} \rightarrow  \infty$ , the Markov chain is
(i) stable if $\sum_{j=1}^{J} \mathds{E}A_{j} S_{j}<
\frac{\rho}{1+\rho}$ , and (ii) unstable if $\sum_{j=1}^{J}
\mathds{E}A_{j} S_{j} > \frac{\rho}{1+\rho}$.  \end{itemize}
\Q
\end{theorem}

Part $(B)$ of Theorem~\ref{th:unequal1} says that, in the limit
$\mathsf{K} \rightarrow \infty$, the upper bound on stable throughput
achievable with $E_{o, j}^{s}\left(\rho, Q_{j} \right)$ defined
in~(\ref{eq:Gaussian service quantum}) is \emph{independent} of message
SNR-s and their distribution.  The stability results for the
\emph{continuous-time} models in~\cite{TelGal-JRN-JSAC}
and~\cite{UtpSan-CONF-ITW2002} coincide with the corresponding result,
stated in Part $(B)$ of Theorem~\ref{th:unequal1}, for the discrete-time
model in the limit of large number of simultaneous transmissions.

\begin{proof}
Part(i) of Part $(A)$ is proved in Lemma~\ref{lemma:unequal1}.  To prove
Part(ii) of Part ($A$), consider $c(\alpha) $ as defined in
Lemma~\ref{lemma:unequal1} and the Lyapunov function $V(\alpha) = 1 -
\theta^{c(\alpha)}$ for $0 < \theta < 1$. We observe that $\sum_{s \in
\mathcal{S}_{\mathsf{K}}} g_{c}(\alpha, s) p^{\omega}_{\alpha}(s)$ can
be uniquely determined for the following two special cases. For  $\alpha
\in \mathcal{H}^{c}_{\mathsf{K}}$,
\begin{eqnarray*}
\sum_{s \in \mathcal{S}_{\mathsf{K}}} g_{c}(\alpha, s) 
p^{\omega}_{\alpha}(s) &=& \left\{
\begin{array}{ll}
1, & \mbox{if}\;\; \mathsf{K} = 1, J \geq 1 \\
\mathsf{K}  & \mbox{if}\;\; \mathsf{K} \geq 2, J=1 \\
\end{array}
\right.
\end{eqnarray*}
By applying Part$(B)$ of Lemma~\ref{lemma:c-regularity single condition}
to $V(\alpha)$, we find that the queueing model is unstable if
$\sum_{j=1}^{J} \mathds{E}A_{j} \left\lceil
\frac{S_{j}}{\underline{\phi}_{j}} \right\rceil > \mathsf{K}$ for either
$\mathsf{K}  = 1$ and $J \geq 1$, or $\mathsf{K} \geq 2$ and $J=1$.

To prove Part ($B$), we first observe that, for $E_{o, j}^{s}\left(\rho,
Q_{j} \right)$ as defined in~(\ref{eq:Gaussian
service quantum}),
\begin{eqnarray*}
\lim_{\mathsf{K} \rightarrow \infty}
\min_{s \in \overline{\mathcal{S}}_{\mathsf{K}}} \sum_{j=1}^{J} 
s_{j} \phi_{j}(s) &=& 
\lim_{\mathsf{K} \rightarrow \infty}
\max_{s \in \overline{\mathcal{S}}_{\mathsf{K}}} \sum_{j=1}^{J} 
s_{j} \phi_{j}(s)
= \frac{\rho}{1+\rho}
\end{eqnarray*}
Also, for $1 \leq j \leq J$, $\lim_{\mathsf{K} \rightarrow \infty} S_{j}
+ \overline{\phi}_{j} = S_{j}$ since $\lim_{\mathsf{K} \rightarrow
\infty} \overline{\phi}_{j} = 0$. Proof now follows from the sufficiency
condition for stability stated in Lemma~\ref{lemma:unequal2} and the
sufficiency condition for unstability stated in Lemma~\ref{lemma:K>2J>2
transience} with $B = \{1, 2, \ldots, J\}$. We observe that the inner
bound stated in Lemma~\ref{lemma:unequal2} and the outer bound stated in
Lemma~\ref{lemma:K>2J>2 transience} coalesce in the limit $\mathsf{K}
\rightarrow \infty$.  
\end{proof}

\begin{figure}[h!]
\centering
\includegraphics[width=\tw,height=10cm]{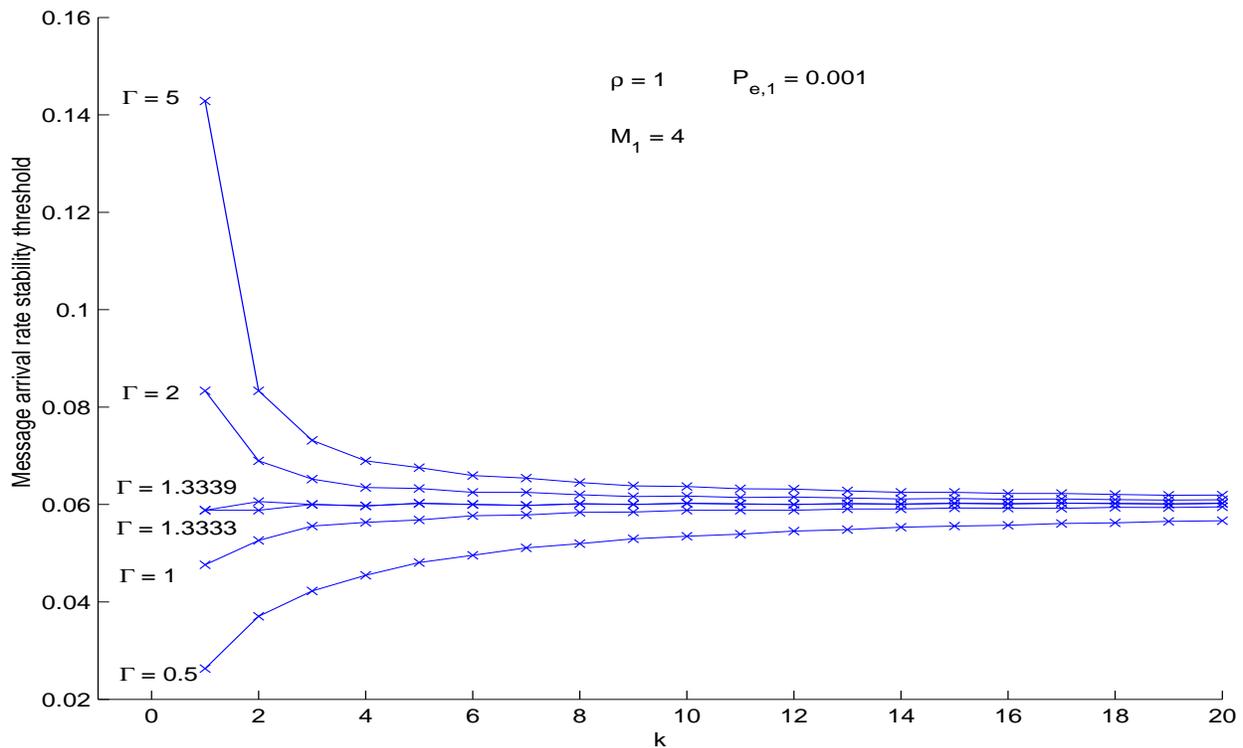}
\caption{Message arrival rate stability threshold versus maximum number of
simultaneous message transmissions, $\mathsf{K}$, in the case $J=1$.}
\label{fig:equal powers finite message length} 
\end{figure}
Figure~\ref{fig:equal powers finite message length} shows plots of
message arrival rate stability threshold versus $\mathsf{K}$, for the
special case $J=1$ and for different values of $\Gamma$ with parameters
$\rho$, $M_{1}$ and $p_{e, 1}$ fixed. From these plots we see that, for
sufficiently small transmit powers, as many simultaneous message
transmissions as possible should be scheduled, i.e., immediate access
should be granted to messages to increase the throughput of the system.
For large transmit power, scheduling many transmissions hurts the system
throughput. This behavior can be explained as follows. For small
transmit powers, the effective noise seen by a transmission arises
mainly from thermal noise, rather than from interference caused by other
ongoing message transmissions.  Thus, interference from other signal
transmissions has insignificant effect on any given transmission, and
scheduling as many transmissions as possible is advantageous from the
stability view point.  For large transmit powers, interference dominates
the effective noise seen by any message transmission.  Hence, limiting
the number of simultaneous transmissions is desirable.

\section{Stability for State-Independent Scheduling Policies}
\label{section:ch1-state-independent}
In this section we consider the class $\Omega^{\mathsf{K}}$ of
stationary state-independent scheduling policies. Before we formally
define a policy $\omega$ in $\Omega^{\mathsf{K}}$, we first introduce
the notions of \emph{sub-schedule} and \emph{maximal sub-schedule}.
\begin{definition}[Sub-Schedule]
For $\mathsf{K} \geq 1$ and $s, s^{\prime} \in
\mathcal{S}_{\mathsf{K}}$, we write $s^{\prime} \preceq s$ if
$s^{\prime}_{j} \leq s_{j}$ for $1 \leq j \leq J$. We then say that
$s^{\prime}$ is a sub-schedule of the schedule $s$.  The maximal
sub-schedule of the schedule $s$ in state $\alpha$ is denoted by
$s^{*}(s, \alpha) \in \mathcal{S}_{\mathsf{K}}$, and is defined by
$s^{*}_{j}(s, \alpha) = \min \left\{ s_{j}, n_{j}(\alpha) \right\}$ for
$1 \leq j \leq J$.  \Q
\end{definition}
We should observe that (i) $s^{*}(s, \alpha)$ for a given schedule $s$
can be the zero schedule in some states $\alpha$, (ii) in a given state
$\alpha$, it is possible that $s^{*}(s, \alpha) = \hat{s}^{*} \left( \hat{s},
\alpha \right)$ for some two schedules $s, \hat{s} \in
\mathcal{S}_{\mathsf{K}}$, and (iii) for $E_{o, j}^{s}\left(\rho,
Q_{j}\right)$ given in~(\ref{eq:Gaussian service quantum}), $\phi_{j}
\left( s^{\prime} \right) \geq \phi_{j}(s)$ for $s^{\prime} \preceq s$
and $1 \leq j \leq J$.

Define $\mathcal{S}_{\mathsf{K}}^{\alpha} = \left\{ t \in
\mathcal{S}_{\mathsf{K}}: t = s^{*}(s, \alpha)\; \mbox{for some}\; s \in
\mathcal{S}_{\mathsf{K}} \right\} \subseteq \mathcal{S}_{\mathsf{K}}$ to
be the set of maximal sub-schedules~\footnote{Every schedule in
$\mathcal{S}_{\mathsf{K}}^{\alpha}$ is a feasible schedule.} in state
$\alpha$.  For an arbitrary probability distribution $\left\{ p^{\omega}(s);
s \in \mathcal{S}_{\mathsf{K}} \right\}$ and $\alpha \in \mathcal{X}$,
define the probability distribution $\left\{ p_{\alpha}^{\omega}(s); \alpha
\in \mathcal{X}\;\mbox{and}\; s \in
\mathcal{S}_{\mathsf{K}}^{\alpha} \right\}$ by defining
\begin{eqnarray}
\label{eq:induced prob measure}
p_{\alpha}^{\omega}(s) &=& \sum_{ \left\{ t \in \mathcal{S}_{\mathsf{K}}: 
t^{*}(t, \alpha) = s \right\} } p^{\omega}(t)
\end{eqnarray}
Formally, a policy $\omega \in \Omega^{\mathsf{K}}$ is defined by a
probability distribution $\left\{ p^{\omega}(s); s \in
\mathcal{S}_{\mathsf{K}} \right\}$ together with the collection of the
random variables $\{ \omega(\alpha); \alpha  \in \mathcal{X} \}$. For
$\alpha \in \mathcal{X}$, the schedule $\omega(\alpha)$ to be
implemented in state $\alpha$ is a random variable that takes values in
the set of maximal sub-schedules $\mathcal{S}_{\mathsf{K}}^{\alpha}$
with the probability measure $\left\{ p_{\alpha}^{\omega}(s); \alpha \in
\mathcal{X}\; \mbox{and}\; s \in \mathcal{S}_{\mathsf{K}}^{\alpha}
\right\}$ defined in~\ref{eq:induced prob measure}. The name
``state-independent'' for the class of policies $\Omega^{\mathsf{K}}$ is
a misnomer . The actual schedule $\omega(\alpha)$ that gets implemented
depends on the state $\alpha$ in conjunction with the probability
measure $\left\{ p_{\alpha}^{\omega}(s); s \in
\mathcal{S}_{\mathsf{K}}^{\alpha} \right\}$. The name
``state-independent'' is used because specification of the
probability measure $\left\{ p^{\omega}(s); s \in
\mathcal{S}_{\mathsf{K}} \right\}$ is independent of the state.
\begin{lemma}
\label{lemma:state-independent independent decoding}
For $\mathsf{K} \geq 1$, $\alpha \in \mathcal{X}$, $\omega \in
\Omega^{\mathsf{K}}$ and $1 \leq j \leq J$, define
$h_{j}^{\omega}(\alpha) = \sum_{k = 1}^{n_{j}(\alpha)} \left( x_{j}(k) +
\overline{\phi}_{j} \right)$, $c(\alpha) = 1 + \sum_{j=1}^{J}
h_{j}^{\omega}(\alpha)$, and
\[
V(\alpha) = \sum_{j}
\frac{\left[ h_{j}^{\omega} (\alpha) \right]^{2}}
{2\left( \sum_{s \in \mathcal{S}_{\mathsf{K}}}
p^{\omega}(s)s_{j}\phi_{j}(s) -  
\left( S_{j} + \overline{\phi}_{j} \right)  \mathds{E}A_{j} \right)}. 
\]
Then the Markov chain is $c$-regular and stable if $\left( S_{j} +
\overline{\phi}_{j} \right) \mathds{E}A_{j} < \sum_{s \in
\mathcal{S}_{\mathsf{K}}} p^{\omega}(s)s_{j}\phi_{j}(s) $ for $1 \leq j
\leq J$.
\Q
\end{lemma}
\begin{proof}
We note here that, in any time-slot, the schedule $s \in
\mathcal{S}_{\mathsf{K}}$ will be chosen with probability
$p^{\omega}(s)$, and the corresponding maximal sub-schedule $s^{*}(s,
\alpha)$ will get implemented in the state $\alpha$. Then $s^{*}_{j}(s,
\alpha)$ class-$j$ messages will be scheduled in state $\alpha$ and each
of them can receive a service quantum up to $\phi_{j} \left( s^{*}(s,
\alpha) \right)$.  Let $a_{j}$ class-$j$ messages arrive in state
$\alpha$ and that the feasible schedule $s$ is implemented in the state
$\alpha$. Assuming that the chain moves to state $\alpha^{\prime}$, we
have
\begin{eqnarray*}
h_{j}^{\omega} \left( \alpha^{\prime} \right) &=& 
h_{j}^{\omega}(\alpha) + f_{j}(a) - 
g_{j}( \alpha, s),\;\mbox{where} \\
f_{j}(a) &=& a_{j} \left( S_{j} + \overline{\phi}_{j} \right), \quad
\mbox{and} \\
g_{j}(\alpha, s) &=& \left\{
\begin{array}{ll}
\sum_{k=1}^{s_{j}} \left[ 
\left( x_{j}(k) + \overline{\phi}_{j}
\right) I_{\left\{ x_{j}(k) \leq \phi_{j}(s) \right\}}
+ \phi_{j}(s) I_{\left\{ x_{j}(k) > \phi_{j} (s) \right\}} \right] &
\mbox{if}\; s_{j} > 0 \\
0 & \mbox{if}\; s_{j} = 0
\end{array}
\right.
\end{eqnarray*}
Let $\left\{ \mathcal{H}_{j}, \mathcal{H}_{j}^{c} \right\}$ be a
partition of $\mathcal{X}$ and define $\mathcal{H}_{j}^{c} = \{\alpha:
n_{j}(\alpha) \geq \mathsf{K}\}$. For $\alpha \in
\mathcal{H}_{j}^{c}$, we have that $s^{*}_{j}(s,\alpha) =
s_{j}$ and 
\[
\sum_{s \in \mathcal{S}_{\mathsf{K}}} g_{j} 
\left( \alpha, s \right) p_{\alpha}^{\omega} (s) 
\stackrel{(a)}{\geq} 
\sum_{s \in \mathcal{S}_{\mathsf{K}}} 
\sum_{ \left\{ t \in \mathcal{S}_{\mathsf{K}} : 
t^{*}(t, \alpha) = s \right\} } 
p^{\omega}(t) s_{j} \phi_{j}(s) \\
= \sum_{s \in \mathcal{S}_{\mathsf{K}}} 
p^{\omega}(s) s_{j} \phi_{j}(s),
\]
where $(a)$ follows from the fact that $g_{j}(\alpha, s) \geq s_{j}
\phi_{j}(s)$.  But the expected increase $\mathds{E}f_{j}$ in
$h_{j}^{\omega}(\alpha)$ is $\left( S_{j} + \overline{\phi}_{j} \right)
\mathds{E}A_{j}$.  Assume that $\left( S_{j} + \overline{\phi}_{j}
\right) \mathds{E}A_{j} < \sum_{s \in \mathcal{S}_{\mathsf{K}}}
p^{\omega}(s) s_{j} \phi_{j}(s) $.  Now applying
Lemma~\ref{lemma:c-regularity} to the functions $c(\alpha)$ and
$V(\alpha)$ as defined in the statement of
Theorem~\ref{lemma:state-independent independent decoding}, we find that
the Markov chain is $c$-regular. Since $c(\alpha) > 1 + \left( \min_{j}
\overline{\phi}_{j} \right) n(\alpha)$ for every $\alpha$, the number of
messages $n(\alpha)$ in the system has finite stationary mean.  Hence
the Markov-chain $\{X_{n}; n\geq 0\}$ is stable.
\end{proof}
\noindent

\section{Interpretation of Information Arrival Rate Stability Region in
terms Information-Theoretic Capacities}
In this section we interpret the information arrival rate stability
region in terms of interference-limited information-theoretic
capacities.  Define $\tilde{A}_{j} = \left( \ln M_{j} \right)
\mathds{E}A_{j}$. Then $\mathds{E}\tilde{A}_{j} = \left( \ln M_{j}
\right) \mathds{E}A_{j}$ and $\mathds{E} \tilde{A} = \left( \mathds{E}
\tilde{A}_{1}, \mathds{E} \tilde{A}_{2}, \ldots, \mathds{E}
\tilde{A}_{J} \right)$ denote the nat arrival rate into queue-$j$ and
the nat arrival rate vector, respectively.  Define $\Gamma_{j} =
\frac{P_{j}}{N_{0}W}$ be the received SNR for a class-$j$ message
transmission,
\begin{theorem}[Capacity Interpretation for $\omega \in
\Omega_{\mathsf{K}}$]
\label{th:capacity interpretation}
Assume Gaussian encoding of messages.
\noindent
\begin{itemize}
\item[(i)]	Let $J=1$.  Then, in the limit $M_{1} \rightarrow
\infty$ and $\rho \rightarrow 0$, the threshold on $\mathds{E}
\tilde{A}_{1}$ approaches a limit that is equal to $\mathsf{K}$ times
the information-theoretic capacity of an AWGN channel with SNR
$\frac{\Gamma_{1}}{(\mathsf{K}-1)\Gamma_{1}+1}$.

\item[(ii)]	Let $J \geq 1$ and $\mathsf{K} \rightarrow \infty$.
Then, in the limit $\min_{j} M_{j} \rightarrow \infty$ and $\rho
\rightarrow 0$, the threshold on $\sum_{j=1}^{J} \mathds{E}
\tilde{A}_{j}$ approaches the limit  1 nat/s/Hz. \Q 
\end{itemize}
\end{theorem}

\begin{proof}
\noindent
{\bf Part (i):}
For $J=1$, we know from Part ($B$) of Theorem~\ref{th:unequal1} that the
system is stable if $\mathds{E}A_{1} \left\lceil
\frac{S_{1}}{\underline{\phi}_{1}} \right\rceil < \mathsf{K}$, or
equivalently, $\mathds{E}\tilde{A}_{1} < \frac{\mathsf{K} (\ln M_{1})
\underline{\phi}_{1}}{\lceil S_{1} \rceil_{\underline{\phi}_{1}}}$.
Since $\lceil S_{1} \rceil_{\underline{\phi}_{1}} = \lceil - \ln p_{e,1}
+ \rho \ln M_{1} \rceil_{\underline{\phi}_{1}} = - \ln p_{e, 1} + \rho
\ln M_{1} + d$, where $0 \leq d < \underline{\phi}_{1}$, we have the
following lower bound and upper bound on nat arrival rate threshold.
\[
\frac{\mathsf{K} \underline{\phi}_{1} \ln M_{1}}{- \ln p_{e, 1} + \rho
\ln M_{1} + \underline{\phi}_{1}} < \frac{\mathsf{K}
\underline{\phi}_{1} \ln M_{1}} {\lceil - \ln p_{e, 1} + \rho \ln M_{1}
\rceil_{\underline{\phi}_{1}}} \leq \frac{\mathsf{K}
\underline{\phi}_{1} \ln M_{1}}{- \ln p_{e, 1} + \rho \ln M_{1} } 
\]
Since $\frac{\mathsf{K} \underline{\phi}_{1} \ln M_{1}}{- \ln p_{e, 1} +
\rho \ln M_{1} + \underline{\phi}_{1}}$ and $\frac{\mathsf{K}
\underline{\phi}_{1} \ln M_{1}}{- \ln p_{e, 1} + \rho \ln M_{1} }$ are
increasing functions of $M_{1}$, and 
\[ 
\lim_{M_{1} \rightarrow \infty}
\frac{\mathsf{K} \underline{\phi}_{1}\ln M_{1}}{- \ln p_{e, 1} + \rho
\ln M_{1} + \underline{\phi}_{1}} = \lim_{M_{1} \rightarrow \infty}
\frac{\mathsf{K} \underline{\phi}_{1}\ln M_{1}}
{- \ln p_{e, 1} + \rho \ln M_{1} } =
\frac{\mathsf{K} \underline{\phi}_{1}}{\rho},
\]
we have that for any given positive integer $M_{1}^{1}$ there exists a
positive integer $M_{1}^{2} > M_{1}^{1}$ such that $\frac{\mathsf{K}
\underline{\phi}_{1} \ln M_{1}^{2}} {\lceil - \ln p_{e, 1} + \rho \ln
M_{1}^{2} \rceil_{\underline{\phi}_{1}}} > \frac{\mathsf{K}
\underline{\phi}_{1} \ln M_{1}^{1}} {\lceil - \ln p_{e, 1} + \rho \ln
M_{1}^{1} \rceil_{\underline{\phi}_{1}}}$, and that $\lim_{M_{1}
\rightarrow \infty} = \frac{\mathsf{K} \underline{\phi}_{1} \ln
M_{1}}{\lceil - \ln p_{e, 1} + \rho \ln M_{1} \rceil_{\phi_{1}}} =
\frac{\mathsf{K} \underline{\phi}_{1}}{\rho}$. Further, for $E_{o,
j}^{s}\left(\rho, Q_{j}\right)$ as defined in~(\ref{eq:Gaussian service
quantum}), $\lim_{\rho \rightarrow 0} \frac{\mathsf{K}
\underline{\phi}_{1}}{\rho} = \mathsf{K} \ln \left( 1 +
\frac{\Gamma_{1}}{(\mathsf{K} -1)\Gamma_{1}+1} \right)$.

\noindent
{\bf Part (ii):}
For $\mathsf{K} \rightarrow \infty$, we know from Part $(C)$ of
Theorem~\ref{th:unequal1} that the system is stable if $\sum_{j=1}^{J}
\mathds{E} A_{j} S_{j} < \frac{\rho}{1+\rho}$, or equivalently,
$\sum_{j=1}^{J} \mathds{E} \tilde{A}_{j} \frac{S_{j}}{\ln M_{j}} <
\frac{\rho}{1+\rho}$.  For positive $p_{e, j}$, we have
$\frac{S_{j}}{\ln M_{j}} \rightarrow \rho$ in the limit $M_{j}
\rightarrow \infty$. Thus we have $\sum_{j=1}^{J} \mathds{E}
\tilde{A}_{j} < \frac{1}{1+\rho}$ in the limit $\min_{j} M_{j}
\rightarrow \infty$. But $\frac{1}{1+\rho} \rightarrow 1$ as $\rho
\rightarrow 0$.  Thus, we have $\sum_{j=1}^{J} \mathds{E} \tilde{A}_{j}
< 1$ in the limit $\min_{j} M_{j} \rightarrow \infty$ and $\rho
\rightarrow 0$.
\end{proof}
For each $s \in \mathcal{S}_{\mathsf{K}}$, define the vector
$\mathcal{C}(s) = \left( \mathcal{C}_{1}(s) , \mathcal{C}_{2}(s),
\ldots, \mathcal{C}_{J}(s) \right)$ of interference-limited capacities
by defining
\begin{eqnarray*}
\mathcal{C}_{j}(s) &=& \left\{
\begin{array}{ll}
s_{j} \ln \left( 1+ \frac{\Gamma_{j}}
{\sum_{i=1}^{J} s_{i}\Gamma_{i} - \Gamma_{j} +1} \right)
&\mbox{if}\;s_{j} > 0 \\
0 &\mbox{if}\;s_{j} = 0
\end{array}
\right.
\end{eqnarray*}
\begin{theorem}[Capacity Interpretation for $\omega \in
\Omega^{\mathsf{K}}$]
Let $J \geq 1$ and $\mathsf{K} \geq 1$. Consider a state-independent
scheduling policy $\omega = \left\{ p^{\omega}(s); s \in
\mathcal{S}_{\mathsf{K}} \right\}$.  Then, for Gaussian encoding of
messages, and in the limit $M_{j} \rightarrow \infty$ and $\rho
\rightarrow 0$, the threshold on $\mathds{E} \tilde{A}_{j}$ approaches
the limit $\sum_{s \in \mathcal{S}_{\mathsf{K}}: s_{j} > 0}
p^{\omega}(s) \mathcal{C}_{j}(s)$.  \Q

\end{theorem}
\begin{proof}
We know from Lemma~\ref{lemma:state-independent independent decoding}
that the queueing model is stable if the nat arrival rate for class-$j$
satisfies the inequality $\mathds{E} \tilde{A}_{j} < \sum_{s \in
\mathcal{S}_{\mathsf{K}}} p^{\omega}(s)s_{j} \phi_{j}(s) \frac{\ln
M_{j}}{S_{j} + \overline{\phi}_{j}}$. But, $\frac{\ln M_{j}}{S_{j} +
\overline{\phi}_{j}}$ increases to $\frac{1}{\rho}$ in the limit $M_{j}
\rightarrow \infty$ and, further, $\frac{s_{j}\phi_{j}(s)}{\rho}
\rightarrow \mathcal{C}_{j}(s) $ in the limit $\rho \rightarrow 0$. Thus
we have $\mathds{E} \tilde{A}_{j} < \sum_{s \in
\mathcal{S}_{\mathsf{K}}} p^{\omega}(s) \mathcal{C}_{j}(s) $ in the limit
$M_{j} \rightarrow \infty$ and $\rho \rightarrow 0$.
\end{proof}
For state-independent policy $\omega$, define the inner bound
\[
\mathcal{R}_{in}^{\omega} = \left\{ \mathds{E}A: \left[ S_{j} +
\overline{\phi}_{j} \right] \mathds{E}A_{j} < \sum_{s \in
\mathcal{S}_{\mathsf{K}}} p^{\omega}(s)s_{j}\phi_{j}(s)\; \mbox{for}\; 1
\leq j \leq J \right\}
\]
to the stability region $\mathcal{R}^{\omega}$ of message arrival rate
vectors $\mathds{E}A$. For $s \in \mathcal{S}_{\mathsf{K}}$, define the
set of vectors $\left\{r^{\prime}(s); s \in \mathcal{S}_{\mathsf{K}}
\right\}$ by defining $r^{\prime}_{j}(s) = \frac{s_{j}
\phi_{j}(s)}{S_{j} + \overline{\phi}_{j}}$. We observe that
$\bigcup_{\omega \in \Omega^{\mathsf{K}}} \mathcal{R}_{in}^{\omega}$ is
the convex hull of the rate vectors $\left\{r^{\prime}(s); s \in
\mathcal{S}_{\mathsf{K}} \right\}$. The interpretation is that the
convex hull of $\left\{r^{\prime}(s); s \in \mathcal{S}_{\mathsf{K}}
\right\}$ represents \emph{a} region of message arrival rate vectors
$\mathds{E}A$ stabilizable by the class $\Omega^{\mathsf{K}}$ of
state-independent scheduling policies. Now we give an interpretation to
the achievable stability region $\bigcup_{\omega \in
\Omega^{\mathsf{K}}} \mathcal{R}^{\omega}$ in terms of
interference-limited information-theoretic capacities.

For $s \in \mathcal{S}_{\mathsf{K}}$, define the sets of vectors
$\left\{\tilde{r}(s); s \in \mathcal{S}_{\mathsf{K}} \right\}$ and
$\left\{\tilde{r}^{\prime}(s); s \in \mathcal{S}_{\mathsf{K}} \right\}$
by defining the components $\tilde{r}_{j}(s) = (\ln M_{j}) r_{j}(s)$ and
$\tilde{r}_{j}^{\prime}(s) = (\ln M_{j}) r_{j}^{\prime}(s)$,
respectively.  In the following corollary, we show that the class
$\Omega^{\mathsf{K}}$ of state-independent scheduling policies achieve
any nat arrival rate vector that is achievable by stationary scheduling
policies in the asymptotic limit $\min_{j} M_{j} \rightarrow \infty$
corresponding to large message lengths.  
\begin{corollary}
\label{coro:in2} 
In the limit $\min_{1 \leq j \leq J} M_{j} \rightarrow \infty$, we have 
\begin{enumerate} 
\item[(i)]	convex hull of $\left\{\tilde{r}^{\prime}(s); s \in
\mathcal{S}_{\mathsf{K}} \right\}$ = convex hull of $
\left\{\tilde{r}(s); s \in \mathcal{S}_{\mathsf{K}} \right\}$ 

\item[(ii)]	in the further limit $\rho \rightarrow 0$ and for
Gaussian encoding of messages, convex hull of $\left\{\tilde{r}(s); s
\in \mathcal{S}_{\mathsf{K}} \right\}$ = convex hull of
$\left\{\mathcal{C}(s); s \in \mathcal{S}_{\mathsf{K}} \right\}$.
\end{enumerate}
\Q
\end{corollary}
\begin{proof}
For a schedule $s$ such that $s_{j} > 0$ and for positive $p_{e, j}$,
$\lim_{M_{j} \rightarrow \infty} \frac{\ln M_{j}}{S_{j} +
\overline{\phi}_{j}} = \frac{1}{\rho}$, and hence
$\tilde{r}^{\prime}_{s}(s) = \tilde{r}_{j}(s) = \frac{s_{j}
\phi_{j}(s)}{\rho}$ in the limit $M_{j} \rightarrow \infty$. Hence the
convex hull of $\left\{\tilde{r}^{\prime}(s); s \in
\mathcal{S}_{\mathsf{K}} \right\}$ = convex hull of $
\left\{\tilde{r}(s); s \in \mathcal{S}_{\mathsf{K}} \right\}$. For
$E_{o, j}^{s}\left(\rho, Q_{j}\right)$ defined in~(\ref{eq:Gaussian
service quantum}), we have $\frac{s_{j} \phi_{j}(s)}{\rho} \rightarrow
\mathcal{C}_{j}(s)$.  Hence the convex hull of $\left\{\tilde{r}(s); s
\in \mathcal{S}_{\mathsf{K}} \right\}$ = convex hull of
$\left\{\mathcal{C}(s); s \in \mathcal{S}_{\mathsf{K}} \right\}$.
\end{proof}
\begin{figure}[t!]
\centering
\includegraphics[width=\tw,height=10cm]{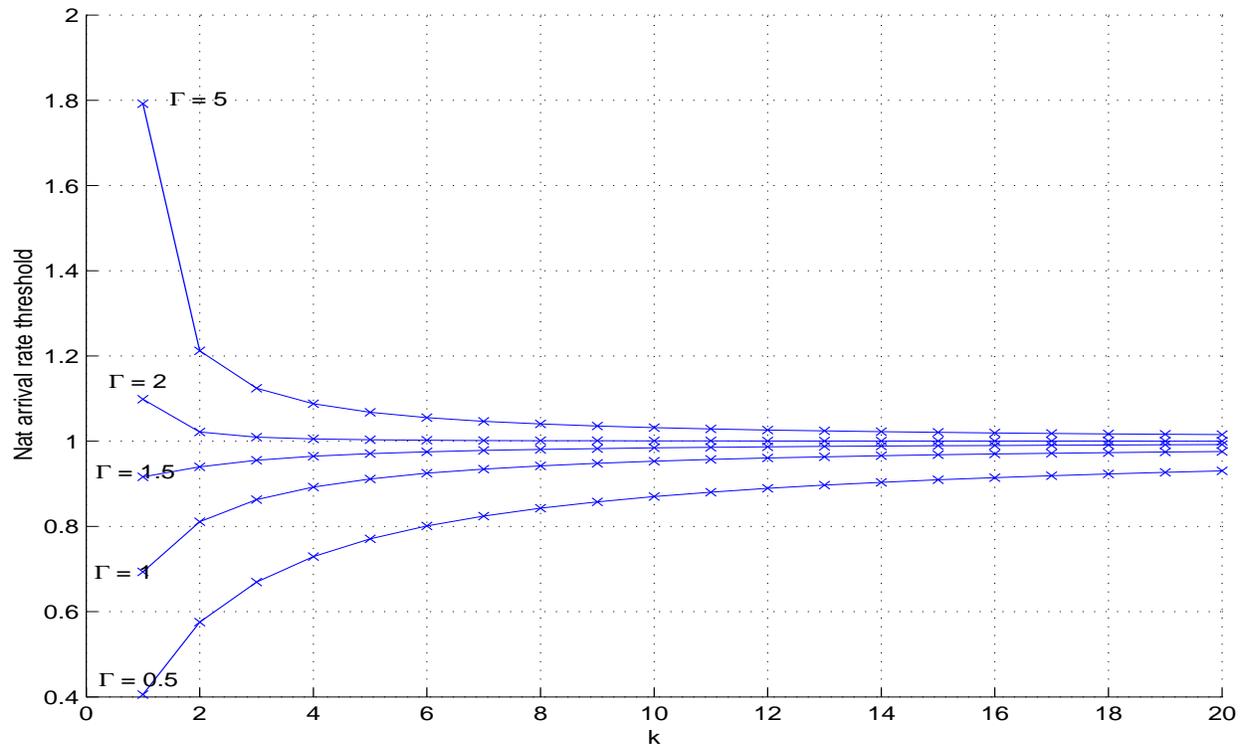}
\caption{Nat arrival rate threshold versus maximum number of
simultaneous message transmissions, $\mathsf{K}$, in the case $J=1$.}
\label{fig:equal powers infinite message length}
\end{figure}

%% file: jointDECODING.tex
\chapter{Multiaccess Communication with Joint Decoding}
\label{ch:joint decoding}

We derive a multiclass discrete-time processor-sharing queueing model
for scheduled message communication over a discrete memoryless
multiaccess channel with joint maximum-likelihood decoding, when
requests for message transmissions arrive at random times. We show that
the stability region of information arrival rate vectors is the
information-theoretic capacity region of a multiaccess channel.


\section{The Information-Theoretic Model}
Consider a discrete stationary memoryless multiple access channel over
which $J$ independent message sources communicate to a receiver. Assume
that source-$j$ has $M_{j} \geq 2$ possible message values to choose
from the message alphabet $\mathcal{M}_{j} = \left\{1, 2, \ldots, M_{j}
\right\}$. Let $M$ denote the vector of source message alphabet sizes
$(M_{1}, M_{2}, \ldots, M_{J})$. For $1 \leq j \leq J$, define the
finite set $\mathcal{X}_{j}$ to be the set of channel input letters into
which the source-$j$ output will be encoded, and $\mathcal{X}_{j}^{(n)}$
be the Cartesian product \footnote{Throughout this chapter the notation
that we use to denote code words has the following interpretation. The
superscript is a positive integer and designates the code word length.
There can be more than one entry in the subscript.  When multiple
entries are included in the subscript, they are separated by commas. The
first entry gives the identification of the source and the second entry
gives the possible message value from that source. For example,
$x^{(n)}_{j, k}(l)$ denotes the $l$th symbol of a $n$-length code word
assigned for the $k$th message value of the $j$th source.  In a slight
abuse of notation we use the notation $x_{j, k}$ in place of
$x^{(1)}_{j, k}(1)$.  In situations when we do not want to be specific
about the particular code word of a given source, we simply ignore the
second entry in the subscript. For example, $x^{(n)}_{j}$ denotes an
$n$-length code word for the $j$th source and $x^{(n)}_{j}(l)$ is its
$l$th symbol.} of $n$ copies of $\mathcal{X}_{j}$. Then $x_{j}^{(N)} \in
\mathcal{X}_{j}^{(N)}$, $N \geq 1$, is an $N$-length sequence of letters
from the set $\mathcal{X}_{j}$.  There is a finite output alphabet
$\mathcal{Y}$ and a channel transition probability assignment $ \left\{
p(y|x_{1}x_{2}\ldots x_{J}); y \in \mathcal{Y}; x_{j} \in
\mathcal{X}_{j} \;\mbox{for}\; 1 \leq j \leq J \right\}$.  The channel
is memoryless in the sense that if $x_{j}^{(N)} = (x_{j}(1), x_{j}(2),
\ldots, x_{j}(N))$ is an $N$-length sequence from the set
$\mathcal{X}_{j}$, then the probability of receiving $y^{(N)} = (y_{1},
y_{2}, \ldots, y_{N})$ for the given set of codewords $x^{(N)} = \{
x_{1}^{(N)}, x_{2}^{(N)}, \ldots, x_{J}^{(N)} \}$ is
\[
p \left( y^{(N)}\left|x^{(N)}\right. \right) = \prod_{n=1}^{N} p \left(
y_{n} \left| x_{1}(n)x_{2}(n) \ldots x_{J}(n) \right.
\right)
\]
Let $m_{j} \in \mathcal{M}_{j}$ and $\hat{m}_{j} \in \mathcal{M}_{j}$ be
two random variables that represent source-$j$ output and its estimate
at the receiver. Define the joint message $m = (m_{1}, m_{2}, \ldots,
m_{J}) \in \times_{j=1}^{J} \mathcal{M}_{j}$.  Consider block encoding
at the respective sources with block length $N$ and using $M_{j}$
codewords for the $j$th source. Let $\left\{ x_{j, k}^{(N)}: 1 \leq k
\leq M_{j} \right\}$ represent the code book for the $j$th source.  We
shall refer to a code $\left\{x_{j, k}^{(N)}; 1 \leq j \leq J; 1 \leq k
\leq M_{j} \right\}$ as an $(N, M)$ code.

Each $N$ units of time and for each $j$, source-$j$ generates an
independent random integer $m_{j}$ uniformly distributed from 1 to
$M_{j}$. The encoders transmit the respective code words $x^{(N)}_{j,
m_{j}} = \left\{ x_{j,m_{j}}(1), x_{j, m_{j}}(2), \ldots, x_{j,
m_{j}}(N) \right\}$, and the corresponding channel output $y^{(N)}$
enters the decoder, and is mapped into a decoded joint message $\hat{m}
= (\hat{m}_{1}, \hat{m}_{2}, \ldots, \hat{m}_{J})$. If $\hat{m} = m$,
i.e., $\hat{m}_{j} = m_{j}$ for each $j$, the decoding is correct,
otherwise, a decoding error occurs.  The probability of decoding error
$p_{e}$ is minimized for each $y^{(N)}$ by a maximum-likelihood decoder
by choosing $\hat{m} = (\hat{m}_{1}, \hat{m}_{2}, \ldots, \hat{m}_{J})$
that maximizes $p\left( y^{(N)} \left| x_{1, \hat{m}_{1}}^{(N)}, x_{2,
\hat{m}_{2}}^{(N)}, \ldots, x_{J, \hat{m}_{J}}^{(N)} \right. \right)$.

For each $j$, define $X_{j} \in \mathcal{X}_{j}$ to be a random variable
and define $Q_{j} = \left\{ Q_{j}(x_{j}); x_{j} \in \mathcal{X}_{j}
\right\}$ to be an arbitrary probability distribution on the set
$\mathcal{X}_{j}$.  Let $S$ denote any \empty{non-empty} subset of the
set of sources $\mathcal{J} = \{1, 2, \ldots, J \}$.  Define the vectors
$x = (x_{1}, x_{2}, \ldots, x_{J})$, $x(S) = (x_{j}; j \in S)$,
$x(S^{c}) = (x_{k}; k \in S^{c})$, $X(S) = \{X_{j}; j \in S\}$, and
$X(S) = \left\{X_{k}; k \in S^{c}\right\}$.  Then define $Q_{S} \left(
x(S) \right) = \prod_{j \in S} Q_{j}(x_{j})$, $Q(S^{c}) \left( x(S^{c})
\right) = \prod_{k \in S^{c}} Q_{k}(x_{k})$ to be probability
distributions on the product alphabets $\left( \times_{j \in S}
\mathcal{X}_{j} \right)$ and $\left( \times_{k \in S^{c}}
\mathcal{X}_{k} \right)$, respectively.  Finally, define the product
probability distribution $Q = \left\{Q(x) = \prod_{j=1}^{J}
Q_{j}(x_{j}): x_{j} \in \mathcal{X}_{j} \right\}$. Consider an ensemble
$(N, M)$ of codes in which each code word $x_{j, m_{j}}^{(N)}$, $1 \leq j
\leq J$ and $1 \leq m_{j} \leq M_{j}$, is independently selected
according to the probability distribution
\begin{eqnarray}
\label{eq:product probability distribution}
Q_{j}^{(N)} \left( x_{j, m_{j}}^{(N)} \right)
 &=& \prod_{n=1}^{N} Q_{j} \left( x_{j, m_{j}}(n) \right) 
\end{eqnarray}  
We state here the following theorem which defines the capacity region
$\mathcal{C}$ of a multiaccess channel.
\begin{theorem}[\cite{CovTho-BOOK}]
For a given product probability distribution $Q$, define the pentagon
$\mathcal{I}(Q)$ to be the set of rate vectors $r = (r_{1}, r_{2},
\ldots, r_{J}) \in \mathbb{R}_{+}^{J}$ satisfying 
\begin{eqnarray*}
\sum_{j \in S} r_{j} &\leq&  I \left( X(S);Y|X\left(S^{c}\right)
\right)
\end{eqnarray*}
for each $S \in \mathcal{P}(\mathcal{J})$. The capacity region
$\mathcal{C}$ is then defined as the convex hull of these pentagons over
all possible product probability distributions $Q$, i.e.,
$\mathcal{C}=\;\mbox{convex hull of}\; \left(\bigcup_{Q} \mathcal{I}(Q)
\right)$.  \Q
\end{theorem}
For each code in the ensemble, the decoder uses maximum-likelihood
decoding, and we wish to upper bound the expected value
$\overline{p}_{e}$ of $p_{e}$ for this ensemble.  Define
$\mathcal{P}(\mathcal{J})$ to be the set of all \emph{non-empty} subsets
of the set $\mathcal{J}$. For a given $S \in \mathcal{P}(\mathcal{J})$,
we define the decoding error event to be of the \emph{type}-$S$ if the
decoded joint message $\hat{m}$ and the original joint message $m$
satisfy: $\hat{m}_{j} \neq m_{j}$ for $j \in S$ and $\hat{m}_{k} =
m_{k}$ for $k \in S^{c}$. Let $\overline{p}_{e, S}$ be the expectation
of the probability of a type-$S$ decoding error event over the ensemble;
obviously $\overline{p}_{e} = \sum_{S \in \mathcal{P}(\mathcal{J})}
\overline{p}_{e,S}$. The following Theorem is stated
in~\cite{HanTse-JRN-ITTRAN1998},~\cite{GueVar-JRN-ITTRAN2000} and the
proof of the Theorem for two sources is given in~\cite{GAL-JRN-ITTRAN}.
\begin{theorem}[\cite{GAL-JRN-ITTRAN}]
\label{th:Slepian-Wolf}
Consider an ensemble $(N, M)$ of block codes in which, for each $j$,
code words $x_{j}^{(N)}$ in the code book are independently chosen
according to~(\ref{eq:product probability distribution}) for a given
probability distribution $Q_{j}$. Then the expected error probability
over the ensemble is $\overline{p}_{e|m} = \sum_{S} \overline{p}_{e,
S|m}$, where for $0 \leq \rho \leq 1$,
\begin{eqnarray*}
\overline{p}_{e, S|m} &\leq& \exp \left[ -N \left[ 
-\rho \sum_{j \in S} R_{j} + E_{o, S}(\rho, Q) \right] \right], \quad
\mbox{and} \\
E_{o, S}(\rho, Q) &=& -\ln \sum_{x(S^{c})} Q_{S^{c}} \left(
x(S^{c}) \right) \sum_{y} \left[ \sum_{x(S)}
Q_{S} \left( x(S) \right)p(y|x)^{\frac{1}{1+\rho}} \right]^{1+\rho} \\
R_{j} &=& \frac{\ln M_{j}}{N} \quad \quad \mbox{for $1 \leq j \leq J$}
\end{eqnarray*}
\Q
\end{theorem}

For future reference, we denote the random coding upper bound on the
expected joint message decoding error probability $\overline{p}_{e}$ by
\begin{eqnarray*}
\chi(\mathcal{J}, N) &=& \sum_{S \in \mathcal{P}(\mathcal{J})} \exp
\left[ -N \left[ -\rho \sum_{j \in S} R_{j} + E_{o, S}(\rho, Q) \right]
\right]
\end{eqnarray*}
We note here that $\chi(\mathcal{J}, N)$ also serves as an upper bound
on the expected \emph{individual message} decoding error probability.
This follows because, for $1 \leq j \leq J$, the expected probability,
over the ensemble, that the $j$th source message is in error satisfies:
$\overline{p} \left( m^{\prime}_{j} \neq m_{j} \right) = \sum_{\left\{ S
\in \mathcal{P}(\mathcal{J}):j \in S \right\}} \overline{p}_{e, S} <
\chi(\mathcal{J}, N)$.  Since no closed form expression exists for $N $,
we derive an upper bound and a lower bound to $N$ in
Lemma~\ref{lemma:bounds on N(s) joint decoding}.
\begin{lemma}
\label{lemma:bounds on N(s) joint decoding}
For a given tolerable joint message decoding error probability $p_{e}$,
let $N$ be the smallest positive integer such that $\chi(\mathcal{J},N)
\leq p_{e}$. Then
\[
\max_{S \in \mathcal{P}(\mathcal{J})} 
\frac{ \left\lceil -\ln p_{e} + \rho \sum_{j \in S} 
\ln M_{j} \right\rceil_{E_{0, S}(\rho, Q)}} {E_{0, S}(\rho, Q)} 
\leq N \leq
\max_{S \in \mathcal{P}(\mathcal{J})} 
\frac{ \left\lceil -\ln \frac{p_{e}}{2^{J}-1}  +
\rho \sum_{j \in S} \ln M_{j} \right\rceil_{E_{0, S}(\rho, Q)}}
{E_{0, S}(\rho, Q)}
\]
\end{lemma} \Q
\begin{proof}
Since $\chi(\mathcal{J}, N) \leq p_{e}$, we have that $\exp [ -N
[ E_{0, S}(\rho, Q) - \rho \sum_{j \in S} R_{j}  ] ]
\leq p_{e}$ for each $S \in \mathcal{P}(\mathcal{J})$. Equivalently,
\begin{eqnarray*}
N &\geq& \frac{ -\ln p_{e} + \rho \sum_{j \in S} \ln M_{j}
}{E_{0, S}(\rho, Q)} \qquad \qquad \forall S \in
\mathcal{P}(\mathcal{J}),  \\
\mbox{i.e.,} \;\;\; N & \geq & \max_{S \in \mathcal{P}(\mathcal{J})} 
\frac{ -\ln p_{e} + \rho \sum_{j \in S} \ln M_{j} }{E_{0, S}(\rho, Q)},\\
\mbox{i.e.,} \;\;\; N & \geq & \max_{S \in \mathcal{P}(\mathcal{J})} 
\frac{\left\lceil  -\ln p_{e} + \rho \sum_{j \in S}
\ln M_{j} \right\rceil_{E_{0, S}(\rho, Q)}}
{E_{0, S}(\rho, Q)}. \\
\end{eqnarray*}
To derive the upper bound, we observe that for at least one subset $S
\in \mathcal{P}(\mathcal{J})$, it is true that $\exp [ -N [ E_{0,
S}(\rho, Q) - \rho \sum_{j \in S} R_{j}]] \geq \frac{p_{e}}{2^{J} - 1}$,
for at least one term in $\chi(\mathcal{J}, N)$ is greater than or equal to
$\frac{p_{e}}{2^{J} - 1}$ when the sum of $2^{J} - 1$ positive terms
equals $p_{e}$. Let $\exp [ -N [ E_{0, S}(\rho, Q) - \rho \sum_{j \in S}
R_{j}]] \geq \frac{p_{e}}{2^{\mathsf{K}} - 1}$ for some subset $S \in
\mathcal{P}(\mathcal{J})$. Then it follows that
\[
N \leq \frac{\left\lceil  -\ln \frac{p_{e}}{2^{J} - 1}
+ \rho \sum_{j \in S} \ln M_{j} \right\rceil_{E_{0,
S}(\rho, Q)}} {E_{0, S}(\rho, Q)} 
\leq \max_{S \in \mathcal{P}(\mathcal{J})} 
\frac{\left\lceil  -\ln \frac{p_{e}}{2^{J} - 1} +
\rho \sum_{j \in S} \ln M_{j} \right\rceil_{E_{0,
S}(\rho, Q)}} {E_{0, S}(\rho, Q)}
\]
Hence the Lemma is proved.
\end{proof}

When a joint message consists of messages from all sources in the set
$\mathcal{J}$, assume that the codeword is of length $N$ in order that
the given tolerable joint message decoding error probability $p_{e}$ is
met. Then, for any subset $S \in \mathcal{P}(\mathcal{J})$ of sources,
when a joint message consists of messages from all sources in the subset
$S$, the codeword needs to be of length at most $N$ in order that the
same tolerable joint message decoding error probability $p_{e}$ is met.  
\begin{lemma}
\label{lemma: sub-schedule}
Let $N$ be the smallest positive integer such that $\chi
\left( \mathcal{J},N \right) \leq p_{e}$, and $S \in
\mathcal{P}(\mathcal{J})$. Then $ \chi \left( S,N \right)
\leq \chi \left( \mathcal{J},N \right)$ \Q
\end{lemma}
\begin{proof}
\begin{eqnarray*}
p_{e} & \geq & \chi (\mathcal{J},N)
= \sum_{S \in \mathcal{P}(\mathcal{J}) } \exp \left[ \rho
\sum_{j \in S} \ln M_{j} -N E_{0, S} \left(
\rho, Q \right) \right]  \\
&\geq& \sum_{S^{\prime} \in
\mathcal{P}(S) } 
\exp \left[ \rho \sum_{j \in S^{\prime}}
\ln M_{j} -N E_{0, S^{\prime}} 
\left( \rho, Q \right) \right]
= \chi ( S,N )
\end{eqnarray*}
\end{proof}

\section{The Queueing-Theoretic Model}
\label{section:queueing theoretic model joint decoding}
We now define a $J$-class discrete-time processor-sharing queueing model
for the $J$ source multiaccess channel with joint maximum-likelihood
decoding considered in the previous section, when requests for message
transmission arrive at random times.

In Section~\ref{section:stability joint decoding}, we consider
stationary scheduling policies that schedule \emph{multiple} messages
with the same message alphabet for simultaneous transmission. Consider
the set $\mathcal{S}_{\mathsf{K}}$ of schedules as defined in
Chapter~\ref{ch:chapter1}.  To interpret Theorem~\ref{th:Slepian-Wolf},
and results from Lemmas~\ref{lemma:bounds on N(s) joint decoding}
and~\ref{lemma: sub-schedule} for the schedule $s \in
\mathcal{S}_{\mathsf{K}}$, it is convenient to view the schedule $s$ as
defining a new multiaccess system that has $\mathcal{J}(s) = \left\{1,
2, \ldots, J(s) \right\}$ as the set of message sources, and message
alphabets $\mathcal{M}_{j}(s)$ for $1 \leq j \leq J(s)$.  Define $m(s) =
\left( m_{1}(s), m_{2}(s), \ldots, m_{J(s)}(s) \right)$, where $m_{j}(s)
\in \mathcal{M}_{j}(s)$ for $1 \leq j \leq J(s)$, to be a joint message
under the schedule $s$.  For the scenario {\bf (S1)} described in
Chapter~\ref{ch:introduction}, we then have $J(s)
= \sum_{j=0}^{J} I_{\{s_{j} > 0\}}$, and the schedule $s$ defines new
message alphabets for message sources that are product versions of their
original message alphabets.  For example, for source-$j$ in
$\mathcal{J}$ and for the schedule $s$ such that $s_{j} > 0$, this
product message alphabet, denoted by $\mathcal{M}_{j}(s) = \left\{1, 2,
\ldots, M_{j}^{s_{j}} \right\}$, is the Cartesian product of $s_{j}$
copies of the original message alphabet $\mathcal{M}_{j}$.  In other
words, we will be encoding $s_{j}$ messages \emph{jointly} under the
schedule $s$. Wth this view point, we redefine the coding rate $R_{k}$
in Theorem~\ref{th:Slepian-Wolf} for the scenario {\bf (S1)} as
$R_{k}(s) = \frac{s_{k}\ln M_{k}}{N(s)}$ thus emphasizing the dependence
of \emph{effective} message alphabet size on schedule $s$.  For the
scenario {\bf (S2)}, we have $J(s) = \sum_{j=1}^{J} s_{j}$, of which
$s_{j}$ sources have $\mathcal{M}_{j}$ as their message alphabet for $1
\leq j \leq J$.  Under this scenario, each message is encoded
\emph{independently}. Let $\mathcal{P}\left(\mathcal{J}(s)\right)$
denote the set of all \emph{non-empty} subsets of the set
$\mathcal{J}(s)$.  In the rest of this chapter, we define $N(s)$ for a
non-empty schedule $s$ to be the smallest positive integer such that
$\chi\left(\mathcal{J}(s), N(s)\right) \leq p_{e}$.

\begin{figure}[h!]
\includegraphics[width=\tw,height=8cm]{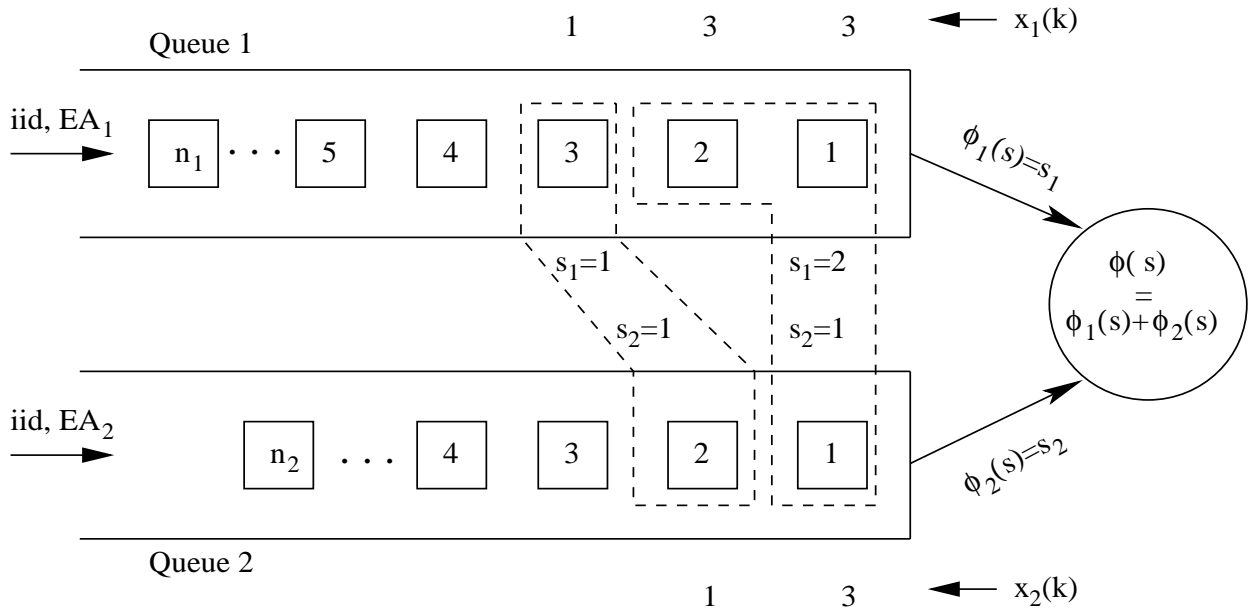}
\caption[An example of the queueing model]{Example of the queueing
model. There are two queues with mean arrival rates $\mathds{E}A_{1}$
and $\mathds{E}A_{2}$, respectively.  Individual messages that are part
of a joint message are shown by encircling them by a dotted line. We can
see that messages $4, 5, \ldots, n_{1}$ in the first queue and messages
$3, 4, \ldots, n_{2}$ in the second queue are not yet assigned to a
joint message of any schedule. Messages 1 and 2 from the first queue,
and the first message from the second queue constitute a joint message
of the schedule (2, 1).  The second joint message conforms to the
schedule (1, 1), and consists of the third message from the first queue
and the second message from the second queue.}
\label{fig:typical queue model}
\end{figure}

The service requirement $N(s)$ of a message depends on the schedule $s$
for which the message is a component message of a joint message.  The
service quantum available to a queue at a discrete-time instant depends
on the schedule employed at that instant. Define
\begin{eqnarray*}
\phi_{j}(s) &=& \left\{
\begin{array}{ll}
1 & \mbox{if}\; s_{j} > 0 \\
0 & \mbox{if}\; s_{j} = 0 \\
\end{array}
\right.
\end{eqnarray*}
to be the service quantum available to a class-$j$ message under the
schedule $s$.  Then the service quantum available to queue-$j$ is
$s_{j}$ units, and the total available service quantum is
$\sum_{j=1}^{J}s_{j}$ units.  We make two observations regarding service
requirement of, and service quantum available to, a message in the case
of independent decoding and joint maximum-likelihood decoding: (i) in
the case of independent decoding, message service requirement
characterization depended only on the message class --- whereas for
joint decoding, it depends on the particular schedule, and (ii) both
$S_{j}$ and $\phi_{j}(s)$ are positive integers for joint decoding ---
whereas they are positive real numbers for independent decoding.
Figure~\ref{fig:typical queue model} shows the queueing model for $J=2$.

We are interested in characterizing an outerbound to the region of
message arrival rate vectors $\mathds{E}A$ for which the queueing model
for joint maximum-likelihood decoding is stable for the class of
stationary scheduling policies.  In the spirit of the discussion given
in Section~\ref{section:The Queueing Model chapter 1} of
Chapter~\ref{ch:chapter1}, define, for $s \in \mathcal{S}_{\mathsf{K}}$,
the rate vector $r(s) = (r_{1}(s), r_{2}(s), \ldots, r_{J}(s))$ by
defining $r_{j}(s) = \frac{s_{j}}{N(s)}$ if $s_{j}
> 0$, and $r_{j}(s) = 0$ if $s_{j} = 0$. With this definition of the
rate vector, Theorem~\ref{th:outer bound chapter 1} can be applied to
the present context except for the following difference: for $1 \leq j
\leq J$ and $s \in \mathcal{S}_{\mathsf{K}}$ such that $s_{j} > 0$,
define $\mathds{E}A_{js}$ as the stationary rate at which messages in
queue-$j$ are assigned to joint messages of the schedule $s$ for
transmission.  Then $\mathds{E} A_{js} N(s) \leq
\pi_{\mathsf{K}}(s)s_{j}$. That is, $\mathds{E} A_{j} \leq \sum_{s \in
\mathcal{S}_{\mathsf{K}}} \pi_{\mathsf{K}}(s) r_{j}(s)$.

\section{Stability for State-Independent Scheduling Policies}
\label{section:stability joint decoding}
In this section we define the class $\Omega^{\mathsf{K}}$ of stationary
state-independent scheduling policies, and then characterize the
stability region $\mathcal{R}^{\omega}$ of message arrival rate vectors
$\mathds{E}A$ for each such policy $\omega \in \Omega^{\mathsf{K}}$.  To
implement a scheduling policy $\omega$, we further classify class-$j$
messages incoming to queue-$j$ based on the particular subclass-$(j, s)$
to be assigned to them.

For $s \in \mathcal{S}_{\mathsf{K}}$ and $1 \leq j \leq J$, we say that
the pair $(j, s)$ defines a subclass if $s_{j} > 0$. For $s_{j} = 0$,
the pair $(j, s)$ does not define a subclass. For each class-$j$ message
arrival, a subclass-$(j, s)$ is chosen independently and at random with
the fixed probability distribution defined later in~(\ref{eq:probability
distribution}), and the message is further classified by assigning the
subclass-$(j, s)$ to it. Then messages from source-$j$ and are stamped
with the subclass-$(j, s)$ are put into the subclass queue-$(j, s)$. For
subclass-$(j, s)$, let $\mathds{E}A_{js}$ denote the mean number of
messages of the subclass-$(j, s)$ that arrive to the system in a
time-slot; obviously $\sum_{\left\{ s \in \mathcal{S}_{\mathsf{K}}:
s_{j}>0 \right\}} \mathds{E}A_{js} = \mathds{E}A_{j}$.  A consequence of
class sub-classification is that messages of subclass-$(j, s)$ will be
required to transmit codewords of length $N(s)$, i.e., service
requirement gets fixed.  The state of the system is defined by the
residual service requirements of messages of each subclass present in
the system. Thus the definition of the system state $\alpha$ in the
present context is essentially the same as defined in
expression~(\ref{eq:state vector definition}) of
Chapter~\ref{ch:chapter1}, except that the state includes a message's
residual service requirement after sub-classification is done. We should
observe here that $n_{js}(\alpha) = 0$ if the pair $(j, s)$ does not
define a subclass.

We now define the notion of a schedule on the set of message subclasses.
We define a subclass schedule by a non-negative integer vector $z =
\left( z_{js}: 1 \leq j \leq J; s \in \mathcal{S}_{\mathsf{K}} \right) $
such that $0 \leq z_{js} \leq s_{j}$.  We define the set
$\mathcal{Z}_{\mathsf{K}} = \left\{ z: 0 \leq \sum_{j=1}^{J} \sum_{
\left\{ s \in \mathcal{S}_{\mathsf{K}} \right\}} z_{js} \leq \mathsf{K}
\right\}$ to be the set of all subclass schedules that schedule at most
$\mathsf{K}$ messages in each time-slot.  We say that schedule $z$ is
feasible in state $\alpha$ if $z_{js} \leq n_{js}(\alpha)$ for all
subclasses-$(j, s)$.  We implement a feasible schedule $z$ by serving
the first $z_{js}$ messages at the head of the subclass queue-$(j, s)$.
We define the \emph{ongoing transmission} of the schedule $s \in
\mathcal{S}_{\mathsf{K}}$ in state $\alpha$ as the schedule
$\eta(\alpha, s) \in \mathcal{Z}_{\mathsf{K}}$, and $\eta(\alpha, s)$ is
defined as follows: for $1 \leq j \leq J$ and $t \in
\mathcal{S}_{\mathsf{K}}$,
\begin{eqnarray*}
\eta_{jt}(\alpha, s) = \left\{
\begin{array}{ll}
\sum_{k=1}^{n_{js}(\alpha)} I_{\left\{ x_{js}(k) < N(s) \right\} } &
\mbox{if}\; t = s \;\mbox{and}\; s_{j} > 0 \\
0 & \mbox{otherwise}
\end{array}
\right.
\end{eqnarray*}
We say that a message is \emph{fresh} if the message has not yet been
scheduled for the first time, i.e., the first code letter of the
corresponding codeword is yet to be transmitted. The number of fresh
messages of subclass-$(j, s)$ in state $\alpha$ is denoted by
$\beta_{js}(\alpha)$, and is given by $\beta_{js}(\alpha) =
\sum_{k=1}^{n_{js}(\alpha)} I_{\left\{ x_{js}(k) = N(s) \right\} }$.

We constrain the operation of the system by requiring that there can be
at most one ongoing transmission for any schedule $s \in
\mathcal{S}_{\mathsf{K}}$ in any state $\alpha$. Since
first-in-first-out service discipline is used to schedule messages in
each subclass queue-$(j, s)$, we can determine whether there is an
ongoing transmission of the schedule $s$ in state $\alpha$ by examining
the residual service requirement of the messages at the head of the
subclass queues-$(j, s)$.  If there is one ongoing, then for at least
one subclass-$(j, s)$ , we have $1 \leq x_{js}(1) \leq N(s)-1$.

Formally, a policy in this class is defined by (i) an arbitrary
probability distribution $\left\{ p^{\omega}(s); s \in
\mathcal{S}_{\mathsf{K}} \right\}$, and (ii) the mapping $\left\{
\omega: \mathcal{X} \rightarrow \mathcal{Z}_{\mathsf{K}} \right\}$.  We
follow the convention that specification of the policy $\omega$ and of
the probability distribution $\left\{ p^{\omega}(s); s \in
\mathcal{S}_{\mathsf{K}} \right\}$ are equivalent. We now define the
notion of maximal schedule $z^{*} (\alpha, s)$ in the set
$\mathcal{Z}_{\mathsf{K}}$ of the schedule $s$ in state $\alpha$.

\begin{definition}[Sub-Schedule]
For $z, z^{\prime} \in \mathcal{Z}_{\mathsf{K}}$, we write $z^{\prime}
\preceq z$ if $z^{\prime}_{js} \leq z_{js}$ for each subclass-$(j, s)$.
We then say that $z^{\prime}$ is a sub-schedule of the schedule $z$.
The maximal schedule $z^{*} (\alpha, s) \in \mathcal{Z}_{\mathsf{K}}$ of
the schedule $s$ in state $\alpha$ is defined as follows: for $1 \leq j
\leq J$ and $t \in \mathcal{S}_{\mathsf{K}}$,
\begin{eqnarray*}
z^{*}_{jt} (\alpha, s) &=& \left\{
\begin{array}{ll}
\min \left\{ s_{j}, n_{js}(\alpha) \right\} & \mbox{if}\; t=s \\
0 & \mbox{otherwise}
\end{array}
\right.
\end{eqnarray*} \Q
\end{definition}
To implement a state-independent policy $\omega $, a schedule $s \in
\mathcal{S}_{\mathsf{K}}$ is chosen \emph{independent} of the state
$\alpha$ in each time-slot with probability $p^{\omega}(s)$. Then the
subclass schedule
\begin{eqnarray*}
\omega(\alpha) &=& \left\{
\begin{array}{ll}
\eta(\alpha, s), & \mbox{if}\; \eta(\alpha, s) \;\mbox{is a non-empty
schedule} \\
z^{*}(\alpha, s), &\mbox{otherwise}
\end{array}
\right.
\end{eqnarray*}
is implemented in state $\alpha$.  For the given probability distribution
$\left\{ p^{\omega}(s); s \in \mathcal{S}_{\mathsf{K}} \right\}$, the
mapping $\left\{ \omega: \mathcal{X} \rightarrow
\mathcal{Z}_{\mathsf{K}} \right\}$ induces the probability distribution
$\left\{ p^{\omega}_{\alpha}(z); z \in \mathcal{Z}_{\mathsf{K}}
\right\}$, which is defined by
\begin{eqnarray*}
p^{\omega}_{\alpha}(z) &=& \left\{
\begin{array}{ll}
p^{\omega}(s) & \mbox{if $z=\eta(\alpha, s)$ where $\eta(\alpha,
s)$ is a non-empty schedule}, \\
& \mbox{or $z=z^{*}(\alpha,s)$ and $\eta(\alpha, s)$ is the empty schedule} \\
0 & \mbox{otherwise}
\end{array}
\right.
\end{eqnarray*}
\begin{lemma}
\label{lemma:joint decoding}
Let $\mathsf{K} \geq 1$, $J \geq 1$ and $\omega \in
\Omega^{\mathsf{K}}$. For $\alpha \in \mathcal{X}$ and for each
subclass-$(j, s)$, define~\footnote{$I_{\{A\}}$ denotes the indicator
function of the
event $A$. $I_{\{A\}} = 1$ if $A$ is true, and 0 if $A$ is false.}
$h^{\omega}_{js}(\alpha) = N(s) \beta_{js}(\alpha) + s_{j}x_{js}(1) 
I_{ \left\{ n_{js}(\alpha) > \beta_{js}(\alpha)\right\}}$,
$c(\alpha) = 1 + \sum_{js} h^{\omega}_{js}(\alpha)$, and $V(\alpha) = \sum_{js}
\frac{\left(h^{\omega}_{js}(\alpha)\right)^{2}}
{2 \left( p^{\omega}(s)s_{j} - N(s) \mathds{E}A_{js} \right)}$.
Then the Markov chain is $c$-regular and stable if $ \mathds{E}A_{js}
N(s) < p^{\omega}(s)s_{j}$ for each subclass-$(j, s)$. \Q
\end{lemma}
\begin{proof}
For each subclass-$(j, s)$, define $\left\{ \mathcal{H}_{js},
\mathcal{H}_{js}^{c} \right\}$ to be a partition of $\mathcal{X}$ such
that $\mathcal{H}_{js}^{c} = \left\{ \alpha \in \mathcal{X}:
\eta(\alpha, s)\; \mbox{is a non-zero schedule, or}\; \beta_{js}(\alpha)
\geq s_{j} \right\}$.  Let $a_{js}$ subclass-$(j, s)$ messages get
generated in state $\alpha$ and that the feasible schedule $z \in
\mathcal{Z}_{\mathsf{K}}$ is implemented in the state $\alpha$. Assuming
that the chain moves to state $\alpha^{\prime}$, we have
\begin{eqnarray*}
h^{\omega}_{js} \left(\alpha^{\prime} \right) &=&
h^{\omega}_{js}(\alpha) + f_{js}(a) - g_{js}(\alpha, z), 
\;\mbox{where} \\
f_{js}(a) &=& a_{js} N(s) \quad \mbox{and} \quad 
g_{js}(\alpha, z) = z_{js}
\end{eqnarray*}
But,
\begin{eqnarray*}
z_{js} &=& \left\{
\begin{array}{ll}
0 & \mbox{if $\alpha$ is the zero state, or $\alpha \in
\mathcal{H}_{js}^{c}$, $z \neq \eta(\alpha, s)$,} \\
& \mbox{and $z \neq
z^{*}(\alpha, s)$} \\
s_{j} & \mbox{if $\alpha \in
\mathcal{H}_{js}^{c}$, and either $z = \eta(\alpha, s)$ or $z =
z^{*}(\alpha, s)$}\\
(n_{js}(\alpha) - s_{j})N(s) + s_{j} & \mbox{if $\alpha \in
\mathcal{H}_{js}$ and $\alpha$ is a non-zero state}
\end{array}
\right.
\end{eqnarray*}
Now consider $\alpha \in \mathcal{H}_{js}^{c}$. Then $g_{js}(\alpha) =
\sum_{z \in \mathcal{Z}_{\mathsf{K}}} g_{js}(\alpha, z) p^{\omega}(z)=
s_{j} p^{\omega}(s)$.  Also, $\mathds{E}f_{js} = N(s) \mathds{E}
A_{js}$.

Assuming $N(s) \mathds{E}A_{js} < p^{\omega}(s)s_{j}$ for each subclass-
$(j, s)$, and then applying Lemma~\ref{lemma:c-regularity} to
$c(\alpha)$ and $V(\alpha)$ as defined in the statement of
Lemma~\ref{lemma:joint decoding}, we find that the queueing model
$\{X_{n}; n \geq 0 \}$ is $c$-regular.  Since there can be at most one
ongoing transmission of any schedule $s$ in any state $\alpha$, we have
$n_{js}(\alpha) \leq \beta_{js}(\alpha) + s_{j}$. By observing that
$x_{js}(k) \geq 1$ and $N(s) \geq 1$, we have
\[
h^{\omega}_{js}(\alpha) \geq \beta_{js}(\alpha) + s_{j}I_{ \left\{
n_{js}(\alpha) > \beta_{js}(\alpha) \right\} } = \left\{
\begin{array}{ll}
n_{js}(\alpha) & \mbox{if}\; n_{js}(\alpha) = \beta_{js}(\alpha) \\
\beta_{js}(\alpha) + s_{j} & \mbox{otherwise}
\end{array}
\right.
\]
Since $h_{js}^{\omega}(\alpha) \geq n_{js}(\alpha)$ for every $\alpha$,
existence of finite stationary mean for $c(\alpha)$ implies existence of
finite stationary mean for $n(\alpha)$.  Hence the queueing model is
stable.
\end{proof}

Let $\mu_{j} = \left( \mu_{js}; s \in \mathcal{S}_{\mathsf{K}}\;
\mbox{and}\; s_{j} > 0 \right)$ be a splitting probability vector
defined by
\begin{eqnarray}
\label{eq:probability distribution}
\mu_{js} &=& \frac{ \frac{p^{\omega}(s)s_{j}}{N(s)}}
{\sum_{ \{ s^{\prime} \in \mathcal{S}_{\mathsf{K}}: s^{\prime}_{j} > 0
\} }
\frac{p^{\omega}\left( s^{\prime} \right) s^{\prime}_{j}}{N \left( s^{\prime}
\right)}}.
\end{eqnarray}
with the interpretation that $\mu_{js}$ is the probability that a
class-$j$ message request is assigned the schedule $s$.
\begin{lemma}
\label{lemma:joint decoding transience}
For $\mathsf{K} \geq 1$ and $J \geq 1$, the Markov chain is unstable if
$N(s) \mathds{E} A_{js} >  p^{\omega}(s) s_{j}$ for at least one
subclass-$(j, s)$. \Q
\end{lemma}
\begin{proof}
For the subclass-$(j, s)$, define $h^{\omega}_{js}(\alpha) =
\sum_{k=1}^{n_{js}(\alpha)} x_{js}(k)$.  Then, we have
\begin{eqnarray*}
h^{\omega}_{js} \left(\alpha^{\prime} \right) &=&
h^{\omega}_{js}(\alpha) + f_{js}(a) - g_{js}(\alpha, z), \;\mbox{where}
\end{eqnarray*} 
$f_{js}(a) = a_{js}N(s)$ and $g_{js}(\alpha, z) = z_{js}$.  Consider the
partition $\left\{ \mathcal{H}_{js}, \mathcal{H}_{js}^{c} \right\}$ of
the state space $\mathcal{X}$ defined by $\mathcal{H}_{js}^{c} = \left\{
\alpha \in \mathcal{X}: n_{js}(\alpha) > 0 \right\}$.   We now consider
$\alpha \in \mathcal{H}_{js}^{c}$. Since $z_{js} \leq s_{j}$, we have
$g_{js}(\alpha) = \sum_{ \left\{ z \in \mathcal{Z}_{\mathsf{K}}
\right\}} g_{js}(\alpha, z) p^{\omega}_{\alpha}(z) \leq
s_{j}p^{\omega}(s)$. Also, $\mathds{E}f_{js} = N(s)\mathds{E}A_{js}$.
By applying Lemma~\ref{lemma:transience} to $V(\alpha) = 1 -
\theta^{h^{\omega}_{js}(\alpha)}$, $0 < \theta < 1$, we find that for
$N(s) \mathds{E} A_{js} > p^{\omega}(s) s_{j}$ the Markov chain
is unstable.
\end{proof}

From Lemma~\ref{lemma:joint decoding} and
Lemma~\ref{lemma:joint decoding transience}, we can easily see that
\begin{eqnarray*}
\mathcal{R}^{\omega} &=& \left\{\mathds{E}A: \mathds{E}A_{j} <
\sum_{s \in \mathcal{S}_{\mathsf{K}}}
p^{\omega}(s) r_{j}(s) \quad \mbox{for}\; 1 \leq j \leq J
\right\},
\end{eqnarray*}
and that the threshold on $\mathds{E}A_{j}$ is a convex combination of
the set of rates $\left\{ r_{j}(s); s \in \mathcal{S}_{\mathsf{K}}
\right\}$. Define $\mathcal{R}\left( \Omega^{\mathsf{K}} \right) =
\bigcup_{\omega \in \Omega^{\mathsf{K}}} \mathcal{R}^{\omega}$. Then
$\mathcal{R}\left( \Omega^{\mathsf{K}} \right)$ is the interior of the
convex hull of the rate vectors $\left\{r(s); s \in
\mathcal{S}_{\mathsf{K}} \right\}$. We denote the interior of the set
$A$ by $A^{o}$.
\begin{corollary}
\label{corollary:inner bound equals outer bound}
$\mathcal{R} \left( \Omega^{\mathsf{K}} \right) =
\mathcal{R}_{out}^{o}$. For any given message arrival rate vector
$\mathds{E}A \in \mathcal{R}^{o}_{out}$, there exists a
state-independent scheduling policy $ \left\{p^{\omega}(s); s \in
\mathcal{S}_{\mathsf{K}}\right\}$ such that the queueing model is
stable.  \Q
\end{corollary}
The significance of this Corollary is that, if the queueing model is
stable for the message arrival processes $\{A_{j}; 1 \leq j \leq J\}$
and an arbitrary stationary scheduling policy, then there exists a
state-independent scheduling policy $\omega \in \Omega^{\mathsf{K}}$
such that the queueing model is stable for the same message arrival
processes $\{A_{j}; 1 \leq j \leq J\}$.

\begin{proof}
Suppose that, for some stationary scheduling policy, the queueing model
$\{X_{n}; n \geq 0\}$ is stable for the message arrival processes
$\{A_{j}; 1 \leq j \leq J\}$.  Let $\left\{\pi_{\mathsf{K}}(s): s \in
\mathcal{S}_{\mathsf{K}}\right\}$ be the induced stationary probability
distribution on the set of schedules $\mathcal{S}_{\mathsf{K}}$. Let
$\pi_{\mathsf{K}}(0) > 0$ be the stationary probability that \emph{no}
schedule is served in a time-slot. Since the queueing model is stable,
the stationary mean residual service for subclass-$(j, s)$ is finite,
and hence $\mathds{E}A_{js} N(s) = \pi_{\mathsf{K}}(s)s_{j}$.

Let us define a state-independent scheduling policy $\left\{
p^{\omega}(s); s \in \mathcal{S}_{\mathsf{K}} \right\}$ as follows. For
non-empty schedule $s \in \mathcal{S}_{\mathsf{K}}$, define
$p^{\omega}(s) = \pi_{\mathsf{K}}(s) + \epsilon_{s}$ where $\epsilon_{s}
> 0$ and $\sum_{s} \epsilon_{s} = \pi_{\mathsf{K}}(0)$.  Then, for each
subclass-$(j, s)$, $\mathds{E}A_{js} N(s) < p^{\omega}(s)s_{j} $. That
is, for the message arrival processes $\{A_{j}, 1 \leq j \leq J\}$, the
state-independent policy $\omega$ makes the queueing model stable.
\end{proof}

\section{Information-Theoretic Interpretation to the Stability Region}
\label{section:capacity}
For a fixed state-independent schedule $s \in \mathcal{S}_{\mathsf{K}}$, i.e.,
$p^{\omega}(s) = 1$, we know from Lemma~\ref{lemma:joint decoding} and
Lemma~\ref{lemma:joint decoding transience} that the queueing model is
stable if $\mathds{E}\tilde{A}_{j} < R_{j}(s)$ for $1 \leq j \leq J$ and
$s_{j} > 0$, and unstable if $\mathds{E}\tilde{A}_{j} > R_{j}(s)$ for at
least one queue $j$ such that $s_{j} > 0$. We remind the reader that
$R_{j}(s) = \frac{s_{j} \ln M_{j}}{N(s)}$.

In this section, we give the information-theoretic interpretation to the
stability region of nat arrival rate vectors $\mathds{E}\tilde{A}$ for
the scenario ({\bf S1}). A formal statement of this interpretation is
made in Theorem~\ref{th:capacity interpretation joint decoding}.  For $s
\in \mathcal{S}_{\mathsf{K}}$, define the code rate vector $R(s) =
\left(R_{1}(s), R_{2}(s), \ldots, R_{J}(s) \right)$.  In
Theorem~\ref{th:capacity interpretation joint decoding}, we show the
following: (i) for a given joint probability distributions $Q$, and
message arrival processes $\left\{ A_{j}; 1 \leq j \leq J \right\}$ such
that $\mathds{E} \tilde{A} = r \in \mathcal{I}^{o}(Q)$, we determine a
schedule $s$, message alphabet size vector $M$, and a value for the
parameter $\rho$ such that the message communication system for $s$,
$M$, $\rho$, and the arrival processes $\left\{ A_{j}; 1 \leq j \leq J
\right\}$ , is stable (i.e., $R_{j}(s) > r_{j}$, $1 \leq j \leq J$);
(ii) for any $s$, $M$, and $\rho$, we show that $R(s) \in
\mathcal{I}^{o}(Q)$. Define $\mathcal{R}(Q) = \left\{ R(s): 0 < \rho
\leq 1; \mathsf{K} \geq 1; s \in \mathcal{S}_{\mathsf{K}}; M \in
\mathbb{Z}_{+}^{J} \right\}$ to be the set of all possible code rate
vectors $R(s)$. 
\begin{theorem}[Information-Theoretic Interpretation]
\begin{eqnarray*}
\mathcal{R}(Q) &=& \mathcal{I}^{o}(Q) 
\end{eqnarray*} \Q
\label{th:capacity interpretation joint decoding}
\end{theorem}
\begin{proof}
We first show that $\mathcal{I}^{o}(Q) \subset \mathcal{R}(Q) $.  Choose
an $r  \in \mathcal{I}^{o}(Q)$. Then there exists an $\epsilon > 0$ such
that $r+\epsilon = (r_{1}+\epsilon, r_{2}+\epsilon, \ldots,
r_{J}+\epsilon) \in \mathcal{I}^{o}(Q)$. For $1 \leq j \leq J$ and a
positive real number $\mathsf{A}$, let us first choose $s_{j}$ and
$M_{j}$ as real numbers such that the product $s_{j}\ln M_{j} =
\mathsf{A}(r_{j}+\epsilon)$.  From Lemma~\ref{lemma:bounds on N(s) joint
decoding},
\begin{eqnarray*}
\min_{S \in \mathcal{P}(\mathcal{J})}
\frac{s_{k} (\ln M_{k}) E_{0, S}(\rho, Q)}
{ \left\lceil -\ln \frac{p_{e}}{2^{J}-1} + \rho \sum_{j \in S}
s_{j} \ln M_{j} \right\rceil_{E_{0, S}(\rho, Q)}}
&\leq& R_{k}(s) \\
& \leq &
\min_{S \in \mathcal{P}(\mathcal{J})}
\frac{s_{k} (\ln M_{k}) E_{0, S}(\rho, Q)} { \left\lceil -\ln p_{e}  + 
\rho \sum_{j \in S} s_{j} \ln M_{j} \right\rceil_{E_{0, S}(\rho, Q)}}
\end{eqnarray*}
We can see that
\begin{eqnarray*}
\lim_{\rho \rightarrow 0} \lim_{\mathsf{A} \rightarrow \infty} 
R_{k}(s) &=& 
\lim_{\rho \rightarrow 0} \lim_{\mathsf{A} \rightarrow \infty}
\min_{S \in \mathcal{P}(\mathcal{J})}
\frac{\mathsf{A} (r_{k} + \epsilon) E_{0, S}(\rho, Q)}
{ \left\lceil -\ln p_{e} + \rho \sum_{j \in S}
\mathsf{A} (r_{j}+\epsilon)  \right\rceil_{E_{0, S}(\rho, Q)}} \\
&=& \lim_{\rho \rightarrow 0} \lim_{\mathsf{A} \rightarrow \infty}
\min_{S \in \mathcal{P}(\mathcal{J})}
\frac{\mathsf{A} (r_{k}+\epsilon) E_{0, S}(\rho, Q)} { \left\lceil -\ln
\frac{p_{e}}{2^{J}-1}  + \rho \sum_{j \in S} \mathsf{A} (r_{j}
+ \epsilon)
\right\rceil_{E_{0, S}(\rho, Q)}} \\
&=& \lim_{\rho \rightarrow 0}  \min_{S \in \mathcal{P}(\mathcal{J})}
\frac{(r_{k} + \epsilon) E_{0, S}(\rho, Q)}
{\rho \sum_{j \in S}(r_{j}+\epsilon)} \\
&\stackrel{(a)}{=} &  \min_{S \in \mathcal{P}(\mathcal{J})}  (r_{k} + \epsilon)
\frac{I \left( X(S); Y | X(S^{c}) \right)}{\sum_{j \in
S}(r_{j}+\epsilon)} \\
&\stackrel{(b)}{>}& r_{k} + \epsilon,
\end{eqnarray*}
where $(a)$ follows from Part (i) of Lemma~\ref{lemma:MAC appendix B},
and $(b)$ follows from the fact that $r + \epsilon  \in
\mathcal{I}^{o}(Q)$ and hence $\sum_{j \in S} (r_{j} + \epsilon)  < I
\left( X(S); Y | X(S^{c}) \right)$.  Denote by $\lim_{\rho \rightarrow
0} \lim_{\mathsf{A} \rightarrow \infty} R_{k}(s) = R^{*}(s)$.

Choose two positive real numbers $\delta_{k}$ and $\delta_{k}^{\prime}$
such that $\epsilon - \delta_{k}
- \delta_{k}^{\prime} > 0$. Then there exists a $\rho \left( \delta_{k}
  \right) < 1$ such that for all $0 < \rho < \rho \left( \delta_{k}
\right)$, we have $\lim_{\mathsf{A} \rightarrow \infty} R_{k}(s) >
R^{*}(s) - \delta_{k} > r_{k} + \epsilon - \delta_{k}$. Now, for a fixed
value $\rho_{k}$ for $\rho$ such that $\rho_{k} < \rho \left( \delta_{k}
\right)$, there exists a $\mathsf{A} \left( \rho_{k},
\delta_{k}^{\prime} \right)$ such that for all $\mathsf{A} > \mathsf{A}
\left( \rho_{k}, \delta_{k}^{\prime} \right)$, we have $R_{k}(s) >
\lim_{\mathsf{A} \rightarrow \infty} R_{k}(s) - \delta_{k}^{\prime} >
r_{k} + \epsilon - \delta_{k} - \delta_{k}^{\prime} > r_{k}$. Choose an
$\mathsf{A}_{k}$ for $\mathsf{A}$ such that $\mathsf{A}_{k} >
\mathsf{A}_{k} \left( \rho_{k}, \delta_{k}^{\prime} \right)$. Define
$\mathsf{A}^{*} = \max_{k} \mathsf{A}_{k}$ and $\rho^{*} = \min_{k}
\rho_{k}$.

Since $s_{j}$ and $M_{j}$ for $1 \leq j \leq J$ have to be positive
integers, one can, for a given $\mathsf{A}^{*}$ choose $s_{j} =
\left\lceil \frac{\mathsf{A}^{*}(r_{j}+\epsilon)}{\ln M_{j}}
\right\rceil$ for a given $M_{j}$, and $M_{j} = \left\lceil \exp\left(
\frac{\mathsf{A}^{*}(r_{j}+\epsilon)}{s_{j}} \right) \right\rceil$ for a
given $s_{j}$, and still have the same limit as above.

Next, we prove $ \mathcal{R}(Q) \subset \mathcal{I}^{o}(Q)$ by showing
that $R(s)$ for each triplet $s$, $\rho$, and $M$ satisfies all the
$2^{J}-1$ constraints that define the set $\mathcal{I}^{o}(Q)$.  From
Lemma~\ref{lemma:bounds on N(s) joint decoding}, we have for any $S \in
\mathcal{P}(\mathcal{J})$ that
\begin{eqnarray*}
\sum_{k \in S} R_{k}(s)
& < & \sum_{k \in S} \min_{S^{\prime} \in \mathcal{P}(\mathcal{J})}
\frac{s_{k} (\ln M_{k}) E_{0, S^{\prime}}(\rho, Q)} { \left\lceil -\ln
p_{e}  + \rho \sum_{j \in S^{\prime}} 
s_{j} \ln M_{j} \right\rceil_{E_{0, S^{\prime}}(\rho, Q)}} \\
& < & \sum_{k \in S}
\frac{s_{k} (\ln M_{k}) E_{0, S}(\rho, Q)} { \left\lceil -\ln
p_{e}  + \rho \sum_{j \in S} 
s_{j} \ln M_{j} \right\rceil_{E_{0, S}(\rho, Q)}} \\
& < & \sum_{k \in S} \frac{s_{k} (\ln M_{k}) E_{0,
S}(\rho, Q)} {\rho \sum_{j \in S} s_{j} \ln M_{j}} \\
&=& \frac{E_{0, S}(\rho, Q)}{\rho} \\
&\stackrel{(c)}{<} &I \left( X(S); Y|X(S^{c}) \right),
\end{eqnarray*} 
where $(c)$ follows from Part (ii) of Lemma~\ref{lemma:MAC appendix B}
Thus, $R(s) \in \mathcal{I}^{o}(Q)$ for each $s$, $\rho$, and $M$.
\end{proof}

%% file: broadcastChannel.tex
\chapter{Communication Over Degraded Broadcast Channels}
\label{ch:degraded broadcast channel}

The primary intention in this chapter is to demonstrate that the
queueing-theoretic model derived for scheduled multiaccess message
communication with joint maximum-likelihood decoding in
Chapter~\ref{ch:joint decoding} can be used to model scheduled message
communication over degraded broadcast channels with random message
arrivals. Due to similarity in the queueing model, we skip queueing
model analysis details wherever and whenever possible.

\section{The Information-Theoretic Model}
\label{section:information theoretic model degraded broadcast channel}
A broadcast channel is one through which one source communicates its
information to two or more receivers. Formally, a discrete-time
stationary memoryless broadcast channel with $J$ receivers is defined by
a finite input alphabet $\mathcal{X}$ and finite output alphabets
$\mathcal{Y}_{j}$, $1 \leq j \leq J$, and a transition probability law
$\{ p(y_{1}, y_{2}, \ldots, y_{J}|x); x \in \mathcal{X} \; \mbox{and} \;
y_{j} \in \mathcal{Y}_{j} \}$. An assumption inherent in this definition
is ``no-collaboration'' among the $J$ receivers. This assumption then
allows us to view a broadcast channel as a collection of $J$ single-user
channels with marginal transition probabilities $p(y_{1}|x), p(y_{2}|x),
\ldots, p(y_{J}|x)$.

For $1 \leq j \leq J$ and integers $M_{j} \geq 2$, let $\mathcal{M}_{j}
= \{ 1, 2, \ldots, M_{j} \}$ denote the message alphabet of the $j$th
source, and define $\mathcal{X}_{j}$ to be a finite set of symbols. For
$1 \leq j \leq J$, define the random variables $X_{j}$ and $Y_{j}$ that
take values in the sets $\mathcal{X}_{j}$ and $\mathcal{Y}_{j}$,
respectively.  Let the $j$th source output be modeled by the random
variable $m_{j}$ that takes values in the set $\mathcal{M}_{j}$. Some
notation specific to this chapter is introduced now. Let $a$ and $b$ be
any two positive integers such that $1 \leq a \leq b \leq J$.  We define
the Cartesian products $\mathcal{M}_{a}^{b} = \mathcal{M}_{a} \times
\mathcal{M}_{a+1} \times \cdots \times \mathcal{M}_{b}$, and similarly
$\mathcal{X}_{a}^{b}$ and $\mathcal{Y}_{a}^{b}$. Then the vectors
$m_{a}^{b} = (m_{a}, m_{a+1}, \ldots, m_{b}) \in \mathcal{M}_{a}^{b}$,
and similarly $x_{a}^{b} \in \mathcal{X}_{a}^{b}$.  Define $Q_{J} =
\left\{ Q_{J}(x_{J}); x_{J} \in \mathcal{X}_{J} \right\}$ to be an
arbitrary probability assignment on $\mathcal{X}_{J}$. For $1 \leq j
\leq J-1$, define $Q_{j} \left(x_{j+1}\right) = \left\{ Q_{j} \left(
x_{j}|x_{j+1} \right); x_{j} \in \mathcal{X}_{j} \right\}$ to be an
arbitrary probability assignment on $\mathcal{X}_{j}$ for each $x_{j+1}
\in \mathcal{X}_{j+1}$ .  Define $Q_{j}^{J} = \left\{ Q_{j}^{J}
\left(x_{j}^{J}\right) = Q_{J} \left( x_{J} \right) \prod_{l=j}^{J-1}
Q_{l} \left(x_{l}|x_{l+1} \right); x_{j}^{J} \in \mathcal{X}_{j}^{J}
\right\}$ to be the product distribution on $\mathcal{X}_{j}^{J}$.

\begin{figure}[h!]
\label{fig:model}
\includegraphics[width=\tw,height=5cm]{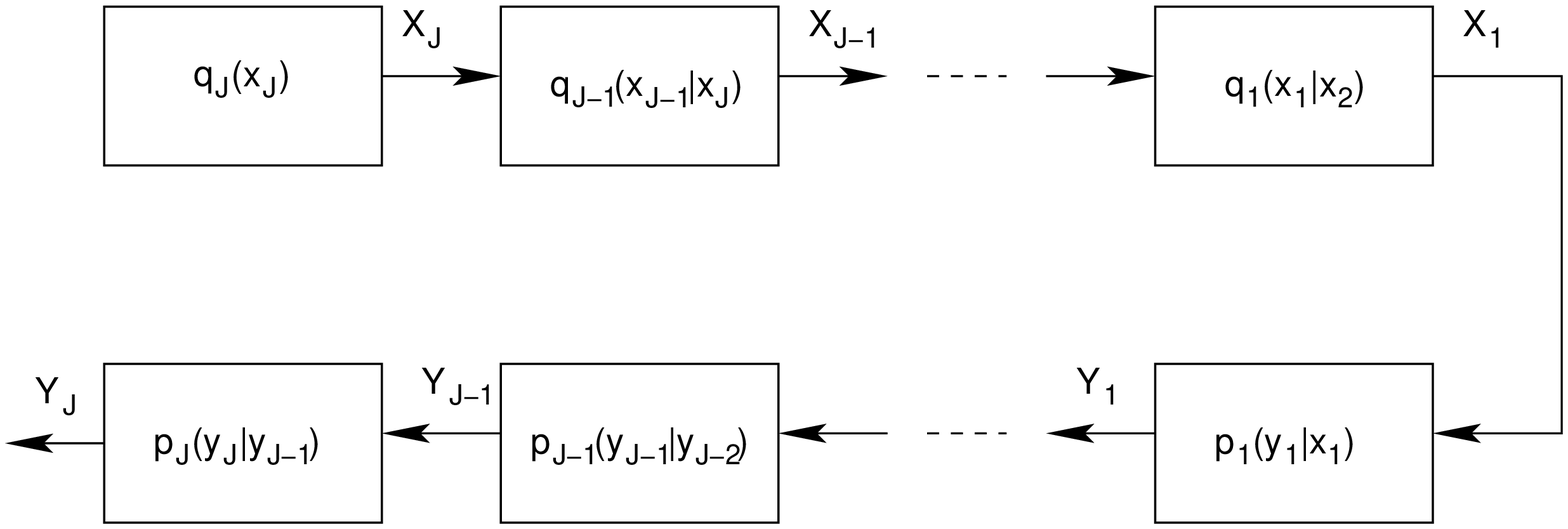}
\caption{Model of Degraded Broadcast Channel}
\end{figure}

A broadcast channel $\{ p(y_{1}, y_{2}, \ldots, y_{J}|x_{1}); x_{1} \in
\mathcal{X}_{1} \; \mbox{and} \; y_{j} \in \mathcal{Y}_{j} \}$ is said
to be degraded if $X_{1} \rightarrow Y_{1} \rightarrow Y_{2} \rightarrow
\cdots \rightarrow Y_{J}$ is a Markov chain, i.e., for $2 \leq j \leq
J$, there exist probability distributions $p_{j}(y_{j}|y_{j-1})$ such
that $p(y_{j}|x_{1}) = \sum_{y_{1}^{j-1}} \left( p(y_{1}|x_{1})
\prod_{l=2}^{j} p_{l}(y_{j}|y_{j-1}) \right)$.  Fig.~\ref{fig:model}
shows a degraded broadcast channel through which $J$ sources communicate
information to the respective receivers.  We note that with
superposition encoding $X_{J}, \ldots, X_{1}, Y_{1}, \ldots, Y_{J}$ is a
Markov chain, and for $2 \leq j \leq J$, the $j$th channel is a degraded
version of the $(j-1)$th channel.

Formally, for an integer $N \geq 1$, the superposition encoder is
defined by the mapping $\left\{ \mathcal{M}_{1}^{J} \rightarrow
\mathcal{X}_{1}^{(N)} \right\}$, and the decoder at the $j$th receiver
is defined by the mapping $\left\{ Y_{j}^{(N)} \rightarrow
\mathcal{M}_{j} \right\}$.  The capacity region for general degraded
broadcast channels, first conjectured in Cover~\cite{Cov-JRN-ITTRAN},
was established by Bergmans~\cite{Ber-JRN-ITTRAN}. The converse was
established by Bergmans~\cite{Ber-JRN-ITTRAN-CONVERSE} and
Gallager~\cite{Gal-JRN-PIT}.
\begin{theorem}[\cite{Ber-JRN-ITTRAN}] 
Consider a degraded broadcast channel consisting of $J$ component
channels (receivers) and represented as the Markov chain
\begin{eqnarray*}
X_{J} \rightarrow X_{J-1} \rightarrow \cdots \rightarrow X_{2}
\rightarrow X_{1} \rightarrow Y_{1} \rightarrow Y_{2} \rightarrow \cdots
\rightarrow Y_{J-1} \rightarrow Y_{J}.
\end{eqnarray*}
For a given joint probability distribution 
\begin{eqnarray*}
Q\left(x_{1}^{J}\right) &=&
Q_{J}(x_{J}) Q_{J-1}(x_{J-1}|x_{J}) \cdots Q_{1}(x_{1}|x_{2})
p(y_{1}y_{2} \cdots y_{J}|x_{1}), 
\end{eqnarray*}
define $\mathcal{I}(Q)$ to be the set of rate vectors $r = (r_{1},
r_{2}, \ldots, r_{J})\in \mathbb{R}_{+}^{J}$ satisfying $r_{j} \leq
I(X_{j}; Y_{j}|X_{j+1})$ for $1 \leq j \leq J-1$, and  $r_{J} \leq
I(X_{J}; Y_{J})$.  The capacity region $\mathcal{C}$ is then defined as
the convex hull of $\bigcup_{Q} \mathcal{I}(Q)$.  \Q
\end{theorem}

Let $X^{(N)}_{m_{j}^{J}} \in \mathcal{X}_{j}^{(N)}$ denote the codeword
chosen for the message vector $m_{j}^{J}$. Then, for $1 \leq l \leq N$,
let $x_{m_{j}^{J}}(l)$ denote the $l$th letter of the codeword.  The
ensemble of broadcast codes we consider here is the same as
Bergmans~\cite{Ber-JRN-ITTRAN} constructed.  The random code ensemble is
generated in $J$ stages as follows.  First, consider the ensemble of
$M_{J}$ code words $\left\{ X^{(N)}_{m_{J}^{J}} \right\}$ in which each
of the $N$ letters in each of the $M_{J}$ code words is independently
selected according to the probability assignment $Q_{J}$. For each of
these code words, we choose $M_{J-1}$ code words with independent
letters from the set $\mathcal{X}_{J-1}$ according the assignment
$Q_{J-1}$.  That is, conditional on $X^{(N)}_{m^{J}_{J}} = \left(
x_{m^{J}_{J}}(l); 1 \leq l \leq N \right)$ being the $m^{J}_{J}$th
code word, $1 \leq m^{J}_{J} \leq M_{J}$, the probability of a code word
$X^{(N)}_{m^{J-1}_{J}} = \left( x_{m^{J-1}_{J}}(l); 1 \leq l \leq
N \right)$, $m^{J-1}_{J} = (m_{J-1}, m_{J})$, $m_{J-1} \in
\mathcal{M}_{J-1}$ and $m_{J} \in \mathcal{M}_{J}$, is
\begin{eqnarray*}
p \left( X^{(N)}_{m^{J-1}_{J}} \left| X^{(N)}_{m^{J}_{J}} 
\right. \right) &=&
\prod_{l=1}^{N} Q_{J-1} \left( x_{m^{J-1}_{J}}(l) \left|
x_{m^{J}_{J}}(l) \right. \right)
\end{eqnarray*}
Continuing this way, we would have, at the beginning of the $j$th stage,
generated $M_{j+1}M_{j+2} \cdots M_{J}$ codewords.  At the end of the
$J$th stage, $M_{1}M_{2} \cdots M_{J}$ code words will be generated.
During the $j$th stage, the process of codeword generation can be
modeled by an \emph{artificial} DMC with transition probability
$Q_{j}(x_{j}|x_{j+1})$. Each of the $N$-length $M_{j+1}M_{j+2} \cdots
M_{J}$ codewords generated so far are passed through the artificial
channel $M_{j}$ times, thus generating a total of $M_{j}M_{j+1} \cdots
M_{J}$ codewords.

A random coding upper bound on message decoding error probabilities for
the two receiver degraded broadcast channel was derived
in~\cite{Gal-JRN-PIT}. Here we extend that result to a degraded
broadcast channel with arbitrary number of receivers.  The objective of
the decoder at the $j$th receiver is to compute an estimate $\hat{m}_{j,
j}$~\footnote{Since, for $j \leq k \leq J$, $k$th source message is
estimated at the $j$th receiver, we denote an estimate of the $k$th
source at the $j$th receiver by $\hat{m}_{k, j} \in \mathcal{M}_{j}$.}
of $m_{j}$. This is achieved by successive decoding, with the $j$th
decoder first decoding and then subtracting the signals intended for the
users with noisier channels before decoding its own.  Let the event
$\left\{ \hat{m}_{k, j} \neq m_{k} \right\}$ be the event that the
decoder at the $j$th receiver makes an error in decoding the $k$th
source.  The probability of error for the $j$th decoder then is $p
\left( \left\{ \hat{m}_{j, j} \neq m_{j} \right\} \right)$.

For $1 \leq j \leq J$ and $j \leq k \leq J$, let $p_{e, k, j}$ denote
the probability of decoding the $k$th source at the $j$th receiver
\emph{incorrectly} conditioned on $k+1, k+2, \ldots, J$th sources being
decoded \emph{correctly}, and $\overline{p}_{e, k, j}$ the expectation
of $p_{e, k, j}$ over the ensemble of broadcast codes. The transition
probability of the \emph{effective} channel between $X_{k}$, $1 \leq k
\leq J$, and $Y_{j}$, $1 \leq j \leq J$, for $x_{k} \in \mathcal{X}_{k}$
and $y_{j} \in \mathcal{Y}_{j}$, is given by 
\begin{eqnarray*}
p_{Y_{j}|X_{k}}^{\prime}(y_{j}|x_{k}) &=& \sum_{x_{1}^{k-1},
y_{1}^{j-1}} \left( \prod_{t=k-1}^{1} Q_{t}(x_{t}|x_{t+1}) \right)
p(y_{1}|x_{1}) \left( \prod_{l=2}^{j} 
p_{l} \left( y_{l}|y_{l-1} \right) \right)
\end{eqnarray*} 
One can then consider $y_{j}$ as being produced by passing $x_{k}$
through a DMC with transition probability law
$p_{Y_{j}|X_{k}}^{\prime}(y_{j}|x_{k})$.  In the following
Theorem~\ref{th:upper bound}, we compute an upper bound on the expected
probability of the event $ \left\{ \hat{m}_{j, j} \neq m_{j} \right\}$. 
\begin{theorem}
\label{th:upper bound}
For $1 \leq j \leq J$, (i) $p \left( \left\{ \hat{m}_{j, j} \neq m_{j}
\right\} \right) \leq \sum_{k=j}^{J} p_{e, k, j}$ and, (ii) for $0 \leq
\rho \leq 1$, the expected probability $\overline{p}_{e, k, j|m}$ given
that the joint message $m$ is encoded is upper bounded as
\begin{eqnarray}
\overline{p}_{e, k, j|m} &\leq& \exp \left( 
-N E_{X_{k}, Y_{j}} (R_{k}) \right) \nonumber \\ 
E_{X_{k}, Y_{j}} (R_{k}) &=& E_{o, X_{k}, Y_{j}}(\rho) - \rho R_{k}
\nonumber \\ 
\label{eq:rate definition broadcast channel}
R_{k} &=& \frac{\ln M_{k}}{N} \\
E_{o, X_{k}, Y_{j}}(\rho) &=& \hspace{-0.25cm}
-\ln \sum_{x_{k+1}^{J}}
Q_{k+1}^{J} \left( x_{k+1}^{J} \right) 
\sum_{y_{j}} \left( \sum_{x_{k}} Q_{k} \left(
x_{k}|x_{k+1} \right) p^{\prime}_{Y_{j}|X_{k}} \left( y_{j}|x_{k}
\right)^{\frac{1}{1+\rho}} \right)^{1+\rho} \nonumber \\
\mbox{for $j \leq k \leq J-1$, and} &&  \nonumber \\
E_{o, X_{J}, Y_{j}}(\rho)  &=& -\ln
\sum_{y_{j}} \left( \sum_{x_{J}} Q_{J} \left( x_{J} \right)
p^{\prime}_{Y_{j}|X_{J}} \left( y_{j}|x_{J} \right)^{\frac{1}{1+\rho}}
\right)^{1+\rho}  \quad \mbox{for $k = J$} \nonumber
\end{eqnarray}
\Q
\end{theorem}
\begin{proof}
To prove Part (i), consider the joint ensemble formed by the random
vectors $\left( \hat{m}_{j, j}, \hat{m}_{j+1, j}, \ldots, \hat{m}_{J, j}
\right)$ and $(m_{j}, m_{j+1}, \ldots, m_{J})$.  Define the event
$E_{j}$ as the set of all sample points in this joint ensemble such that
$\hat{m}_{j, j} \neq m_{j}$.  We now show that $E_{j}$ can be expressed
as a union of $J-j+1$ mutually exclusive and collectively exhaustive
events.  For $j \leq l \leq J$, define the events $E_{j}(l)$ as follows:
$E_{j}(j) = \left\{ \hat{m}_{j, j} \neq m_{j}; \hat{m}_{k, j} = m_{k} \;
\mbox{for} \; j+1 \leq k \leq J \right\}$, and for $j+1 \leq l \leq
J-1$, $E_{j}(l) = \left\{ \hat{m}_{j, j} \neq m_{j}; \hat{m}_{l, j} \neq
m_{l}; \hat{m}_{k, j} = m_{k} \; \mbox{for} \; l+1 \leq k \leq J
\right\}$, and finally $E_{j}(J) = \left\{ \hat{m}_{j, j} \neq m_{j};
\hat{m}_{J, j} \neq  m_{J} \right\}$.  Then $E_{j} = \cup_{l=j}^{J}
E_{j}(l)$. But, for $j \leq l \leq J-1$, we have that $E_{j}(l)
\subset \left\{ \hat{m}_{l, j} \neq m_{l}; \hat{m}_{k, j} = m_{k} \;
\mbox{for} \; l+1 \leq k \leq J \right\}$, and for $l = J$, $E_{j}(J)
\subset \left\{ \hat{m}_{J, j} \neq  m_{J} \right\}$. Hence, for $j \leq
l \leq J-1$, we have
\begin{eqnarray*}
p \left( E_{j}(l) \right) &\leq& p \left( \left\{ \hat{m}_{l, j} \neq
m_{l}; \hat{m}_{k, j} = m_{k} \; \mbox{for} \; l+1 \leq k \leq J \right\}
\right) \\
&\leq& p \left( \left\{ \hat{m}_{l, j} \neq m_{l} | \hat{m}_{k, j} = m_{k} \;
\mbox{for} \; l+1 \leq k \leq J \right\} \right) \\
&=& p_{e, l, j},
\end{eqnarray*}
and for $l = J$, $p\left( E_{j}(J) \right) < p \left( \left\{
\hat{m}_{J, j} \neq  m_{J} \right\} \right) = p_{e, J, j}$.  Hence $p
\left( \left\{ \hat{m}_{j, j} \neq m_{j} \right\} \right) \leq
\sum_{k=j}^{J} p_{e, k, j}$.

Next we prove Part (ii). Proof of Part (ii) is a straightforward
extension of the proof given in~\cite{Gal-JRN-PIT}. To derive an upper
bound on $\overline{p}_{e, k, j|m}$, we first condition the event of
this type of error upon the code word $X_{m^{J}_{k+1}}^{(N)}$ chosen for
the message vector $m^{J}_{k+1}$. Let $p_{e, k, j} \left(
X_{m^{J}_{k+1}}^{(N)} \right)$ be the probability of this error event.
That is, $p_{e, k, j} \left( X_{m^{J}_{k+1}}^{(N)} \right)$ is the
probability that 
\[
p^{\prime}_{Y_{j}|X_{k}} \left( y_{j}^{(N)}
\left|X_{{m^{\prime}}_{k}^{J}}^{(N)} \right.  \right) \geq
p^{\prime}_{Y_{j}|X_{k}} \left( y_{j}^{(N)}
\left|X_{m_{k}^{J}}^{(N)} \right.  \right)
\]
for some ${m^{\prime}}_{k}^{J}$  such that ${m^{\prime}}_{k+1}^{J} =
m_{k+1}^{J}$ and $m^{\prime}_{k} \neq m_{k}$ in the conditional ensemble,
and $X_{m_{k}^{J}}^{(N)}$ is independently chosen with the probability
assignment $Q_{k}\left( X_{m_{k}^{J}}^{(N)} \left| X_{m_{k+1}
^{J}}^{(N)} \right. \right)$. The coding Theorem 5.6.1~\cite{Gal-BOOK}
applies to this situation, yielding
\begin{eqnarray*}
p_{e, k, j} \left( X_{m^{J}_{k+1}}^{(N)} \right) &\leq&
\left(M_{k}-1 \right)^{\rho} \prod_{n=1}^{N} \sum_{y_{j}} \left(
\sum_{x_{k}} Q_{k} \left(x_{k}|x_{k+1}\right) 
p_{Y_{j}|X_{k}}^{\prime}(y_{j}|x_{k})^{\frac{1}{1+\rho}}
\right)^{1+\rho}
\end{eqnarray*}
Next, $\overline{p}_{e, k, j|m}$ is the expected value of $p_{e, k, j}
\left( X_{m^{J}_{k+1}}^{(N)} \right)$ over $m^{J}_{k+1}$ and
$X_{m^{J}_{k+1}}^{(N)}$. Since the bound is independent of $m^{J}_{k+1}$,
we average only over $X_{m^{J}_{k+1}}^{(N)}$.
\begin{eqnarray*}
\overline{p}_{e, k, j|m} &=& \sum p_{e, k, j} \left( X_{m^{J}_{k+1}}^{(N)} \right)
Q_{k+1}^{J} \left( X_{m^{J}_{k+1}}^{(N)} \right) \\
&\leq & \left(M_{k}-1 \right)^{\rho} \left[ \sum_{x_{k+1}^{J}} Q_{k+1}^{J} 
\left( x_{k+1}^{J}\right) \sum_{y_{j}} \left( \sum_{x_{k}} Q_{k} 
\left(x_{k}|x_{k+1}\right) 
p_{Y_{j}|X_{k}}^{\prime}(y_{j}|x_{k})^{\frac{1}{1+\rho}}
\right)^{1+\rho} \right]^{N}
\end{eqnarray*}
For $k = J$, $\overline{p}_{e, J, j |m}$ is the expected probability of
error in decoding source-$J$ when it is communicated over a discrete
memoryless channel with transition probability law
$p_{Y_{j}|X_{J}}^{\prime}(y_{j}|x_{J})$ and channel input distribution
$Q_{J}$. Thus, the probability of decoding error is bounded above by the
usual results for decoding on a DMC (Theorem 5.6.1~\cite{Gal-BOOK}).
\end{proof}

\section{The Queuing-Theoretic Model}
The queuing-theoretic model for a $J$ receiver degraded broadcast
channel that we derive is similar to the queuing-theoretic model we
derived for the $J$ source multiaccess channel with joint
maximum-likelihood decoding for the scenario ({\bf S1}) in
Chapter~\ref{ch:introduction}.
The similarity can be seen as follows: we maintained a queue for each
source in the case of the multiaccess channel, whereas we maintain a
queue for each receiver at the transmitter in the case of the degraded
broadcast channel. Hence, messages that arrive at the transmitter and
are intended for receiver-$j$ are put into queue $j$. For $\mathsf{K}
\geq 1$, let $\mathcal{S}_{\mathsf{K}}$ (as defined in
Chapter~\ref{ch:chapter1}) be the set of schedules that encode at most
$\mathsf{K}$ messages for transmission.

Under the scenario {\bf (S3)}, a schedule $s \in
\mathcal{S}_{\mathsf{K}}$ defines product message alphabets
$\mathcal{M}_{j}(s) = \left\{1, 2, \ldots, M_{j}^{s_{j}} \right\}$ such
that $s_{j} \neq 0$ for each of the $J$ receivers. Hence we need to
redefine the coding rate $R_{k}$ (Eq.~(\ref{eq:rate definition broadcast
channel}) in Theorem~\ref{th:upper bound}) for receiver-$k$ as $R_{k}(s)
= \frac{s_{k} \ln M_{k}}{N(s)}$, thus emphasizing the dependence of
\emph{effective} message alphabet size on schedule $s$.  Let $\chi_{j}
\left( s, N_{j}(s) \right)$ denote the random coding upper bound
$\sum_{k=j}^{J} \exp \left( -N_{j}(s) E_{X_{k}, Y_{j}} (R_{k}(s))
\right)$ for the $j$th receiver under the schedule $s$, and $\{ p_{ej};
1 \leq j \leq J \}$ the set of tolerable message decoding error
probabilities.  For $1 \leq j \leq J$ and $s \in
\mathcal{S}_{\mathsf{K}}$ such that $s_{j} > 0$, define $N_{j}(s)$ to be
the smallest positive integer such that $\chi_{j}(s, N_{j}(s)) \leq
p_{ej}$. In the following Lemma~\ref{lemma:bounds on N(s) superposition
encoding}, we derive an upper bound and a lower bound on $N_{j}(s)$.
\begin{lemma} 
\label{lemma:bounds on N(s) superposition encoding}
\begin{eqnarray*} \max_{j \leq k \leq J} \frac{\left\lceil -\ln p_{ej} +
\rho s_{k} \ln M_{k} \right\rceil_{E_{o, X_{k},Y_{j}}}}{E_{o,
X_{k},Y_{j}}} \leq N_{j}(s) \leq \max_{j \leq k \leq J} \frac{\lceil
-\ln \frac{p_{ej}}{J-j+1} + \rho s_{k} \ln M_{k} \rceil_{E_{o,
X_{k},Y_{j}}}} {E_{o, X_{k},Y_{j}}} 
\end{eqnarray*} \Q 
\end{lemma}
\begin{proof} The arguments leading to the above bounding are similar to
the arguments given in the proof of Lemma~\ref{lemma:bounds on N(s)
joint decoding}.  Hence we skip the detailed proof.  
\end{proof}
\begin{lemma} Let $s^{\prime} \preceq s$.  Then $N_{j}\left( s^{\prime}
\right) \leq N_{j}(s)$ for $1 \leq j \leq J$. \Q 
\end{lemma}
\begin{proof} 
Since $s_{k}^{\prime} \leq s_{k}$, we first note that
$R_{k} \left( s_{k}^{\prime} \right) \leq R_{k}(s_{k})$. Since, $E_{o,
X_{k}, Y_{j}}(\rho)$ is independent of $M_{k}$ and $s_{k}$, we conclude
that $E_{X_{k}, Y_{j}} \left(R_{k} \left( s_{k}^{\prime}\right)\right)
\geq E_{X_{k}, Y_{j}}\left(R_{k} (s_{k}) \right) $.  Now, we observe
that 
\[ p_{ej} \geq \sum_{k=j}^{J} \exp \left( -N_{j}(s) E_{X_{k},
Y_{j}}\left(R_{k} (s_{k}) \right) \right) \geq \sum_{k=j}^{J} \exp
\left( -N_{j}(s) E_{X_{k}, Y_{j}}\left(R_{k} (s_{k}^{\prime}) \right)
\right) 
\] 
Thus, $\chi_{j} \left( s^{\prime}, N_{j}(s) \right) \leq
p_{ej}$.  Since $N_{j}(s^{\prime})$ is the smallest positive integer $N$
such that \newline $\chi_{j} \left( s^{\prime}, N_{j} \left( s^{\prime} \right)
\right) \leq p_{ej}$, we have that $N_{j} \left(s^{\prime}\right) \leq
N_{j}(s)$.  
\end{proof} 
Define $N(s) = \max_{j} N_{j}(s)$. Then $N(s)$ is the smallest positive
integer such that $\chi_{j} \left(s, N(s) \right) \leq p_{ej}$ for $1
\leq j \leq J$.  

At this point we should observe that, at the beginning of each
time-slot, we need to inform the receivers about the schedule $s$ that
will be implemented in that time-slot. This is achieved by assuming that
synchronized common randomness is available at the transmitter and
receivers to generate schedules with the distribution $p^{\omega}$ (and
also the code books). Then, only those receivers-$j$ such that $s_{j} >
0$ will decode their respective received signals. But, in a particular
time-slot, it may happen that a schedule $s$ is chosen for transmission
and enough messages of each class required by the schedule $s$ are
\emph{not} present  in the system.  To resolve this problem, we can
substitute each such ``missing message'' by a message with null value,
thus embedding control information in information from sources.
Inclusion of the null message in $\mathcal{M}_{j}$ increases the
cardinality $M_{j}$ by one and may have the effect of increasing $N(s)$
accordingly, thus reducing the throughputs achievable for finite message
lengths. But this effect disappears in the asymptotic limit $M_{j}
\rightarrow \infty$.  

However, in the following we assume that this control information is
passed to the receivers over an error-free control channel, so that the
queueing model analysis presented in Chapter~\ref{ch:joint decoding}
applies in the present context without modifications.

Define the service requirement $N(s)$ of a message under schedule $s$,
and the service quantum available to queue $j$ at a discrete-time
instant, as in Chapter~\ref{ch:joint decoding}.  Then (i) the notion of
rate vectors $\left\{r(s); s \in \mathcal{S}_{\mathsf{K}}\right\}$ and
the outer bound $\mathcal{R}_{out}$ derived in
Section~\ref{section:queueing theoretic model joint decoding} on the
stability region of message arrival rate vectors $\mathds{E}A$
achievable by stationary scheduling policies, and (ii) the definition of
state-independent scheduling policies and their stability analysis
described in Section~\ref{section:stability joint decoding} and the
following Corollary~\ref{corollary:inner bound equals outer bound}
therein, apply verbatim to the queueing model for the degraded broadcast
channel.

\section{Information-Theoretic Interpretation to the Stability Region}
For a fixed state-independent schedule $s \in \mathcal{S}_{\mathsf{K}}$,
i.e., $p^{\omega}(s) = 1$, we know from Theorem~\ref{lemma:joint
decoding} and Theorem~\ref{lemma:joint decoding transience} that the
queueing model is stable if $\mathds{E}\tilde{A}_{j} < R_{j}(s)$ for $1
\leq j \leq J$, and unstable if $\mathds{E}\tilde{A}_{j} > R_{j}(s)$ for
at least one queue $j$.

In this section, we give the information-theoretic interpretation to the
stability region of nat arrival rate vectors $\mathds{E}\tilde{A}$ for
the scenario ({\bf S3}). A formal statement of this interpretation is
made in Theorem~\ref{th:capacity interpretation broadcast channel}.  For $s
\in \mathcal{S}_{\mathsf{K}}$, define the code rate vector $R(s) =
\left(R_{1}(s), R_{2}(s), \ldots, R_{J}(s) \right)$.  In
Theorem~\ref{th:capacity interpretation broadcast channel}, we show the
following: (i) for a given joint probability distributions $Q$, and
message arrival processes $\left\{ A_{j}; 1 \leq j \leq J \right\}$ such
that $\mathds{E} \tilde{A} = r \in \mathcal{I}^{o}(Q)$, we determine a
schedule $s$, message alphabet size vector $M$, and a value for the
parameter $\rho$ such that the message communication system for $s$,
$M$, $\rho$, and the arrival processes $\left\{ A_{j}; 1 \leq j \leq J
\right\}$ , is stable (i.e., $R_{j}(s) > r_{j}$, $1 \leq j \leq J$);
(ii) for any $s$, $M$, and $\rho$, we show that $R(s) \in
\mathcal{I}^{o}(Q)$. Define $\mathcal{R}(Q) = \left\{ R(s): 0 < \rho
\leq 1; \mathsf{K} \geq 1; s \in \mathcal{S}_{\mathsf{K}}; M \in
\mathbb{Z}_{+}^{J} \right\}$ to be the set of all possible code rate
vectors $R(s)$.
\begin{theorem}[Information-Theoretic Interpretation]
\label{th:capacity interpretation broadcast channel}
\begin{eqnarray*}
\mathcal{R}(Q) &=& \mathcal{I}^{o}(Q)
\end{eqnarray*} \Q
\end{theorem}
\begin{proof}
We first show that $\mathcal{I}^{o}(Q) \subset \mathcal{R}(Q)
$.  Choose an $r  \in \mathcal{I}^{o}(Q)$. Then there exists an
$\epsilon > 0$ such that $r+\epsilon = (r_{1}+\epsilon, r_{2}+\epsilon,
\ldots, r_{J}+\epsilon) \in \mathcal{I}^{o}(Q)$. For $1 \leq j \leq J$
and a positive real number $\mathsf{A}$, let us first choose $s_{j}$ and
$M_{j}$ as real numbers such that the product $s_{j}\ln M_{j} =
\mathsf{A}(r_{j}+\epsilon)$.  From Lemma~\ref{lemma:bounds on N(s)
superposition encoding},
\begin{eqnarray*}
\min_{1 \leq j \leq J} \min_{j \leq k \leq J}
\frac{s_{i} (\ln M_{i}) E_{o, X_{k}, Y_{j}}}
{ \left\lceil -\ln \frac{p_{ej}}{J-j+1} + \rho
s_{k} \ln M_{k} \right\rceil_{E_{o, X_{k}, Y_{j}}}}
&\leq& R_{i}(s) \\
& \leq &
\min_{1 \leq j \leq J} \min_{j \leq k \leq J}
\frac{s_{i} (\ln M_{i}) E_{o, X_{k}, Y_{j}}} { \left\lceil -\ln
p_{ej}  + \rho s_{k} \ln M_{k}
\right\rceil_{E_{o, X_{k}, Y_{j}}}}
\end{eqnarray*}
We can see that
\begin{eqnarray*}
\lim_{\rho \rightarrow 0} \lim_{\mathsf{A} \rightarrow \infty}
R_{i}(s) &=&
\lim_{\rho \rightarrow 0} \lim_{\mathsf{A} \rightarrow \infty}
\min_{1 \leq j \leq J} \min_{j \leq k \leq J}
\frac{\mathsf{A} (r_{i} + \epsilon) E_{o, X_{k}, Y_{j}}}
{ \left\lceil -\ln p_{ej} + \rho 
\mathsf{A} (r_{k}+\epsilon)  \right\rceil_{E_{o, X_{k}, Y_{j}}}} \\
&=& \lim_{\rho \rightarrow 0} \lim_{\mathsf{A} \rightarrow \infty}
\min_{1 \leq j \leq J} \min_{j \leq k \leq J}
\frac{\mathsf{A} (r_{i}+\epsilon) E_{o, X_{k}, Y_{j}}} { \left\lceil -\ln
\frac{p_{ej}}{J-j+1}  + \rho \mathsf{A} (r_{k}
+ \epsilon)
\right\rceil_{E_{o, X_{k}, Y_{j}}}} \\
&=& \lim_{\rho \rightarrow 0} \min_{1 \leq j \leq J} \min_{j \leq k \leq J}
\frac{r_{i} + \epsilon}{r_{k}+\epsilon} \frac{E_{o, X_{k}, Y_{j}}}{\rho} \\
&\stackrel{(a)}{=}& \min_{1 \leq j \leq J} \min_{j \leq k \leq J} 
\frac{r_{i} + \epsilon}{r_{k} + \epsilon}
I \left( X_{k};Y_{j}|X_{k+1} \right) \\
&\stackrel{(b)}{>}& (r_{i} + \epsilon) \min_{1 \leq j \leq J} 
\min_{j \leq k \leq J} \frac{I\left( X_{k};Y_{j}|X_{k+1} \right)}
{I\left( X_{k};Y_{k}|X_{k+1} \right)} \\
&\stackrel{(c)}{\geq}& r_{i} + \epsilon,
\end{eqnarray*}
where $(a)$ follows from Part (i) of Lemma~\ref{lemma:DBC appendix B},
$(b)$ follows from the fact that $r + \epsilon  \in \mathcal{I}^{o}(Q)$
and hence $r_{k} + \epsilon  < I \left(X_{k}; Y_{k}| X_{k+1} \right)$,
and $(c)$ follows from the data processing inequality (Theorem 2.8.1
in~\cite{CovTho-BOOK}) applied to the Markov chain $X_{J} \rightarrow
\cdots X_{1} \rightarrow Y_{1} \cdots \rightarrow Y_{J}$.  Denote by
$\lim_{\rho \rightarrow 0} \lim_{\mathsf{A} \rightarrow \infty} R_{k}(s)
= R^{*}(s)$.

Choose two positive real numbers $\delta_{i}$ and $\delta_{i}^{\prime}$
such that $\epsilon - \delta_{i}
- \delta_{i}^{\prime} > 0$. Then there exists a $\rho \left( \delta_{i}
  \right) < 1$ such that for all $0 < \rho < \rho \left( \delta_{i}
\right)$, we have $\lim_{\mathsf{A} \rightarrow \infty} R_{i}(s) >
R^{*}(s) - \delta_{i} > r_{i} + \epsilon - \delta_{i}$. Now, for a fixed
value $\rho_{i}$ for $\rho$ such that $\rho_{i} < \rho \left( \delta_{i}
\right)$, there exists a $\mathsf{A} \left( \rho_{i},
\delta_{i}^{\prime} \right)$ such that for all $\mathsf{A} > \mathsf{A}
\left( \rho_{i}, \delta_{i}^{\prime} \right)$, we have $R_{i}(s) >
\lim_{\mathsf{A} \rightarrow \infty} R_{i}(s) - \delta_{i}^{\prime} >
r_{i} + \epsilon - \delta_{i} - \delta_{i}^{\prime} > r_{i}$. Choose an
$\mathsf{A}_{i}$ for $\mathsf{A}$ such that $\mathsf{A}_{i} >
\mathsf{A}_{i} \left( \rho_{k}, \delta_{i}^{\prime} \right)$. Define
$\mathsf{A}^{*} = \max_{i} \mathsf{A}_{i}$ and $\rho^{*} = \min_{i}
\rho_{i}$.

Since $s_{j}$ and $M_{j}$ for $1 \leq j \leq J$ have to be positive
integers, one can, for a given $\mathsf{A}^{*}$ choose $s_{j} =
\left\lceil \frac{\mathsf{A}^{*}(r_{j}+\epsilon)}{\ln M_{j}}
\right\rceil$ for a given $M_{j}$, and $M_{j} = \left\lceil \exp\left(
\frac{\mathsf{A}^{*}(r_{j}+\epsilon)}{s_{j}} \right) \right\rceil$ for a
given $s_{j}$, and still have the same limit as above.

Next, we prove $\mathcal{R}(Q) \subset \mathcal{I}^{o}(Q)$ by showing
that $R(s) \in \mathcal{I}^{o}(Q)$ for each $s$, $\rho$, and $M$.  From
Lemma~\ref{lemma:bounds on N(s) superposition encoding},
\begin{eqnarray*}
R_{i}(s) & \leq & \min_{1 \leq j \leq J} \min_{j \leq k \leq J}
\frac{s_{i} (\ln M_{i}) \; E_{0, X_{k}, Y_{j}}} { \left\lceil -\ln
p_{ej}  + \rho s_{k} \ln M_{k} 
\right\rceil_{E_{0, X_{k}, Y_{j}}}} \\
& \leq & \min_{1 \leq j \leq J} 
\frac{s_{i} (\ln M_{i}) \; E_{0, X_{j}, Y_{j}} } {\rho s_{j} \ln M_{j}}
\leq \frac{E_{0, X_{i}, Y_{i}}}{\rho} \\
& \stackrel{(d)}{<} & \left\{
\begin{array}{ll}
I \left( X_{i}; Y_{i} | X_{i+1} \right) & \mbox{for} \quad 
1 \leq i \leq J-1 \\
I \left( X_{J}; Y_{J} \right) & \mbox{for} \quad i = J
\end{array}
\right.
\end{eqnarray*}
where $(d)$ follows from Part (ii) of Lemma~\ref{lemma:DBC appendix B}.
Thus, $R(s) \in \mathcal{I}^{o}(Q)$ for each $s$, $\rho$, and $M$.
\end{proof}

%% file: conclusion.tex
\chapter{Conclusion}
We have developed a unified framework, namely, multiclass discrete-time
processor-sharing queueing model of Chapter~\ref{ch:chapter1}, to
analyze stability of scheduled message communication over multiaccess
channels with either independent decoding or joint decoding, and over
degraded broadcast channels.  Under this framework, we modeled both the
random message arrivals and the subsequent reliable communication by
suitably combining techniques from queueing theory and information
theory.

For scheduled message communication over a multiaccess channel with
independent maximum-likelihood decoding, we showed the following.
\begin{enumerate}
\item   For finite message lengths, inner bounds and outer bounds to the
message arrival rate stability region are derived. For arrival rates
within the inner bounds, we show finiteness of the stationary mean for
the number of messages in the system and hence for message delay. For
the case of equal received signal powers, with sufficiently large SNR,
the stability threshold increases with decreasing maximum number of
simultaneous transmissions (see Fig.~\ref{fig:equal powers finite
message length}).

\item   When message lengths are large, the information arrival rate
stability region has an interpretation in terms of interference-limited
information-theoretic capacities. For the case of equal received powers,
this stability threshold is the interference-limited
information-theoretic capacity.

\item   We propose a class of stationary policies called
state-independent scheduling policies, and then show that they achieve
this asymptotic information arrival rate stability region.

\item   In the asymptotic limit corresponding to immediate access, the
stability region for Gaussian encoding and non-idling scheduling
policies is shown to be identical irrespective of received signal
powers. This observation essentially shows that transmit power control
is not needed.  We show that, in the asymptotic limit corresponding to
immediate access and large message lengths, a spectral efficiency of 1
nat/s/Hz is achievable with non-idling scheduling policies (see
Fig~\ref{fig:equal powers infinite message length}).
\end{enumerate}

For scheduled message communication over  multiaccess channels with
joint maximum-likelihood decoding, we derived an outer bound to the
stability region of message arrival rate vectors achievable by the class
of stationary scheduling policies.  Then we showed for any message
arrival rate vector that satisfies the outer bound, that there exists a
stationary ``state-independent'' policy that results in a stable system
for the corresponding message arrival processes. Finally, we showed that
for any achievable rate vector in the capacity region of a multiaccess
channel, there exists a scheduling strategy and message lengths for that
rate vector such that the message system with random message arrivals is
stable.

We showed that the queueing model derived in the case of multiaccess
channels with joint maximum-likelihood decoding can be used to model
scheduled message communication over degraded broadcast channels with
superposition encoding and successive decoding. We then showed that the
results obtained from stability analysis of multiaccess channels apply
verbatim to the degraded broadcast channels. We also showed that for any
achievable rate vector in the capacity region of a degraded broadcast
channel , there exists a scheduling strategy and message lengths for
that rate vector such that the broadcast message system with random
message arrivals is stable.

%% file: Appendix-A.tex
\chapter{Drift theorems for Positive Recurrence and Transience}
\label{appendix:drift theorems}
Drift theorems for classification (in terms of transience, positive recurrence 
and null recurrence) of discrete-time Markov chains taking values in a
general state space have been stated in ~\cite{MeyTwe-BOOK}. We rewrite the
theorems here for discrete-time Markov chains taking
values in a countable state space.
\par
\indent
~\cite{MeyTwe-BOOK} defines a measure, $\psi$, on the state space, $X$, which is
called the `maximal irreducibility measure'. For an irreducible Markov
chain taking values in a countable state space, the measure $\psi$ is 
generated by a
counting measure on $X$ [p. 88, ~\cite{MeyTwe-BOOK}]. Hence, an irreducible
Markov chain taking values in a countable state space, $X$, is
$\psi$-irreducible with $\psi(\alpha)=1 \quad \forall \alpha \in X$, where
$\alpha$ denotes a state in the countable state space.
\par
\indent
~\cite{MeyTwe-BOOK} defines the set, $B(X)$, [p. 55, ~\cite{MeyTwe-BOOK}] as some 
$\sigma$-algebra on the general state space, $X$. 
For a countable state space, without loss of
generality, we take this $\sigma$-algebra as the set of all subsets of $X$.
Then, $B^{+}(X)$ defined as $B^{+}(X)=\{A\in B(X) : \psi(A)>0\}$, in the
case of irreducible Markov chains on a countable state space, becomes the
set of all non-empty subsets of $X$, i.e., $B^+(X)=B(X)-\phi$.  
\par
\indent
We now rewrite the theorems stated in ~\cite{MeyTwe-BOOK} for discrete-time Markov 
chains taking values in a countable state space.
\section{Theorem for Positive Recurrence}
Theorem 11.3.4  stated in ~\cite{MeyTwe-BOOK} can be written as follows. Suppose 
$C \in B(X)$ and $C$ is petite, and an everywhere finite function 
$V:X \rightarrow [0,\infty)$ satisfies \\ $\Delta V(x)
\footnote{$\Delta V(x)$ is the expected drift of the function, $V$, in state 
$x$.} \le -1 + b\ 1_C(x),\ x\in X,\ b<\infty$, then 
$\mathbf \Phi$ is positive Harris recurrent.   
\par
\indent
Finite sets in a countable state space are petite [p.192, ~\cite{MeyTwe-BOOK}].
A chain ${\mathbf \Phi}$ is Harris recurrent if every set in $B(X)$ is Harris
recurrent [p.200, ~\cite{MeyTwe-BOOK}]. 
If a set is Harris recurrent then it is recurrent [p.201, ~\cite{MeyTwe-BOOK}].
Positivity for a $\psi$-irreducible Markov chain means existence of an
invariant probability measure for the chain [p.230, ~\cite{MeyTwe-BOOK}]. Thus,
for an irreducible Markov chain on a countable state space positive Harris
recurrence implies positive recurrence. 
The theorem stated above, together with the interpretations made before, can 
then be written in the context of Markov chains in a countable state space as 
follows. 
\par
\indent
An irreducible Markov chain, $\mathbf \Phi$, is positive recurrent
if there exists a finite subset $C$ of $X$ and an everywhere finite function 
$V:X \rightarrow [0,\infty)$ bounded on $C$ such that \\
a. $\Delta V(x)\le -1,\ x\in C^c$, and \\
b. $\Delta V(x)\le -1+b, b<\infty,\mbox{ for } x\in C$.

\section{Theorem for Transience}
Theorem 8.0.2 (i) stated in ~\cite{MeyTwe-BOOK} is as follows : Suppose $\Phi$ is
a $\psi$-irreducible chain. The chain, $\Phi$, is transient iff there exists
a bounded non-negative function, $V$, and a set $C \in B^{+}(X)$ such that
for all $x \in C^c$, $\Delta V(x) \ge 0$, and $D=\{x:\ V(x) > \sup_{y \in C}
V(y)\} \in B^{+}(X)$.  
\par
\indent
Thus, the theorem stated above, together with the interpretations made before, 
can be written in the context of Markov chains in a countable state space as
follows :
\par
\indent
An irreducible Markov chain, $\mathbf \Phi$, is transient iff there exists a
bounded non-negative function, $V$, and a non-empty set $C \subset X$ such
that for all $x \in C^c$, $\Delta V(x) \ge 0$, and $\exists x \in C^c$ such
that $V(x)> \sup_{y\in C}V(y)$.

%% file: Appendix-B.tex
\chapter{Two properties of $E_{o, S}(\rho, Q)$ and $E_{o, X_{k},
Y_{j}}$}

In this appendix, we state two properties of $E_{o, S}(\rho, Q)$ and
$E_{o, X_{k}, Y_{j}}$ used in the random coding upper bounds on expected
decoding error probabilities for joint maximum-likelihood decoding for
the multiaccess channel (Chapter~\ref{ch:joint decoding}) and the
degraded broadcast channel (Chapter~\ref{ch:degraded broadcast
channel}), respectively. These properties are derived by a straightforward
application of Theorem 5.6.3. in~\cite{Gal-BOOK} for the respective
communication channels.

For a finite set $\mathcal{Z}$, define a random variable $Z$ that takes
values in the set $\mathcal{Z}$ with the probability distribution $Q_{Z}
= \{Q_{Z}(z); z \in \mathcal{Z}\}$. Let $X$ and $Y$ be the input and
output of a DMC, and for $z \in \mathcal{Z}$  consider the input
distributions $Q_{X}^{z} = \{Q_{X}^{z}(x); x \in \mathcal{X}\}$ and the
transition probability law $\{p^{z}(y|x); x \in \mathcal{X}, y \in
\mathcal{Y}\}$.  Then $\left\{p^{z}(y); y \in \mathcal{Y} \right\}$ is
the probability distribution induced on the output alphabet
$\mathcal{Y}$.  Define $g(\rho, z) = \sum_{y} \left( \sum_{x}
Q_{X}^{z}(x) p^{z}(y|x)^{\frac{1}{1+\rho}} \right)^{1+\rho}$, and then
$E_{o}(\rho) = -\ln \sum_{z} Q_{Z}(z) g(\rho, z)$.
\begin{lemma}
\label{lemma:error exponent appendix B}
$\lim_{\rho \rightarrow 0} \frac{E_{o}(\rho)}{\rho} = I(X; Y|Z)$, and
$\frac{E_{o}(\rho)}{\rho}$ is a decreasing function for $\rho \in (0,
1]$.  \Q
\end{lemma}
\begin{proof}
Define $G(\rho) = \exp \left( -E_{o}(\rho) \right) = \sum_{z} Q_{Z}(z)
g(\rho, z)$. It is easy to observe that $g(0, z) = 0$. We have
\begin{eqnarray*}
\frac{\partial g(\rho, z)}{\partial \rho} &=& \sum_{y} \left(
\left[ \sum_{x} Q_{X}^{z}(x)
p^{z}(y|x)^{\frac{1}{1+\rho}} \right]^{1+\rho} \left\{
\ln \sum_{x} Q_{X}^{z}(x)
p^{z}(y|x)^{\frac{1}{1+\rho}} \right. \right.\\
&& + \left. \left. \frac{1+\rho}{\sum_{x} Q_{X}^{z}(x)
p^{z}(y|x)^{\frac{1}{1+\rho}}} \left( \sum_{x} Q_{X}^{z}(x)
p^{z}(y|x)^{\frac{1}{1+\rho}} \frac{-1}{(1+\rho)^{2}} \ln p^{z}(y|x) \right)
\right\} \right) \\
\left. \frac{\partial g(\rho, z)}{\partial \rho} \right|_{\rho = 0} 
&=& \sum_{y} p^{z}(y) \ln  p^{z}(y) + \sum_{y} \sum_{x} Q_{X}^{z}(x)
p^{z}(y|x) \ln \frac{1}{p^{z}(y|x)} \\
&=& -H(Y|Z=z ) + H(Y|X, Z=z) \\
&=& -I(X; Y|Z=z), \quad \mbox{and}
\end{eqnarray*}
\[
\left. \frac{dG(\rho)}{d\rho} \right|_{\rho = 0} = \sum_{z} Q_{Z}(z) 
\left.  \frac{\partial g (\rho, z)}{\partial \rho} \right|_{\rho = 0} = - 
\sum_{z} Q_{Z}(z) I(X; Y|Z=z) = - I(X; Y|Z)
\]
Since $E_{o}(0) = 0$, and $\frac{dG(\rho) } {d\rho} = -
\frac{dE_{o}(\rho)}{d\rho} \exp \left( -E_{o}(\rho) \right)$, we
therefore have $\left.  \frac{dE_{o}(\rho)}{d\rho} \right|_{\rho = 0} =
I (X; Y|Z)$.

Define $f(\rho) = \frac{E_{o}(\rho)}{\rho}$ for $\rho \in (0, 1]$. Then
$\frac{df(\rho)}{d\rho} = \frac{\rho \frac{dE_{o}(\rho)}{d\rho} -
E_{o}(\rho)}{\rho^{2}}$. Now, define $v(\rho) = \rho
\frac{dE_{o}(\rho)}{d\rho} - E_{o}(\rho)$ for $\rho \in [0, 1]$. Then we
can see that $v(0) = 0$, and $\frac{dg(\rho)}{d\rho} = \rho
\frac{d^{2}E_{o}(\rho)}{d\rho^{2}} \leq 0$ for $\rho \geq 0$ (Theorem
5.6.3 in~\cite{Gal-BOOK}). So we conclude that $v(\rho)$ is a decreasing
function in $\rho$, and since $v(0) = 0$ we have that $v(\rho) < 0$ for
$\rho \in (0, 1]$. Equivalently, $\frac{df(\rho)}{d\rho} < 0$ for $\rho
\in (0, 1]$. This establishes that $f(\rho)$ is a decreasing function in
$\rho$ and that $\frac{E_{o}(\rho)}{\rho} < I(X;Y|Z)$ for $\rho \in (0, 1]$.
\end{proof}

We now state the following two Lemmas. Lemma~\ref{lemma:MAC appendix B}
results when Lemma~\ref{lemma:error exponent appendix B} is applied to
$E_{0,S}(\rho, Q)$ in Theorem~\ref{th:Slepian-Wolf}.  Part (i) and (ii)
of Lemma~\ref{lemma:DBC appendix B} result if we apply
Lemma~\ref{lemma:error exponent appendix B} to $E_{o, X_{k}, Y_{j}}$ in
Theorem~\ref{th:upper bound}.
\begin{lemma}
\label{lemma:MAC appendix B}
Consider a $J$ source multiple-access channel with joint
maximum-likelihood decoding. Then (i) $\lim_{\rho \rightarrow 0}
\frac{E_{o, S}(\rho, Q)}{\rho} = I \left( X(S);Y|X\left(S^{c}\right)
\right)$ for $S \in \mathcal{P}(\mathcal{J})$, and (ii) $\frac{E_{o,
S}(\rho, Q)}{\rho} < I \left( X(S);Y|X\left(S^{c}\right) \right)$ for
$\rho \in (0, 1]$.          \Q
\end{lemma}
\begin{lemma}
\label{lemma:DBC appendix B}
Consider a $J$-receiver degraded broadcast channel represented as the
Markov chain $X_{J} \rightarrow X_{J-1} \rightarrow \cdots \rightarrow
X_{1} \rightarrow Y_{1} \rightarrow Y_{2} \rightarrow \cdots \rightarrow
Y_{J}$.  Then (i) $\lim_{\rho \rightarrow 0} \frac{E_{o, X_{k},
Y_{j}}}{\rho} = I (X_{k}; Y_{j}|X_{k+1}X_{k+2} \ldots X_{J}) = I (X_{k};
Y_{j}|X_{k+1})$, (ii) $\frac{E_{o, X_{k},Y_{j}}}{\rho} < I (X_{k};
Y_{j}|X_{k+1})$ for $\rho \in (0, 1]$. 
\Q
\end{lemma}

%